\documentclass[12pt]{article}

\usepackage{color}
\usepackage{amsmath,amssymb}
\usepackage{latexsym}
\usepackage[dvips]{graphicx,psfrag}
\usepackage{cite}
\usepackage{wrapfig}

\allowdisplaybreaks[1]

\newcommand{\eq}[1]{(\ref{#1})}
\newcommand{\nn}{\nonumber}
\newcommand{\ds}{\displaystyle}

\newcommand{\vev}[1]{\left\langle #1 \right\rangle}

\newcommand{\del}{\partial}

\newcommand{\rmq}{{\rm q}}

\newcommand{\bP}{\boldsymbol{P}}
\newcommand{\bQ}{\boldsymbol{Q}}

\newcommand{\bL}{\boldsymbol{L}}

\newcommand{\bGamma}{\boldsymbol{\Gamma}}

\DeclareMathOperator{\tr}{tr}

\DeclareMathOperator{\diag}{diag}

\renewcommand{\thefootnote}{\fnsymbol{footnote}}

\makeatletter
\@addtoreset{equation}{section}

\begin{document}


\begin{titlepage}
\thispagestyle{empty} 
\begin{flushright}
arXiv:0909.1197 \\
\vspace{0.2cm}
September 2009
\end{flushright}

\vspace{2.3cm}

\begin{center}
\noindent{\Large\textbf{
Macroscopic Loop Amplitudes \\ 
\vspace{.3cm}
in the Multi-Cut Two-Matrix Models 
}}
\end{center}

\vspace{1cm}

\begin{center}
\noindent{Chuan-Tsung Chan\footnote{ctchan@thu.edu.tw}$^{,p}$, 
Hirotaka Irie\footnote{irie@phys.ntu.edu.tw}$^{,q}$, 
Sheng-Yu Darren Shih\footnote{s.y.darren.shih@berkeley.edu}$^{,q,r}$ and 
Chi-Hsien Yeh\footnote{d95222008@ntu.edu.tw}$^{,q}$}
\end{center}
\vspace{0.5cm}
\begin{center}
{\it 
$^{p}$Department of Physics, Tunghai University, Taiwan, 40704\\
\vspace{.3cm}
$^{q}$Department of Physics and Center for Theoretical Sciences, \\
National Taiwan University, Taipei 10617, Taiwan, R.O.C \\
\vspace{.3cm}
$^{r}$Department of Physics, University of California, Berkeley, CA 94720-7300%
\footnote{Address after Sept.~1, 2009.}
}
\end{center}

\vspace{1.5cm}

\begin{abstract}
Multi-cut critical points and their macroscopic loop amplitudes are studied 
in the multi-cut two-matrix models, 
based on an extension of the prescription developed by Daul, Kazakov and Kostov. 
After identifying possible critical points and potentials in the multi-cut matrix models, 
we calculate the macroscopic loop amplitudes in the $\mathbb Z_k$ symmetric background. 
With a natural large $N$ ansatz for the matrix Lax operators, 
a sequence of new solutions for the amplitudes in 
the $\mathbb Z_k$ symmetric $k$-cut two-matrix models are obtained, 
which are realized by the Jacobi polynomials. 
\end{abstract}

\end{titlepage}

\newpage

\renewcommand{\thefootnote}{\arabic{footnote}}
\setcounter{footnote}{0}


\section{Introduction and summary}

Non-critical string theory \cite{Polyakov,KPZ,DDK} provides a simple and tractable toy model 
to understand various aspects of string theory. 
One of the fascinating formulation of the theory is given by the description with the matrix model. 
This realizes a non-perturbative definition of string theory \cite{DSL,TwoMatString,GrossMigdal2}
and provides an explicit example of gauge/string duality \cite{MatrixReloaded,GaiottoRastelli}. 

Recently, our knowledge about the duality in non-critical string theory 
has extended to various kind of string theories. 
The first non-trivial extension was to the correspondence between 
the two-cut matrix models \cite{GrossWitten,PeShe,DSS,Nappi,CDM,HMPN} 
and type 0 superstring theory \cite{TT,NewHat,UniCom}; 
later, the multi-cut matrix models \cite{MultiCut} were proposed as a  
description of fractional superstring theory \cite{irie2}. 
Despite of many qualitative discussions, 
little is known quantitatively about the nature 
of multi-cut matrix models. 
Therefore, in this paper we try to give explicit answers to the following questions: 
how do critical points and potentials look like and 
what is the corresponding geometry appearing at the vicinity of critical points?

To address these issues, {\em macroscopic loop amplitudes} 
\cite{BIPZ,KazakovSeries,Kostov1,Kostov2,BDSS,Kostov3,MSS,DKK} 
have played important roles. 
The one-point function of macroscopic loop operator, which is also known as resolvent, 
is defined as
\begin{align}
Q(x) \equiv 
\frac{1}{N} \Bigl<\tr \frac{1}{x-X}\Bigr> \equiv \frac{1}{N} \int dXdY\, 
e^{-N\tr w(X,Y)}
\Bigl[\tr \frac{1}{x-X}\Bigl]. \label{introResolvent}
\end{align}
Geometrically, the macroscopic loop amplitudes describe the probability amplitudes for 
creating holes on the two-dimensional surface as induced from the Feynman graph of 
matrix integral. Analytically, we can view the macroscopic loop amplitudes as a generating 
function for correlation functions of local operators (in $X$) in the matrix models. 
Most importantly, it enables us to probe eigenvalue distribution of a matrix $X$ \cite{BIPZ} 
and spacetime geometry in the weak coupling region \cite{fy3,MMSS}. 
Thus, the macroscopic loop amplitude is of fundamental importance in the theory of matrix models. 
In order to calculate the macroscopic loop amplitudes, one powerful way is to use the method of 
orthogonal polynomials \cite{DKK}. Through a useful identity (see Appendix \ref{RelationResolventOrthPoly}), we can express the resolvent $Q$ and the spectral parameter $x$ 
as a pair of conjugate variables acting on the space of orthogonal polynomials, 
whose commutations relation gives rise to the Douglas (string) equation. In terms of the 
differential operators, the string equation turns into a differential equation on the macroscopic 
loop amplitudes and we can solve it without directly evaluating the matrix integral 
\eq{introResolvent}. 

Right at the critical points, the resolvent has specific critical behavior. 
Conversely, with some proper choice of the critical resolvent, 
one can obtain the critical potentials. 
In the two-matrix-model cases, this approach is especially useful for finding critical points 
\cite{DouglasGeneralizedKdV,TadaYamaguchiDouglas}
and has been explicitly formulated for general $(p,q)$ critical points \cite{DKK}. 
Along this line, we explicitly construct the critical points and potentials in the multi-cut two-matrix models. Our results provide further insight into the understanding of the relationship \cite{HMPN}
between the existence of critical points and the the choice of hermiticity of critical potentials.

Macroscopic loop amplitudes at the vicinity of critical points 
(i.e. with perturbation of the cosmological constant) 
have been investigated in \cite{Kostov1,Kostov2,Kostov3,MSS,DKK,fim}. The seminal formula for general $(p,q)$ critical points 
with cosmological constant $\mu$ was first found by Kostov \cite{Kostov1} and can be expressed as 
\begin{align}
x=   \sqrt{\mu}\,  T_p(z),
\qquad Q= \mu^{q/2 p}\,  T_q(z), \label{IntroBosonicKostov}
\end{align}
with $T_p(\cosh \tau)=\cosh p\tau$, {\em the Chebyshev polynomials of the first kind},
and $z= \cosh\tau$. 
In the case of $(\hat p,\hat q)$ minimal superstring theory,%
\footnote{In the $k(>1)$-cut cases, there are two kinds of indexes: $(p,q)$ and $(\hat p,\hat q)$. 
The former corresponds to the standard CFT labeling of critical points; The latter corresponds to 
order of differential operators $\bP$ and $\bQ$. 
These labellings are different in general \cite{fi1}\cite{irie2}. 
Since the Liouville continuum formulation of $\mathbb Z_k$ 
symmetric critical points has not been known so far, we only use $(\hat p,\hat q)$ labeling 
except for the bosonic $(k=1)$ cases. }
the corresponding formula was also found on the Liouville side \cite{SeSh}. 
The non-trivial solution%
\footnote{There are two solutions in this system, that is, different phases (above and below the tip of a critical point) in the matrix models \cite{UniCom}.  One solution is essentially the same as 
the bosonic case \eq{IntroBosonicKostov} with $(p,q)=(\hat p,\hat q)$, which is called one-cut solution. 
The other is essentially different from the one-cut solution which is called two-cut solution \eq{IntroSuperSeSh}.}
can be expressed%
\footnote{Here we neglect normalization factors and phases 
in front of polynomials which are irrelevant in this section. }
as 
\begin{align}
x=   \sqrt{\mu}\,  U_{\hat p}(z) \, \sqrt{z^2-1},
\qquad Q= \mu^{\hat q/2 \hat p}\,  U_{\hat q}(z) \,\sqrt{z^2-1}, \label{IntroSuperSeSh}
\end{align}
with $U_{\hat p}(\cosh \tau)=\sinh \hat p\tau/\sinh\tau$, {\em the Chebyshev polynomials of the second kind},
and $z= \cosh\tau$.
This general formula has been utilized in various studies of $(p,q)$ (or $(\hat p,\hat q)$) 
critical points \cite{SeSh,KazakovKostov,SatoTsuchiya,
SeSh2,fis,iky,fim,fi1}, even in the D-instanton effects of the string theory \cite{David,GinspargZinnJustin,EynardZinnJustin,fy12,fy3}. 
Moreover, in comparison with worldsheet CFT, e.g.~with the FZZT-brane or ZZ-brane amplitudes in Liouville theory \cite{FZZT,ZZ}, 
it is essentially the above formula which is realized in the calculation of $(p,q)$ (or $(\hat p,\hat q)$) 
minimal (super)string theory \cite{SeSh}. 

One of the significances of this formula is that it provides us further checks of the string duality 
(e.g. matching of correlators) beyond matching of operator contents and critical exponents. 
Since this comparison is necessary especially in the correspondence 
between the multi-cut matrix models and fractional superstring theory \cite{irie2}, 
discovery of the general formula for the multi-cut two-matrix models is one of the most important stage in understanding the conjectured duality as well as the system itself. 

In this paper, 
the generalization of these formulae are explored especially 
in the $\mathbb Z_k$ symmetric potential $w(x,y)$, 
\begin{align}
w(\omega^n x, \omega^{-n}y)=w(x,y),\qquad\omega=e^{2\pi i/k},\qquad  (n \in\mathbb Z),
\end{align}
of the general $k$-cut two-matrix models $(k=3,4,\cdots)$. 
Identifying a proper ansatz for the non-polynomial part 
of the amplitudes (generalizing the square-root of \eq{IntroSuperSeSh}), 
we explicitly show that there is a one-parameter set of solutions 
$(l=0,1,2,\cdots,k-1)$ in the ``unitary'' series $\hat q=\hat p$ 
which can be tersely expressed as 
\begin{align}
x&=   
\mu^{\frac{\hat p}{\hat p+\hat q-1}}
P^{(\frac{2l-k}{k},-\frac{2l-k}{k})}_{\hat p-1}(z) \, 
\sqrt[k]{\bigl(z-1\bigr)^l\bigl(z+1\bigr)^{k-l}}, \\
Q&=
\mu^{\frac{\hat q}{\hat p+\hat q-1}}
P^{(-\frac{2l-k}{k},\frac{2l-k}{k})}_{\hat q-1}(z) \, 
\sqrt[k]{\bigl(z-1\bigr)^{k-l}\bigl(z+1\bigr)^{l}},
\end{align}
in terms of {\em the Jacobi polynomials}%
\footnote{The definition and basic properties of the Jacobi polynomials are listed in Appendix \ref{JacobiP}. } 
$P_{n}^{(\alpha,\beta)}(z)$. 

This ansatz moreover enables us to explicitly describe the geometry of off-critical amplitudes. 
For instance, our ansatz for the $(\hat p,\hat q;k)=(1,1;3)$ 
case gives the following algebraic curve:
\begin{align}
F(x,Q)=Q^3-3\mu xQ -x^3 =0,
\end{align}
which contains precisely three symmetric cuts corresponding to the eigenvalue distributions 
in the three-cut two matrix model. 

The organization of this paper is following: 
Some basics of the multi-cut two-matrix models are reviewed in section
\ref{subsecPreliminary}. 
The scaling ansatz for the orthogonal polynomials  
and its relation to the $\omega^{1/2}$-rotated/real potentials are discussed 
in section \ref{secRotatedVSReal}. 
The solutions for critical resolvents and potentials are studied 
in section \ref{secCritiPotentialsResolvents} 
(some examples of critical potentials are listed in Appendix \ref{ListOfCriticalPotentials}). 
The Daul-Kazakov-Kostov prescription \cite{DKK} is reviewed 
in section \ref{secRevOfDKKOff} and 
the two-cut cases are discussed in section \ref{TwoCutRevisited}. 
The $\mathbb Z_k$ symmetric $k$-cut cases 
are studied in section \ref{KCutSymmetricJacobi}. 
Its algebraic structure is also discussed in section \ref{secGeometryMultiCut}. 
Section \ref{secDiscussion} is devoted to conclusion and discussion. 

Some of the results for the critical potentials are listed in Appendix \ref{ListOfCriticalPotentials}. 
Direct evaluation of Douglas equation in the three-cut cases is given 
in Appendix \ref{DirectEvaluationThreeCutStringEq}. 
Several useful formulae about Jacobi polynomials are collected
in Appendix \ref{JacobiP}.

\section{Critical potentials in multi-cut critical points \label{MultiCutCriticalResolventSection}}

\subsection{Preliminary of the multi-cut two-matrix models \label{subsecPreliminary}}
We start with a brief review of some elementary facts about the orthogonal polynomial system 
in the multi-cut matrix models, which also serves to fix the notation used in this paper. 

As is the usual two-matrix model \cite{Mehta}, 
the multi-cut two-matrix models are characterized 
by the orthonormal polynomial system, 
\begin{align}
\alpha_n(x)= \frac{1}{\sqrt h_n}\Bigl( x^n + \cdots \Bigr), \qquad 
\beta_n(y) =\frac{1}{\sqrt h_n}\Bigl( y^n + \cdots \Bigr),  \label{ONPMETHOD}
\end{align}
with an inner product, 
\begin{align}
\bigl<\alpha_n(x), \beta_m(y)\bigr>&\equiv 
\int_{\mathcal C_x \times \mathcal C_y} d x dy\, 
e^{-Nw(x,y)}\,\alpha_n(x)\,\beta_m(y) = \delta_{n,m},  \label{Orth}
\end{align}
defined with respect to the two-matrix potential $w(x,y) =V_1(x)+V_2(y)-\theta xy$. 
We will use the normalization of the potential, $\theta=1$, throughout the main text. 
The degrees of the potentials are chosen to be multiples of $k$,
\begin{align}
\deg V_1(x)\equiv k m_1,\qquad  \deg V_2(y)\equiv k m_2.  \label{TwoMatPotentialOrder}
\end{align}
Here $k$ is the number of cuts and $m_1,m_2$ are arbitrary positive integers. 
A distinct feature of the multi-cut matrix models 
is that the integration domain $\mathcal C_x\times \mathcal C_y$ 
is defined as a contour along radial directions specified by the $Z_k$ symmetry of the system in the 
complex plane \cite{MultiCut}, 
\begin{align}
\mathcal C_x \,(= \mathcal C_y)= \bigcup_{n=0}^{k-1}\, \omega^n\,\mathbb R \subset \mathbb C,\qquad \omega^k=1,
\end{align}
where $\omega$ is a root of unity, 
which is chosen to be 
\begin{align}
\omega\equiv e^{\frac{2\pi i }{k}}.
\end{align}

The $k$-cut matrix models possess a $\mathbb Z_k$ charge conjugation, 
which is a generalization of the reflection transformation for the matrix ($M \to - M$) in the 
two-cut phases of the one-matrix models \cite{TT,NewHat,UniCom,SeSh2}. 
The $\mathbb Z_k$ charge conjugation in the $k$-cut two-matrix models is given \cite{fi1} as
\begin{align}
(x,y) \quad \to \quad  (\omega^j x, \omega^{-j} y) \qquad (j=0,1,2,\cdots, k-1), 
\end{align}
which preserves the interaction term, $xy$, in the two-matrix potential. 
In general, if the potential $w(x,y)$ is a $\mathbb Z_k$ symmetric potential, 
\begin{align}
w(\omega^j x, \omega^{-j}y)=w(x,y) \qquad (j \in \mathbb Z), 
\end{align}
the system is said to be $\mathbb Z_k$ symmetric. The $\mathbb Z_k$ charge conjugation 
naturally defines a charge conjugation of the orthonormal polynomial as 
\begin{align}
\alpha_n^{(j)}(x) 
\equiv  \omega^{-n j} \alpha_n(\omega^{j} x),\qquad 
\beta_n^{(j)}(y) 
\equiv \omega^{n j} \alpha_n(\omega^{-j} y),\qquad (j=0,1,\cdots,k-1). \label{ChargeConjPols22}
\end{align}
In the $\mathbb Z_k$ symmetric cases, 
the orthonormal polynomials are invariant under this charge conjugation, equivalently we have
\begin{align}
\alpha_n(\omega^j x)= \omega^{nj} \alpha_n(x),\qquad 
\beta_n(\omega^{-j} y)=\omega^{-nj}\beta_n(y),\qquad (j\in \mathbb Z). \label{ZkSymPol}
\end{align}
That is, the orthonormal polynomials are eigenfunctions of this $\mathbb Z_k$ transformation. 

The orthonormal polynomial system is characterized by the recursive relation, which can be written as 
\begin{align}
x\, \alpha_n(x)= A(n, e^{-\del_n}) \cdot \alpha_n(x),& \qquad 
N^{-1}\frac{\del}{\del x}\, \alpha_n(x)= B (n,e^{-\del_n}) \cdot \alpha_n(x), \label{OrthRecAB}\\ 
y\, \beta_n(y)= C(n, e^{-\del_n}) \cdot \beta_n(y),& \qquad 
N^{-1}\frac{\del}{\del y}\, \beta_n(y)=  D (n,e^{-\del_n}) \cdot \beta_n(y), \label{OrthRecCD}
\end{align}
where
\begin{align}
A(n,Z) \equiv  \sqrt{R({n+1})} \sum_{s=-1}^{km_2-1} A_s(n)\, Z^{s}, &\qquad 
B(n,Z) \equiv \frac{1}{\sqrt{R(n)}} \sum_{s=1}^{(km_1-1)(km_2-1)} B_s(n)\, Z^{s}, \\
C(n,Z) \equiv  \sqrt{R({n+1})} \sum_{s=-1}^{km_1-1} C_s(n)\, Z^{s},  &\qquad 
D(n,Z) \equiv  \frac{1}{\sqrt{R(n)}}\sum_{s=1}^{(km_1-1)(km_2-1)} D_s(n)\, Z^{s},
\end{align}
and $R(n)\equiv h_n/h_{n-1}$. 
Here $Z\equiv e^{-\del_n}$ is the index-shift operator 
\cite{GrossMigdal2}\cite{Martinec}\cite{DKK} that shifts 
every index $n$ by one unit on its right-hand side:
\begin{align}
Z^m \cdot \alpha_{n+i}(x) = \alpha_{n-m+i}(x),\qquad Z^l \circ A_s(n) = A_s(n-l) \circ Z^{l},
\end{align}
and $Z^{\rm T} = Z^{-1}$. 
Note that, by construction, $A_{-1}(n) = C_{-1}(n) =(N/n)\,B_1(n)=(N/n)\,D_1(n)=1$
and $R(n)$ cannot be zero, 
\begin{align}
R(n)\neq 0 \qquad (n = 0,1,2, \cdots,\infty),
\end{align}
including $n\rightarrow\infty$. 
In the $\mathbb Z_k$ symmetric case, invariance of the system asserts that the expansion coefficients $A_s(n), B_s(n),C_s(n)$ and $D_s(n)$ must be zero except for 
\begin{align}
A_{mk-1}(n),\qquad B_{mk+1}(n),\qquad C_{mk-1}(n),\qquad D_{mk+1}(n),
\end{align}
with $m\in \mathbb Z$ and $n\in \mathbb Z$. 

From the recursive relation \eq{OrthRecAB} and \eq{OrthRecCD}, one can check that the operator pairs $(A,B)$ and $(C,D)$ satisfy the canonical commutation relations, 
\begin{align}
[A,B]=[C,D]=N^{-1}.  \label{primitiveCCR}
\end{align}
Furthermore, taking inner product \eq{Orth} between $\del_x \alpha_n(x)$ and $\beta_m(y)$ and then integration by part leads to the following identity, 
\begin{align}
B(n,Z) = V_1'\bigl(A(n,Z)\bigr)- C^{\rm T}(n,Z), \quad 
D(n,Z) = V_2'\bigl(C(n,Z)\bigr)-A^{\rm T}(n,Z).  \label{StEqMatrix}
\end{align}
Combine these equations together, 
it can be seen that there is only one independent canonical commutation relation, 
\begin{align}
\bigl[A,\bigl(-C^{\rm T}\bigr)\bigr]=\bigl[C,\bigl(-A^{\rm T}\bigr)\bigr] = N^{-1}. \label{TwoMatCCR}
\end{align}
One can also extract 
equations for $A$ and $C$ from eq. \eq{StEqMatrix}:
\begin{align}
A(n, Z)&= \Bigl[V_2'\bigl(C^{\rm T}(n,Z)\bigr)\Bigr]_{\geq -1} 
- Z^{-1}\frac{n/N}{\sqrt{R(n)}}, \label{TwoMatA}\\
C(n, Z)&= \Bigl[V_1'\bigl(A^{\rm T}(n,Z)\bigr)\Bigr]_{\geq -1} 
- Z^{-1}\frac{n/N}{\sqrt{R(n)}}. \label{TwoMatC}
\end{align}
Here the bracket is defined as 
\begin{align}
\bigl[Z^s\bigr]_{\geq n}\equiv 
\left\{
\begin{array}{ll}
Z^s  &  (s\geq n) \cr
0      &  (s< n).
\end{array}
\right.
\end{align}
The system of equations was originally used to extract the string equations 
\cite{DSL,TwoMatString} and can be considered as the fundamental equations dictating 
the double scaling limit of the two-matrix models. 
Once we solve $A$ and $C$, eq.~\eq{StEqMatrix} leads to the solution of $B$ and $D$ in straightforward manner.


The strategy for solving critical points in two-matrix models was proposed 
in \cite{DouglasGeneralizedKdV, TadaYamaguchiDouglas}. 
The observation is this;  
the operators $A$ and $C$ have the following scaling behaviors at 
$(p,q)$ critical points (determined by critical potentials): 
\begin{align}
A(n,Z) \sim a^{p/2}\del^p +\cdots, \qquad 
C^{\rm T}(n,Z) \sim a^{q/2}\del^q+\cdots, \qquad (a\to 0), \label{TwoMatCriti}
\end{align}
where the lattice spacing $a$ is introduced to satisfy
\begin{align}
&N^{-1} = g_{str}\, a^{\frac{p+q}{2}} \to 0,  \qquad
\frac{n}{N} = \exp(-t\,a^{\frac{p+q-1}{2}})  \to 1,\nn\\[3pt]
&\del_n = -a^{1/2} g_{str}\del_t \equiv -a^{1/2}\del \to 0,\qquad (a\to 0),
\label{ScalingLattice}
\end{align}
and the operator $\del_n$ is the index-derivative operator in the index-shift operator $Z=e^{-\del_n}$. 
Conversely, if one assumes this critical behavior \eq{TwoMatCriti}, 
then there exist critical points. That is, we can obtain critical potentials which satisfy eq's
\eq{TwoMatA} and \eq{TwoMatC} \cite{DouglasGeneralizedKdV, TadaYamaguchiDouglas}. 

Explicit solutions to this problem have been given in \cite{DKK} 
for the general $(p,q)$ critical points in the one-cut cases as follows:
For each pair of potentials with degrees $(m_1,m_2)$ ($k=1$ of eq.~\eq{TwoMatPotentialOrder}), 
the maximal order of the differential operators \eq{TwoMatCriti} is $(p,q)=(m_2,m_1)$ 
and the operator $A$ and $C$ is given as
\begin{align}
A(n,Z) \to A(*,Z)= \sqrt{R_*}\,\frac{(1-Z)^{m_2}}{Z},\quad 
C(n,Z) \to C(*,Z)= \sqrt{R_*}\,\frac{(1-Z)^{m_1}}{Z}, \label{1CutSol}
\end{align}
in the large $N$ limit. Here $R_*$ is the critical value of $R(n)$, 
\begin{align}
R(n) \ \to \  R_*,\qquad (n\to \infty). 
\end{align}
Then it can be shown that, with these critical forms of $A$ and $C$, 
we can solve for the critical potential $w(x,y)$ 
which satisfies eq.~\eq{TwoMatA} and \eq{TwoMatC}. 

In the rest of this section, 
we generalize this idea to the multi-cut setting. 

\subsection{Smooth functions in multi-cut critical points}
The procedure of section \ref{subsecPreliminary} implies that the orthonormal polynomials become
smooth functions of the index $n$ at the vicinity of the critical points, 
especially at eq.~\eq{ScalingLattice}. 

The scaling ansatz for the multi-cut cases is to find a sequence of orthonormal polynomials $\tilde \alpha_n(x)$ 
(and $\tilde \beta_n(y)$) which are smooth continuous functions in not only $x$ but also the index $n$. 
First of all, since the orthonormal polynomials behave under $\mathbb Z_k$ transformation as 
\begin{align}
\alpha_n(x) \quad
\mapsto\quad \alpha_n(\omega x) = \omega^n \alpha_n^*(x) \sim  \omega^n \,\alpha_n(x) +\cdots,
\end{align}
there should be $k$ distinct smooth scaling functions, each of which has a distinct eigenvalue of the $\mathbb Z_k$ transformation. 
This is obvious in the $\mathbb Z_k$ symmetric background \eq{ZkSymPol}. 
Hence, we separate the polynomials into the following $k$ sets of polynomials
\begin{align}
\bigl\{\alpha_{kl+j}(x)\bigr\}_{l\in \mathbb Z}\qquad (j=0,1,\cdots,k-1), 
\end{align}
with respect to mod $k$. 
For each set labeled by $j$, we assume that there exists a smooth function in $l$ which can be regarded as the continuum limit of the sequence.
The proper choice of the smooth functions was first proposed in \cite{MultiCut} 
and we further write it as
\begin{align}
\psi_{l}(x) \equiv 
\begin{pmatrix}
\psi_{l,0}(x) \cr
\psi_{l,1}(x) \cr
\vdots \cr
\psi_{l,k-1}(x)
\end{pmatrix},\qquad 
\chi_{l}(y) \equiv 
\begin{pmatrix}
\chi_{l,0}(y) \cr
\chi_{l,1}(y) \cr
\vdots \cr
\chi_{l,k-1}(y)
\end{pmatrix}
\qquad (l \in \mathbb Z), \label{SmoothFunPsiChi}
\end{align}
with 
\begin{align}
\psi_{l,j}(x) \equiv \omega^{-\frac{kl+j}{2}} \alpha_{kl+j}(x),\qquad 
\chi_{l,j}(y) \equiv \omega^{\frac{kl+j}{2}} \beta_{kl+j}(y).  \label{SmoothFunPsiChiCOMP}
\end{align}
The phase factors indicate that the polynomials are approximated 
by smooth functions in the directions of $x = \omega^{m+1/2}\, \tilde x \, \, (\to \omega^{m+1/2} \infty)$ 
and $y = \omega^{-(m+1/2)}\, \tilde y \, \, (\to \omega^{-(m+1/2)} \infty)$:
\begin{align}
\psi_{l,j}\bigl(\omega^{m+1/2}\tilde x\bigr) \sim \frac{\omega^{mj}}{\sqrt{h_n}}\,\tilde x^{kl+j} + \cdots,\qquad 
\chi_{l,j}\bigl(\omega^{-(m+1/2)}\tilde y\bigr) \sim \frac{\omega^{-mj}}{\sqrt{h_n}}\,\tilde y^{kl+j} + \cdots,
\end{align}
since the phase does not depend on $l$ here. These directions are actually the Liouville 
direction \cite{fy3,MMSS}\cite{SeSh2}\cite{fi1}.
The main reason for this ansatz is that the critical point of eigenvalues, $(x_*,y_*)$, 
is at the origin $(x_*,y_*)=(0,0)$ when we consider $\mathbb Z_k$ symmetric potentials 
$w(\omega x,\omega^{-1} y) = w(x,y)$.%
\footnote{One can explicitly check this fact from the critical potentials we list 
in Appendix \ref{ListOfCriticalPotentials}. }

The consequence is that the operators acting on the vector-valued polynomials $\psi_l(x)$ (and $\chi_l(y)$) 
become smooth functions of $l$. The operators in this basis are expressed as 
\begin{align}
x\, \psi_l(x)= \omega^{1/2} \mathcal A(l, Z) \cdot \psi_l(x),& \qquad 
N^{-1}\frac{\del}{\del x}\, \psi_l(x)= \omega^{-1/2}\mathcal B (l,Z) \cdot \psi_l(x), \label{TildeRecAB}\\
y\, \chi_l(y)= \omega^{-1/2}\mathcal C(l, Z) \cdot \chi_l(y),& \qquad 
N^{-1}\frac{\del}{\del y}\, \chi_l(y)= \omega^{1/2} \mathcal D (l,Z) \cdot \chi_l(y), \label{TildeRecCD}
\end{align}
with explicit expressions:
\begin{align}
&\mathcal A(l,Z) \equiv  \sqrt{\mathcal R_{kl+1}} 
\sum_{r=0}^{k-1} 
\ \underset{0\leq j\leq k-1}{\diag} \Bigl(\sum_{s \in k\mathbb Z+ r} A_s(kl+j)\, \omega^{-\frac{s+1}{2}}\, Z^s\Bigr)
\times  \Gamma(Z)^{k-r}, \label{AopTilde}\\
&\mathcal B(l,Z) \equiv \frac{1}{\sqrt{\mathcal R_{kl}}} 
\sum_{r=0}^{k-1} 
\ \underset{0\leq j\leq k-1}{\diag} \Bigl(\sum_{s \in k\mathbb Z+ r} B_s(kl+j)\, \omega^{-\frac{s-1}{2}}\, Z^s\Bigr)
\times   \Gamma(Z)^{k-r}, 
\label{BopTilde} \\
&\mathcal C(l,Z) \equiv  \sqrt{\mathcal R_{kl+1}} 
\sum_{r=0}^{k-1} 
\ \underset{0\leq j\leq k-1}{\diag} \Bigl(
\sum_{s \in k\mathbb Z+ r} C_s(kl+j)\, \omega^{\frac{s+1}{2}}\, Z^s
\Bigr)
\times  \Gamma(Z)^{k-r},  \label{CopTilde}\\
&\mathcal D(l,Z) \equiv  \frac{1}{\sqrt{\mathcal R_{kl}}}
\sum_{r=0}^{k-1} 
\ \underset{0\leq j\leq k-1}{\diag} \Bigl(\sum_{s \in k\mathbb Z+ r} D_s(kl+j)\, \omega^{\frac{s-1}{2}}\, Z^s\Bigr)
\times  \Gamma(Z)^{k-r},
\label{DopTilde}
\end{align}
where 
\begin{align}
\sqrt{\mathcal R_{kl+1}}&
\equiv \underset{0\leq j\leq k-1}{\diag} \Bigl( \sqrt{R({kl+1+j})}\Bigr),\quad
\frac{1}{\sqrt{\mathcal R_{kl}}}
\equiv \underset{0\leq j\leq k-1}{\diag} \Bigl(\frac{1}{ \sqrt{R({kl+j})}}\Bigr), \label{DefMatR} 
\end{align}
with
\begin{align}
\underset{0\leq j\leq k-1}{\diag}\Bigl(\Lambda_j\Bigr) \equiv 
\begin{pmatrix}
  \Lambda_0           &                &              &                          \cr
                             &       \Lambda_1      &                        \cr
                               &    &   \ddots     &     \cr
                             &               &    &   \Lambda_{k-1}               \cr
\end{pmatrix},
\end{align}
and 
\begin{align}
\Gamma(Z) \equiv 
\begin{pmatrix}
  0           &     Z           &              &                   &       \cr
               &   0            &       Z      &                   &     \cr
               &                &  \ddots  &   \ddots     &     \cr
               &                &               &       0         &   Z   \cr
\dfrac{1}{Z^{k-1}}          &                &               &                   &   0
\end{pmatrix}, 
\qquad 
\Gamma(Z)^k=I_k.
\end{align}
Here $I_k$ is the identity matrix of $k\times k$. 
Now eq.~\eq{StEqMatrix} becomes
\begin{align}
\mathcal B = \omega^{1/2}\,V_2'\Bigl(\omega^{1/2} \mathcal A\Bigr) -\mathcal C^{\rm T},\qquad 
\mathcal D = \omega^{-1/2}\,V_1'\Bigl(\omega^{-1/2}\mathcal C\Bigr) - \mathcal A^{\rm T}, \label{tildeStEqMatrix}
\end{align}
in terms of the $k\times k$ matrix-valued operators $\mathcal A,\mathcal B,\mathcal C$ and $\mathcal D$. 

Note that the smoothness of polynomials $\psi_l(x)$ and $\chi_l(y)$ in the index $l$ only implies 
smoothness of sequence of recursion coefficients $\{A_s(kl+j)\}_{l=0}^{\infty}$ in the index $l$. This means that 
each recursion coefficient sequence (labeled by $j$) can be approximated by a smooth function with the following Taylor expansion with respect to the scaling parameter $a^{\frac{1}{2}}$;
\begin{align}
A_s(kl+j) = A_s(k\infty+j) + a^{1/2}\, f_s^{(j)}(t) + a\, g_s^{(j)}(t)+\cdots. \label{SmoothApproxA}
\end{align}
Here the lattice spacing $a$ and scaling parameter $t$ are defined as
\begin{align}
&N^{-1} = g_{str}\, a^{\frac{\hat p+\hat q}{2}} \to 0,  \qquad
\frac{n}{N} = \exp(-t\,a^{\frac{\hat p+\hat q-1}{2}})  \to 1,\nn\\[3pt]
&\del_n = -a^{1/2} g_{str}\del_t \equiv -a^{1/2}\del \to 0,\qquad (a\to 0),
\label{ScalingLattice2}
\end{align}
for the $(\hat p,\hat q)$ critical points of the multi-cut matrix models. 

\subsection{The $\omega^{1/2}$-rotated v.s. real potentials \label{secRotatedVSReal}}

It has been argued in the two-cut matrix models \cite{HMPN} that 
$\mathbb Z_2$ symmetry breaking critical points can only be realized 
in the potentials $V(x)$ which has an imaginary number $i$ in the following way:
\begin{align}
V(x) = V_0(-ix). \label{AntiHermitian}
\end{align}
Here the potential $V_0(x)$ is polynomial in $x$ with real coefficients. 
This consideration has a natural correspondence in the Liouville theory because 
the Liouville boundary cosmological constant $\zeta$ is related to the eigenvalue $x$ 
with an imaginary number $i$ \cite{UniCom} as 
\begin{align}
x= i \zeta.
\end{align}

A natural extension of this consideration is the $\omega^{1/2}$-rotated potential 
\eq{RotatedPotentials}. 
From our results,%
\footnote{See Appendix \ref{ListOfCriticalPotentials}.} 
within the prescription reviewed in section \ref{subsecPreliminary}, 
one can also find real-potential solutions for the $\mathbb Z_2$ symmetry breaking critical points. 
Therefore, we consider both cases in this section. We will show in the next subsection that, 
at least in the sense of critical potentials, 
the map between two potentials is just an analytic continuation.

\subsubsection{The $\omega^{1/2}$-rotated-potential models}
The $\omega^{1/2}$-rotated potentials are defined as
\begin{align}
w(x,y)= w_0(\omega^{-1/2} x, \omega^{1/2} y), \label{RotatedPotentials}
\end{align}
with a real-coefficient potential $w_0(x,y)$. 
If one puts $k=2$, then this turns out to be eq.~\eq{AntiHermitian} given in \cite{HMPN}.
From the definition, this potential satisfies
\begin{align}
w^*(x,y) = w\bigl(\omega x, \omega^{-1} y\bigr), 
\end{align}
and consequently one can show that%
\footnote{It can be similarly shown that $h_n$ are real functions. }
\begin{align}
\delta_{m,n} &= \bigl< \beta_m(y), \alpha_n(x)\bigr> 
= \bigl< \beta_m(y), \alpha_n(x)\bigr> ^* \nn\\
&= \int_{\mathcal C\times \mathcal C} dx^*dy^* \, e^{-Nw^*(x^*,y^*)}\, \beta_m^*(y^*)\, \alpha_n^*(x^*)  
= \int_{\mathcal C^*\times \mathcal C^*} dxdy \, e^{-Nw(\omega x,\omega^{-1}y)}\, \beta_m^*(y)\, \alpha_n^*(x) \nn\\
&= \int_{\mathcal C\times \mathcal C} dxdy \, e^{-Nw(x,y)}\, \beta_m^*(\omega y)\, \alpha_n^*(\omega^{-1}x) \nn\\
&= \bigl< \omega^{-m}\beta_m^*(\omega y), \omega^n \alpha_n^*(\omega^{-1}x)\bigr>,
\end{align}
therefore the uniqueness of the orthonormal polynomials implies
\begin{align}
\alpha_n^*( x) = \omega^{-n} \alpha_n(\omega x),\qquad 
\beta_n^*( y) = \omega^{n} \beta_n(\omega^{-1} y).
\end{align}
Here we have used the invariance of contour $\mathcal C^*=\mathcal C$ and the $\mathbb Z_k$ invariance of the measure,
$d(\omega x)d(\omega^{-1}y) = dxdy$. 
This complex conjugation corresponds to the $\mathbb Z_k$ charge conjugation of the polynomials
\eq{ChargeConjPols22}, 
\begin{align}
\alpha_n^*(x)= \alpha^{(1)}_n(x),\qquad \beta_n^*(y) = \beta^{(1)}_n(y). 
\end{align}
The hermiticity of coefficients in the operators $A$, $B$, $C$ and $D$
is also obtained in the same way as 
\begin{align}
A_s(n)^* = \omega^{-s-1} A_s(n) 
\quad &\Leftrightarrow \quad 
\bar A_s(n) \equiv  \omega^{-\frac{s+1}{2}}\,A_s(n) \in \mathbb R, \nn\\
B_s(n)^* = \omega^{-s+1} B_s(n) 
\quad &\Leftrightarrow \quad 
\bar B_s(n)\equiv\omega^{-\frac{s-1}{2}}\, B_s(n) \in \mathbb R, \nn\\
C_s(n)^* = \omega^{s+1} C_s(n) 
\quad &\Leftrightarrow \quad 
\bar C_s(n)\equiv \omega^{\frac{s+1}{2}}\,C_s(n) \in  \mathbb R, \nn\\
D_s(n)^* = \omega^{s-1} D_s(n) 
\quad &\Leftrightarrow \quad 
\bar D_s(n) \equiv \omega^{\frac{s-1}{2}}\,D_s(n)  \in  \mathbb R. \label{BarABCD}
\end{align}
Comparing this with the expression \eq{AopTilde}, \eq{BopTilde}, \eq{CopTilde} and \eq{DopTilde}, 
one can see that the coefficients of the matrix-valued operators $\mathcal A, \mathcal B,\mathcal C,\mathcal D$ 
are all real number. In this sense, 
eq.~\eq{tildeStEqMatrix} can have critical solutions 
with this $\omega^{1/2}$-rotated potential, 
including the $\mathbb Z_k$ symmetry breaking critical points. 

\subsubsection{The real-potential models}
The real-potential models are defined by real two-matrix potentials $w(x,y)$ 
with all coefficients real:
\begin{align}
w(x,y) \in \mathbb R,\qquad (x,y \in \mathbb R).  \label{RealPotentials}
\end{align}
The same discussion as above shows that 
the polynomials $\alpha_n(x)$ and $\beta_n(y)$, 
and the operators $A,B,C$, and $D$ are all real functions. 

The relation between eq.~\eq{StEqMatrix} and the smooth operators becomes simpler if 
one considers another convention of smooth orthonormal polynomials:%
\footnote{The two-cut matrix models are studied almost with this convention in literature. }
\begin{align}
\psi_{l,j}^{(\rm real)}(x) \equiv (-1)^{[\frac{kl+j}{k}]} \alpha_{kl+j}(x),\qquad 
\chi_{l,j}^{(\rm real)}(y) \equiv (-1)^{[-\frac{kl+j}{k}]} \beta_{kl+j}(y),
\end{align}
instead of eq.~\eq{SmoothFunPsiChiCOMP}. 
The corresponding smooth operators 
$\mathcal{A}^{(\rm real)},\mathcal{B}^{(\rm real)},
\mathcal{C}^{(\rm real)}$ and $\mathcal{D}^{(\rm real)}$ are then given as
\begin{align}
x\, \psi^{(\rm real)}_l(x)
= \mathcal{A}^{(\rm real)}(l, Z) \cdot \psi^{(\rm real)}_l(x),& \qquad 
N^{-1}\frac{\del}{\del x}\, \psi_l^{(\rm real)}(x)
= \mathcal{B}^{(\rm real)} (l,Z) \cdot \psi^{(\rm real)}_l(x), \\
y\, \chi^{(\rm real)}_l(y)
= \mathcal{C}^{(\rm real)}(l, Z) \cdot \chi^{(\rm real)}_l(y),& \qquad 
N^{-1}\frac{\del}{\del y}\, \chi^{(\rm real)}_l(y)
=  \mathcal{D}^{(\rm real)} (l,Z) \cdot \chi^{(\rm real)}_l(y).
\end{align}
Here $\psi^{(\rm real)}_l(x)$ and $\chi^{(\rm real)}_l(y)$ are the counterpart of 
$\psi_l(x)$ and $\chi_l(y)$ given in eq.~\eq{SmoothFunPsiChi}.
One can easily see that the operator $\mathcal{A}^{(\rm real)}$ can be obtained from 
$\mathcal{A}$ by just replacing the phase $\omega^{-(s+1)/2}$ in eq.~\eq{AopTilde} with $(-1)^{[-(s+1)/k]}$ 
(the phases in the other operators $\mathcal B,\mathcal C,\mathcal D$ 
are also similarly replaced: the phase $\omega^{\Theta /2}$ is replaced by $(-1)^{[\Theta/k]}$) 
and by replacing the matrix $\Gamma(Z)$ with the matrix $\Gamma^{(\rm real)}(Z)$,
\begin{align}
\Gamma(Z)=
\begin{pmatrix}
  0           &     Z           &              &                   &       \cr
               &   0            &       Z      &                   &     \cr
               &                &  \ddots  &   \ddots     &     \cr
               &                &               &       0         &   Z   \cr
\dfrac{1}{Z^{k-1}}          &                &               &                   &   0
\end{pmatrix}
\quad \rightarrow \quad \Gamma^{(\rm real)}(Z)\equiv 
\begin{pmatrix}
  0           &     Z           &              &                   &       \cr
               &   0            &       Z      &                   &     \cr
               &                &  \ddots  &   \ddots     &     \cr
               &                &               &       0         &   Z   \cr
\dfrac{-1}{Z^{k-1}}          &                &               &                   &   0
\end{pmatrix},
\end{align}
which satisfies $(\Gamma^{(\rm real)}(Z))^k=-I_k$. 
Here $[-(s+1)/k]$ is the Gauss symbol, which expresses the maximal integer less than or equal to $-(s+1)/k$.
Then eq.~\eq{tildeStEqMatrix} is equivalent to 
\begin{align}
\mathcal{B}^{(\rm real)} = V_2'\bigl( \mathcal{A}^{(\rm real)}\bigr) 
-{\mathcal{C}^{(\rm real)}}^{\rm T},\qquad 
\mathcal{D}^{(\rm real)} = V_1'\bigl( \mathcal{C}^{(\rm real)}\bigr) 
- {\mathcal{A}^{(\rm real)}}^{\rm T}. \label{tildetildeStEqMatrix}
\end{align}
Since everything is real here, one can have critical points (even $\mathbb Z_k$ symmetry breaking ones) in the same sense as the $\omega^{1/2}$-rotated models.

\subsection{Multi-cut critical potentials and resolvents \label{secCritiPotentialsResolvents}}

The large $N$ solutions to eq.~\eq{tildeStEqMatrix} are given by critical values of 
the matrix-valued operators 
$\mathcal A$ and $\mathcal C$ (defined in eq.~\eq{AopTilde} and eq.~\eq{CopTilde}), 
\begin{align}
\mathcal A(n,Z) \to \mathcal A(*,Z) \sim \del^{\hat p}+\cdots,\qquad 
\mathcal C^{\rm T}(n,Z) \to \mathcal C^{\rm T}(*,Z) \sim \del^{\hat q}+\cdots, \label{ACCriticalMultiCut}
\end{align}
and the critical potentials $w_*(x,y)$. 
At the vicinity of critical points, the coefficients of operators $\mathcal A$ (and $\mathcal C$) 
should be approximated by 
the following smooth functions;
\begin{align}
A_s(kl+j) = A_s(k\infty +j) + a^{1/2}\, f_s^{(j)}(t) + a\, g_s^{(j)}(t)+\cdots, 
\qquad (l \to \infty).  \label{ScalingFunctionsAC}
\end{align}
In principle, the leading coefficients $A_s(k\infty +j)$ (and $C_s(k\infty +j)$) 
can depend on the index $j$ and the critical operators of $\mathcal A$ (and $\mathcal C^{\rm T}$) are
expanded by these coefficients. 
For a while, let us see what happens in some special cases.%
\footnote{The calculations in this section have been carried out 
mostly with the help of {\em Mathematica}${}^{\rm TM}$.} 
That is, we consider the $(\hat p,\hat q)=(m_2,m_1)$ critical points realized in the potentials of  
degree $(km_1,km_2)$  (See eq.~\eq{TwoMatPotentialOrder}).

\begin{figure}[htbp]
 \begin{center}
  \includegraphics[scale=0.7]{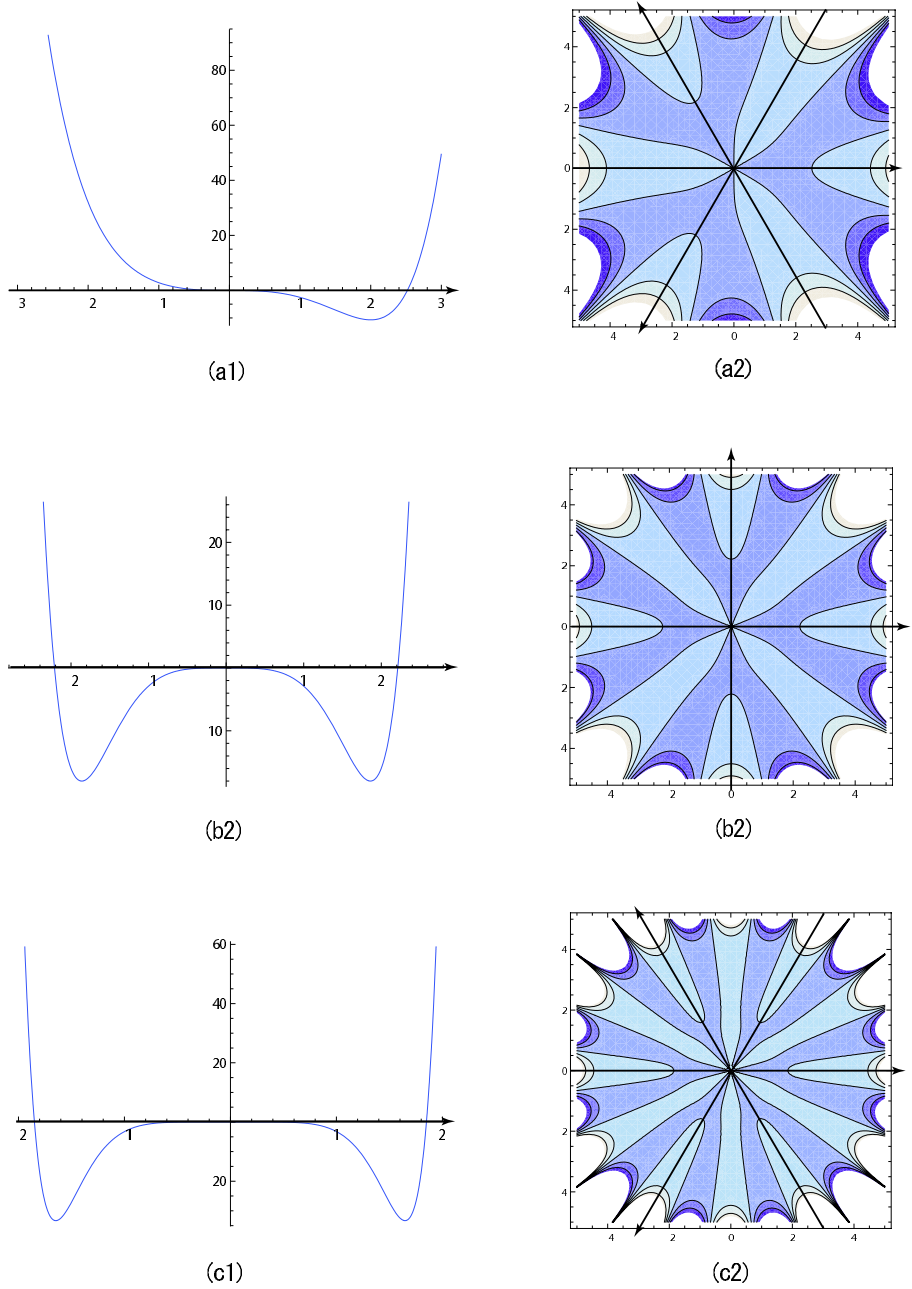}
 \end{center}
 \caption{\footnotesize 
The critical potentials of (a) $(\hat p,\hat q;k)=(2,2;3)$, $\ds V_1(x)=V_2(x)=
-\frac{8 x^3}{3}+\frac{x^6}{6}$;
(b) $(\hat p,\hat q;k)=(2,2;4)$, $\ds V_1(x)=V_2(x)=-3 x^4+\frac{x^8}{8}$;
(c) $(\hat p,\hat q;k)=(2,2;6)$, $\ds V_1(x)=V_2(x)=-\frac{10 x^6}{3}+\frac{x^{12}}{12}$.
(a1), (b1) and (c1) are plots in the real axes. 
The figures (a2), (b2) and (c2) are plots of the real part of the potentials ${\rm Re}(V_1(x))$
on the complex sheet of $x$. The straight lines with arrows on them correspond 
to the contour $\mathcal C_x$ of integration. Darker color means lower height. }
 \label{criticalpotential223}
\end{figure}

First we solve the condition \eq{ACCriticalMultiCut}.
Since $\mathcal A$ and $\mathcal C^{\rm T}$ are almost similar, the operator $\mathcal A$ is only considered here. 
By solving eq.~\eq{ACCriticalMultiCut} in this critical point $(\hat p,\hat q)=(m_2,m_1)$,%
\footnote{
By using the expression $Z=e^{-\del_n}$, one performs the Taylor expansion of $\mathcal A(*,Z)$ 
in terms of $\del_n$ and requires $\mathcal A(*,Z) = 
(\text{const.}) \times \del_n^{\hat p}+ O(\del_n^{\hat p+1})$. 
Then one can obtain constraints on the coefficients $A_s(k\infty+j)$. The result is eq.~\eq{SolDerivAAA}}
one can obtain the critical operator of $\mathcal A$ as
\begin{align}
\mathcal A(*,Z) = 
\sqrt{\mathcal R_{k\infty +1}} \, \frac{(1-Z^k)^{\hat p} }{Z}\, \Gamma(Z).  \label{SolDerivAAA}
\end{align}
The matrix $\sqrt{\mathcal R_{k\infty +1}}$ is defined in eq.~\eq{DefMatR} 
and the elements $R(k\infty +1+j)$ can depend on the index $j$ in principle. 
From eq.~\eq{AopTilde}, therefore, one can obtain
\begin{align}
\frac{1}{Z}\sum_{s \in k \mathbb Z-1} A_s(k\infty +j)\, \omega^{-\frac{s+1}{2}} Z^{s+1}
=\frac{(1-Z^k)^{\hat p} }{Z}
\end{align}
Here each term contains a phase factor $\omega^{-\frac{s+1}{2}}$. Since $s+1$ is a multiple of $k$
(say $s+1=kl$) and $\omega^{-\frac{s+1}{2}}=e^{\pi i l}=(-1)^l$, we get 
\begin{align}
\frac{1}{Y}\sum_{s+1 \in k \mathbb Z} A_s(k\infty +j)\, Y^{s+1} = \frac{(1+Y^k)^{\hat p}}{Y}\qquad 
(Y^k=-Z^k). 
\end{align}
That is, the operator $A$ (and similarly the operator $C$) is given as 
\begin{align}
A(k\infty +j,Z) &=\sqrt{R(k\infty +1+j)} \, \frac{(1+Z^k)^{\hat p}}{Z}, \nn\\
C(k\infty +j,Z) &=\sqrt{R(k\infty +1+j)} \, \frac{(1+Z^k)^{\hat q}}{Z}. \label{CritiACWithoutT}
\end{align}
We then consider eq.~\eq{TwoMatA} and eq.~\eq{TwoMatC}. 
The above operators $A(k\infty+j)$ and $C(k\infty +j)$ ($j=0,1,\cdots,k-1$) 
satisfy eq.~\eq{TwoMatA} and eq.~\eq{TwoMatC} for each $j$. By solving these equations, 
one can obtain the critical potentials and the solutions are realized when one has 
\begin{align}
R(k\infty+1)=R(k\infty+2)=\cdots = R(k\infty+k) \equiv R_*.  \label{RareEqual}
\end{align}
Therefore, the critical behavior of matrix-valued operators $\mathcal A$ and $\mathcal C$ is finally given as 
\begin{align}
\mathcal A(*,Z) = \sqrt{R_*} \frac{(1-Z^k)^{\hat p}}{Z} \Gamma(Z), \qquad 
\mathcal C^{\rm T}(*,Z) = \sqrt{R_*} \frac{(1-Z^{-k})^{\hat q}}{Z^{-1}} \Gamma(Z)^{k-1}. 
\label{CriticalCoeff}
\end{align}
Several examples of critical potentials calculated by using the above procedure are listed
in Appendix \ref{ListOfCriticalPotentials}
and some critical potentials are drawn in Fig.~\ref{criticalpotential223}. 
Note that all the parameters here, including $R_*$, are fixed in this procedure.%
\footnote{The counting of redundant degree of freedom is the same as in the one-cut cases \cite{DKK}. 
It can be seen as follows: 
Originally we have four redundant degrees of freedom, $(x,y) \to (a x+b,cy+d)$, but 
the critical points of eigenvalues are located at the origin, $(x_*,y_*)=(0,0)$, 
we have chosen the scale, $\theta =1$, and $(x,y)\to (a x,a^{-1}y)$ was fixed by choosing the 
convention of overall coefficient in the orthonormal polynomials \eq{ONPMETHOD}. 
In this sense, there is no free parameter here. 
One can also say that one can freely tune the parameter $R_*$ by turning on $\theta$. 
This procedure is a redundant deformation which does not change the result, and we will use this fact
in Appendix \ref{ListOfCriticalPotentials}. }

The result \eq{RareEqual} implies that all the $k$ coefficient sequences $A_s(k\infty +j)$ ($j=0,1,\cdots,k-1$) 
have the same asymptotic behavior at $a\rightarrow 0$ limit, i.e. at critical point, the coefficients satisfy 
\begin{align}
A_s(k\infty+j) \equiv  A_s(*),\qquad C_s(k\infty +j) \equiv C_s(*)
\end{align}
even though $A_s(n)$ can be approximated by $k$ different scaling functions (e.g.~$f^{(j)}(t)$) 
in the sub-leading terms of eq.~\eq{ScalingFunctionsAC}. 
This was just an assumption in the previous discussion 
of the two-cut two-matrix models \cite{fi1} but here 
we have checked that this is generally correct in the multi-cut two-matrix models. 


Next consider the critical points of $(\hat p,\hat q)=(m_2,m_1-1)$ realized in 
the potentials of degree $(km_1,km_2)$ (See \eq{TwoMatPotentialOrder}). 
This contains $\mathbb Z_k$ symmetry breaking critical points $(\hat p,\hat q)$. 
The critical operators of $\mathcal A$ and $\mathcal C$ are given as 
\begin{align}
\mathcal A(*,Z) &= \sqrt{R_*} \frac{(1-Z^k)^{\hat p}}{Z} \Gamma(Z), \\
\mathcal C^{\rm T}(*,Z) &=\sqrt{R_*} \frac{(1-Z^{-k})^{\hat q}(1-C_{km_1-1}(*) \,Z^{-k})}{Z^{-1}} \Gamma(Z)^{k-1} + \nn\\
&\quad + \sqrt{R_*} \sum_{r=0}^{k-2} \bar C_r(*) \frac{(1-Z^{-k})^{\hat q}}{Z^{r}} \Gamma(Z)^{r},
\end{align}
with $\bar C_r(*) \equiv \omega^{\frac{r+1}{2}} C_r(*)$ (see eq.~\eq{BarABCD}). 
In particular, from eq.~\eq{CopTilde}, one can see the following relations:
\begin{align}
\frac{1}{Z}\sum_{s+1 \in k\mathbb Z } C_s(*)\, \omega^{\frac{s+1}{2}}\, Z^{s+1} &= 
\frac{(1-Z^{k})^{\hat q}(1- C_{km_1-1}(*) \,Z^k)}{Z} \nn\\
\omega^{\frac{r+1}{2}}Z^r \sum_{s -r \in k\mathbb Z } C_s(*)\, \omega^{\frac{s-r}{2}}\, Z^{s-r} &= 
\bar C_r(*) (1-Z^{k})^{\hat q}Z^{r}. 
\end{align}
Therefore, the operators $A$ and $C$ are also obtained in the similar manner as eq.~\eq{CritiACWithoutT}, and given as
\begin{align}
A(*,Z) &= \sqrt{R_*} \frac{(1+Z^k)^{\hat p}}{Z} \nn\\
C(*,Z) &= \sqrt{R_*}\, \frac{\bigl(1+Z^k\bigr)^{\hat q}(1+C_{km_2-1}(*)\,Z^k)}{Z}+
\sqrt{R_*}\, \sum_{r=0}^{k-2} C_{r}(*)\cdot \bigl(1+Z^k\bigr)^{\hat q} Z^{r}. \label{ACcritiBreaking1}
\end{align}
The critical points and potentials in this cases are also obtained with these operators and 
eq's \eq{TwoMatA} and \eq{TwoMatC}, which are also listed in Appendix \ref{ListOfCriticalPotentials}. 
Note that the critical potentials now depend on the parameters $\{C_r(*)\}_{r=0}^{k-2}$ and $C_{km_2-1}(*)$. 

Note that the difference between the $\omega^{1/2}$-rotated potentials \eq{RotatedPotentials} 
and the real potentials \eq{RealPotentials} is almost the complex phase of $C_r(*) = \bar C_r(*) \,e^{iu(r)}$ 
$(\bar C_r(*) \in \mathbb R; \, r=0,1,\cdots,k-2)$ in eq.~\eq{ACcritiBreaking1}. 
Since we obtain the critical potentials by solving eq's \eq{TwoMatA} and \eq{TwoMatC} 
by using eq.~\eq{ACcritiBreaking1}, 
the map from real critical potentials to $\omega^{1/2}$-rotated ones 
is also given by just analytically continuing the parameters $\{\bar C_r(*)\}_{r=0}^{k-2}$ 
in the real potentials to the complex values $\{\bar C_r(*)\, \omega^{-(r+1)/2}\}_{r=0}^{k-2}$.
This means that all the $\mathbb Z_k$ symmetric $(\hat p,\hat q)$ critical points and 
$\mathbb Z_k$-symmetry breaking $(\hat p,\hat q)$ critical points 
can be realized within both real potentials and $\omega^{1/2}$-rotated potentials 
of $k$-cut two-matrix models, and also that these real/$\omega^{1/2}$-rotated critical potentials 
are related to each other by just the above analytic continuation. 


One can then realize the canonical operators in terms of the Lax pair, and they are given by
\begin{align}
\mathcal A(n,Z) = a^{\hat p/2} \bP(t,\del),\qquad \mathcal C^{\rm T}(n,Z) = -a^{\hat q/2} \bQ (t,\del),
\qquad x\,\omega^{-1/2} = a^{\hat p/2} \zeta \label{ToLax1}
\end{align}
with 
\begin{align}
\bP(t,\del) &= \sqrt{R_*}(-k)^{\hat p}\times \Gamma \, \del^{\hat p} + \sum_{n=0}^{\hat p-1}H_n(t)\, \del^{\hat p-n}, \\
\bQ(t,\del) &= -\sqrt{R_*}\, k^{\hat q}\times \Bigl[\bigl(1+\bar C_{k\hat q-1}(*)\bigr)\Gamma^{k-1}  
+  \sum_{r=0}^{k-2} \bar C_r(*) \Gamma^{r}\Bigr] \del^{\hat q} 
+ \sum_{n=0}^{\hat q-1}\widetilde H_n(t)\, \del^{\hat q- n},\label{ToLax2}
\end{align}
and the canonical commutation relation \eq{TwoMatCCR} becomes the Douglas equation, $[\bP,\bQ]=g_{str}I_k$. 
The functions $H_n(t)$ and $\widetilde H_n(t)$ are some combination of 
scaling functions in \eq{ScalingFunctionsAC}. 
The parameter $a$ is a lattice spacing of $\del_n = - a^{1/2} g_{str} \del_t$. 
Here we define 
\begin{align}
\Gamma \equiv \Gamma(1)=
\begin{pmatrix}
  0   & 1  &    &            &     \cr
     &  0  &  1 &            &     \cr
     &    &  \ddots  &  \ddots  &     \cr
     &    &     &       0     &  1  \cr
1 &   &     &             &0
\end{pmatrix}, 
\qquad \Gamma^k=I_k.
\end{align}
Here $H_n(t)$ and $\widetilde H_n(t) \in \mathbb R$ are all real functions. 
Since we can erase some parts of sub-leading scaling functions $H_n(t)$ and $\widetilde H_n(t)$ 
by changing normalization of orthonormal polynomials (some similarity transformation), 
we can always put
\begin{align}
0= \sum_{i=1}^{k}\bigl[H_1(t)\bigr]_{l+i,i}, \label{HoneConst}
\end{align}
for $l=0,1,\cdots,k-1$. Here we extend the meaning of indices as 
$\bigl[H_1(t)\bigr]_{i+lk,j+mk} \equiv  \bigl[H_1(t)\bigr]_{i,j}$

It is also convenient to use the diagonalized basis of $\Gamma$ which is useful for analyzing  
multi-component KP hierarchy \cite{fi1}:
\begin{align}
\Omega \equiv U^\dagger \Gamma U 
=
\begin{pmatrix}
 1  &               &    &            &     \cr
     & \omega &    &            &     \cr
     &               &  \omega^2  &       &     \cr
     &               &                   &    \ddots        &   \cr
     &               &                   &                      &   \omega^{k-1}
\end{pmatrix}, \label{DefOmega}
\end{align}
and the matrix elements of $U$ are given by 
\begin{align}
U_{ij}=\frac{\omega^{(i-1)(j-1)}}{\sqrt{k}}. 
\end{align}
In this basis, the relation \eq{HoneConst} is simply given as 
\begin{align}
\bigl[U^\dagger H_1(t) U  \bigr]_{i,i} = 0\qquad  (i=1,2,\cdots,k).
\end{align}
However, from the analysis we have shown in section \ref{secRotatedVSReal}, 
the component functions in the operators 
$\mathcal A^{(\rm KP)}, \mathcal B^{(\rm KP)},
\mathcal C^{(\rm KP)} ,\mathcal D^{(\rm KP)}$ 
$(\mathcal X^{(\rm KP)} \equiv U^\dagger \mathcal X U)$ 
are complex-valued functions now. 

Here we comment on the ``unitary'' series which is defined by the condition 
of $V_1(x)=V_2(x)$.%
\footnote{This is also called ``$\mathbb Z_2$ symmetric'' case in literature, 
in the sense of exchanging the matrices $X \leftrightarrow Y$. 
Since the terminology of ``$\mathbb Z_2$ symmetric'' is dedicated 
to $\mathbb Z_k$ symmetry in this paper, 
we will use the term ``unitary'' referring to the $V_1(x)=V_2(x)$ case. 
Of course, the terminology of ``unitary'' should come 
from the worldsheet CFT of dual string theory, and therefore 
our terminology ``unitary'' should be understood as a formal notion.
In the case of one- and two-cut matrix models, this situation always gives unitary sequences
although obtaining unitary sequences does not necessarily require this condition. } 
We should note that the unitary series of the $\mathbb Z_k$ symmetric background 
is given by $(\hat p,\hat q)=(\hat p,\hat p)$ instead of $(\hat p,\hat p+1)$. 
This can be understood as follows:
In general, the ``unitary'' condition gives 
\begin{align}
\mathcal A(*,Z) = \sqrt{R_*} \frac{(1-Z^k)^{m}}{Z} \Gamma(Z), \qquad 
\mathcal C^{\rm T}(*,Z) = \sqrt{R_*} \frac{(1-Z^{-k})^{m}}{Z^{-1}} \Gamma(Z)^{k-1},
\end{align}
in the $\mathbb Z_k$ symmetric background. Here $m=m_1=m_2$. 
In the superstring ($k=2$) cases, since the matrix $\Gamma(Z)$ 
satisfies 
\begin{align}
\Gamma(Z)^{k-1}= \Gamma(Z), 
\end{align}
the following combination of the operators $\mathcal A$ and $\mathcal C^{\rm T}$
\begin{align}
\frac{1}{2}\Bigl(\mathcal A(*,Z)+(-1)^m\mathcal C^{\rm T}(*,Z) \Bigr)&
= a^m\times \sqrt{R_*}(-k)^m \,\Gamma\, \del^m + \cdots,\nn\\
\frac{1}{2}\Bigl(\mathcal A(*,Z)-(-1)^m\mathcal C^{\rm T}(*,Z) \Bigr)&
= -a^{m+1}\times \sqrt{R_*}(-k)^{m+1} \,\Gamma\, \del^{m+1} + \cdots,
\end{align}
gives the $(\hat p,\hat q)=(m,m+1)$ system. 
The same argument also applies to the bosonic ($k=1$) cases. 
For the $k\,(>2)$-cut $\mathbb Z_k$ symmetric background case, 
the matrices $\Gamma(Z)$ and $\Gamma(Z)^{-1}$ are essentially different matrices; 
hence, the previous argument fails. This means that the unitary series should be 
$(\hat p,\hat q)=(m,m)$. 

On the other hand, if we consider the $\mathbb Z_k$ 
symmetry breaking critical points which correspond to minimal fractional superstring theory \cite{irie2}, 
\begin{align}
\mathcal A(*,Z) &= \sqrt{R_*} \frac{(1-Z^k)^{\hat p}}{Z} \Gamma(Z) 
= a^{\hat p}\times \sqrt{R_*}(-k)^{\hat p} \,\Gamma\, \del^{\hat p} + \cdots,\nn\\
\mathcal C^{\rm T}(*,Z) &= \sqrt{R_*} \frac{(1-Z^{-k})^{\hat q}}{Z} \Gamma(Z)
+\sqrt{R_*} \frac{(1-Z^{-k})^{\hat q+1}}{Z^{-1}}\Gamma(Z)^{k-1} \nn\\[2pt]
&= a^{\hat q}\times \sqrt{R_*}\,k^{\hat q} \,\Gamma\, \del^{\hat q} + \cdots,
\end{align}
the matrices of the leading terms in $\mathcal A$ and $\mathcal C^{\rm T}$ are 
both given by $\Gamma(Z)$. That is, the situation is similar to the super(bosonic)string cases. 
In fact, unitary minimal fractional superstring theory corresponds to
the $(\hat p,\hat q)=(m,m+1)$ critical points, instead of $(\hat p,\hat q)=(m,m)$. 

\section{Macroscopic loop amplitudes \label{SecOffCriticalResolvents}}

In this section, we consider off-critical behavior of the critical points we identified in the multi-cut matrix models. 
The prescription we adopt for studying off-critical amplitudes was 
developed in the one-cut two-matrix models \cite{DKK}, 
which we refer to as the DKK prescription. In the following, 
we first review their procedure then consider the extension to the multi-cut cases. 

\subsection{The one-cut cases: a review of \cite{DKK} \label{secRevOfDKKOff}}

At the $(p,q)$ one-cut critical points, 
the canonical pair of $A$ and $C$ operators was given as 
\begin{align}
A(*,Z)= \frac{(1-Z)^p}{Z},\qquad C^{\rm T}(*,Z)= \frac{(1-Z^{-1})^q}{Z^{-1}}.
\end{align}
Here we choose $R_*=1$. 
Considering the dependence on $n/N\, (= \exp(-t\,a^{\frac{p+q-1}{2}}))$ 
in the operators $A$ and $C$ 
is equivalent 
to turning on the coupling of the most relevant operator, $t$. 
Since we have dimensionful parameter $t$, 
we can make a dimensionless combination with $\del_t$ as 
\begin{align}
z \equiv \kappa\, t\,\del_t,\qquad [z,t^n] = n \kappa t^n, \label{DefOfsmallz}
\end{align}
with $\kappa \equiv  g_{str}/t^{\frac{p+q}{p+q-1}}$. 
Then the off-critical amplitudes of the canonical operators are generally expressed as 
\begin{align}
A(n,Z)=a ^{\frac{p}{2}}\,\lambda^p \, \pi_{p}(z),\qquad 
-C^{\rm T}(n,Z)=a^{\frac{q}{2}}\lambda^q \, \xi_q(z), \qquad 
\lambda\equiv t^{\frac{1}{p+q-1}}.
\end{align}
Note that the leading behavior of the polynomials $\pi_p$ and $\xi_q$ are 
\begin{align}
\pi_p(z)= (-1)^p\, z^p+\cdots,\qquad 
\xi_q(z) = - z^q+\cdots. 
\end{align}
So the commutation relation becomes 
\begin{align}
N^{-1} \, 
(=\kappa\, a^{\frac{p+q}{2}}\lambda^{p+q})
=[C^{\rm T}, A] = N^{-1} \,
\frac{q\,\pi'_p(z)\, \xi_q(z)-p\,\xi'_q(z)\,\pi_p(z)}{p+q-1},
\end{align}
and 
\begin{align}
q\,\pi'_p(z)\, \xi_q(z)-p\,\xi'_q(z)\,\pi_p(z)=(p+q-1).  \label{DKKeq}
\end{align}
This was found in \cite{EynardZinnJustin,DKK} and is one of the main equation for off-critical amplitudes 
in the two-matrix models. 

It is known that this system has a simple solution for conformal background 
which is given by the Chebyshev polynomials of the first kind, 
$T_p(\cosh(\tau))=\cosh(p\tau)$ \cite{Kostov1}. 
In the case of unitary series $(p,q)=(p,p+1)$, 
the equation \eq{DKKeq} becomes the addition formula for the hyperbolic sine and cosine functions:
\begin{align}
q T_p'(z) T_q(z)-p T_q'(z) T_p(z) &
= pq \, \frac{\sinh (p \tau )\cosh (q\tau)- \sinh(p\tau)\cosh(q\tau)}{\sinh\tau} \nn\\
&= -pq \, \frac{\sinh(q-p)\tau}{\sinh\tau} = -pq \qquad (q=p+1).
\end{align}
Consequently, the solution for $q=p+1$ is 
\begin{align}
A(n,Z)&
=a ^{\frac{q}{2}}\times 2\, \Bigl(-\frac{c}{2} \lambda\Bigr)^p \, T_{p}\bigl(z/c \bigr) = 
a ^{\frac{q}{2}}\times 2\, \Bigl(-\frac{c}{2} \lambda\Bigr)^p \, \cosh(p \tau),\\
C^{\rm T}(n,Z)&
=a^{\frac{q}{2}}\times 2\, \Bigl(\frac{c}{2} \lambda\Bigr)^q \, T_q\bigl(z/c\bigr) 
=a^{\frac{q}{2}}\times 2\, \Bigl(\frac{c}{2} \lambda\Bigr)^q \, \cosh(q\tau), 
\end{align}
with the normalization factor $c$ which satisfies $2 (-1)^{p}(c/2)^{p+q-1} = (p+q-1)/pq$ and 
$z= c \cosh(\tau)$. 

For the general $(p,q)$ cases, the Chebyshev-polynomial is not a solution to eq.~\eq{DKKeq}. 
However, aside from the unitary series, the most relevant operator and cosmological constant are different in general; 
considering this fact here, it means that there can be a Jacobian factor in the eq.~\eq{DKKeq} in general. 
Therefore, we will utilize this degree of freedom to deform the eq.~\eq{DKKeq};  
it was proposed in \cite{DKK} that the simple solutions with the Chebyshev polynomials can be derived from   
\begin{align}
p\, T'_{\frac{q}{q-p}}(w) T_{\frac{p}{q-p}}(w)
-q\, T_{\frac{q}{q-p}}(w) T'_{\frac{p}{q-p}}(w)  
&= pq \, \frac{\sin (q \tau )\cos (p\tau)- \sin(p\tau)\cos(q\tau)}{(q-p)\sin(q-p)\tau}  \nn\\
&=\frac{pq}{q-p}.  \label{ChebPQeq}
\end{align}
Here $T_{\frac{q}{q-p}}(w)=T_q(z(w))$ with $w=\cos (q-p)\tau = T_{q-p}(z)$. 
This procedure corresponds to utilizing the canonical pair $(w,\mu)$ 
of the cosmological constant $\mu$ \cite{DKK},
\begin{align}
[w, \mu^n ] = \kappa \, n\, \mu^n,
\end{align}
instead of the canonical coordinates $(z, t)$ of the most relevant operator $t$,
\begin{align}
[z, t^n] = \kappa\, n\, t^n.
\end{align}
So in general the equation we have is 
\begin{align}
\frac{1}{q}\,\pi'_q(z)\, \xi_p(z)-\frac{1}{p}\,\pi_q(z)\,\xi'_p(\hat z)= w_{p,q}'(z). \label{DeformedDKKeq}
\end{align}

In this procedure, the requirement of matrix model is that the functions of $z$ in the operators $A$ and $C^{\rm T}$ 
are all polynomials in $z$. One can expect that $w_{p,q}(z)$ should be polynomials in general background. 

\subsection{The two-cut cases: revisited \label{TwoCutRevisited}}

The solution for conformal background in the superstring cases was first obtained in \cite{SeSh}. 
There are two classes of the solutions. The first class is called the one-cut-phase solution:
\begin{align}
\zeta = \sqrt{\mu} \cosh (\hat p \tau) =\sqrt{\mu}\, T_{\hat p}(z),\qquad 
\del_\zeta \mathcal Z(\zeta(z))=\mu^{\frac{\hat q}{2\hat p}} \cosh (\hat q \tau) 
=\mu^{\frac{\hat q}{2\hat p}}\, T_{\hat q}(z).
\end{align}
The second is called the two-cut-phase solution:
\begin{align}
\zeta = \sqrt{\mu} \sinh (\hat p \tau) = \sqrt{\mu} U_{\hat p}(z) \sqrt{z^2-1},\quad 
\del_\zeta \mathcal Z(\zeta(z))= \mu^{\frac{\hat q}{2\hat p}} \sinh (\hat q \tau) 
=  \mu^{\frac{\hat q}{2\hat p}} \,U_{\hat q}(z)\sqrt{z^2-1}, \label{2CutSolSeSh}
\end{align}
Here we use $z= \cosh(\tau)$ and $\mathcal Z(\zeta)$ is the FZZT disk amplitude. 
Realization of this solution in the pure-supergravity critical point was shown in \cite{UniCom}. 
Realization of $(\hat p,\hat q)$ solutions in the two-cut two-matrix models was shown in \cite{fi1}, 
with the SFT formulation \cite{fkn}\cite{fy12,fy3} and its prescription utilizing the $W_{1+\infty}$ constraints \cite{fim}. 

In the DKK prescription, what we need to consider is the Douglas equation 
for the matrix-valued Lax operators $\mathcal A(n,Z)$ and $\mathcal C^{\rm T}(n,Z)$ of eq's \eq{AopTilde} and \eq{CopTilde}.  
Since $\mathcal A(n,Z)$ and $\mathcal C^{\rm T}(n,Z)$ 
can generally be expanded in terms of Pauli matrices $\sigma_i$, 
one can decompose the string equation in the following form:
\begin{align}
0= [\mathcal C^{\rm T},\mathcal A]\, a^{-\frac{\hat p+\hat q}{2}}\lambda^{-(\hat p+\hat q)} 
-\kappa I_2 = \text{(CCR0)}\, \kappa^0 + \text{(CCR1)} \,\kappa I_2
+ \text{(CCR2)}\, \kappa \sigma_3+O(\kappa^2). \label{Doug2Cut}
\end{align}
The one-cut solution is trivially the solution found in the bosonic case. 
The two-cut solution is less trivial but can be realized as 
\begin{align}
\mathcal A &= a^{\hat p/2}\, \Bigl\{c^{-1}\bigl(-c \lambda\bigr)^{\hat p}, U_{\hat p-1}(z/c ) 
\begin{pmatrix}
0 & z +c \cr
z -c & 0 \cr
\end{pmatrix}
\Bigr\}, \nn\\
\mathcal C^{\rm T} &= a^{\hat q/2}\Bigl\{c^{-1}\bigl(c \lambda\bigr)^{\hat q}, U_{\hat q-1}(z/c)
\begin{pmatrix}
0 & z +c \cr
z-c  & 0 \cr
\end{pmatrix}
 \Bigr\}, \label{2CutSolDKK}
\end{align}
with $4(-1)^{\hat p} c^{\hat p+\hat q-1} = (\hat p+\hat q-1)/\hat p\hat q$. 
One can easily show that this satisfies the Douglas equations \eq{Doug2Cut}. 
The curly bracket $\{,\}$ denotes anti-commutator which is necessary to satisfy (CCR2). 

The implications of this system are the following:
\begin{itemize}
\item The zero-th order equation of the Douglas equation, (CCR0), indicates that 
$\mathcal A$ commutes with $\mathcal C^{\rm T}$. That is, $\mathcal A$ and $\mathcal C^{\rm T}$ can be diagonalized simultaneously.%
\footnote{It has been shown in \cite{fi1} that each pair of eigenvalues, $(P_i,Q_i) \ (i=1,2)$, 
corresponds to the FZZT brane amplitude with each different R-R charge. }
\item Eigenvalues of the matrices $\mathcal A$ and $\mathcal C^{\rm T}$ satisfy the same equation \eq{DKKeq} 
as in the bosonic case. The important difference is now the eigenvalues can include non-polynomial parts (e.g.~the square root in eq.~\eq{2CutSolSeSh}), 
as long as all the matrix elements remain to be polynomials in the matrix-model realization \eq{2CutSolDKK}. 
\item Usually, some naive expression of the matrix-model realization (e.g. eq.~\eq{2CutSolDKK} 
without the anti-commutator) cannot satisfy (CCR2). 
In this case, we need to add proper first order corrections to the operators $\mathcal A$ and $\mathcal C^{\rm T}$.%
\footnote{In this case, all we need to do is just to change the ordering of $\lambda$ and $z$. 
But in the higher-cut cases, 
we need to add purely first order corrections to the scaling functions 
in the operators $\mathcal A$ and $\mathcal C^{\rm T}$. See section \ref{DirectAnalysis11Cases}.}
\end{itemize}

\subsection{The $k$-cut cases: $\mathbb Z_k$ symmetric background \label{KCutSymmetricJacobi}}

Now we can extend the previous discussion to the $k$-cut matrix models. 
Here we will focus on the general considerations;   
more detailed investigation of the three-cut string equation can be found in Appendix 
\ref{DirectEvaluationThreeCutStringEq}.
For the sake of simplicity, we consider the models with the $\mathbb Z_k$ symmetric background. 
In this case, the operators $\mathcal A(n,Z)$ and $\mathcal C^{T}(n,Z)$ are restricted to 
\begin{align}
\mathcal A(n,Z) = & a^{\hat p/2} \lambda^{\hat p}
\begin{pmatrix}
0           & F_1(z) &             &   &   \cr
           & 0 & F_2(z)           &    &  \cr
           &     & \ddots &\ddots & \cr
           &      &             &    0       &  F_{k-1}(z) \cr
F_k(z)            &      &            &           &  0
\end{pmatrix}
,\\
\mathcal C^{\rm T}(n,Z)= &a^{\hat q/2} \lambda^{\hat q}
\begin{pmatrix}
0           &  &             &   &   G_k(z) \cr
G_1(z)         & 0 &         &    &  \cr
           &  G_2(z)   & \ddots &  & \cr
           &      &     \ddots        &    0       &   \cr
           &      &            &      G_{k-1}(z)     &  0
\end{pmatrix}.
\end{align}
The matrix-model requires that all the functions $F_i(z)$ and $G_i(z)$ are polynomials in $z$. 
Thus the diagonalization is just given by 
\begin{align}
\mathcal A(n,Z) \simeq a^{\hat p/2} \lambda^{\hat p}\, \pi_{\hat p}(z)\, \Omega ,\qquad 
\mathcal C^{\rm T}(n,Z) \simeq -a^{\hat q/2} \lambda^{\hat q}\, \xi_{\hat q}(z)\,\Omega^{-1}. \label{DiagAC}
\end{align}
with 
\begin{align}
\pi_{\hat p}(z) \equiv  \bigl(h_{\hat p}(z)\bigr)^{1/k} \equiv \rho_{\hat p} (z) \,\bigl(u_{\hat p}(z)\bigr)^{1/k},\qquad
\xi_{\hat q}(z) \equiv  \bigl(\widetilde h_{\hat q}(z)\bigr)^{1/k} \equiv \eta_{\hat q} (z) \,\bigl(v_{\hat q}(z)\bigr)^{1/k},
\end{align}
and $h_{\hat p}(z) = \prod_i F_i(z)$ and $\widetilde h_{\hat q}(z)= \prod_i G_i(z)$. 
Here $\rho_{\hat p}(z), \eta_{\hat q}(z), u_{\hat p}(z)$ and $v_{\hat q}(z)$ are all polynomials and the equality ``$\simeq$'' indicates diagonalization. 
Then eq.~\eq{DKKeq} is expressed as 
\begin{align}
\hat q\, h_{\hat p}'(z)\, \widetilde h_{\hat q}(z)-\hat p\, \widetilde h_{\hat q}'(z)\, h_{\hat p}(z) 
= k(\hat p+\hat q-1)\, \bigl(h_{\hat p}(z)\, \widetilde h_{\hat q}(z)\bigr)^{\frac{k-1}{k}}.
\end{align}
Since the left-hand side of the equation is a polynomial, we need to resolve the $k$-th root $(...)^{(k-1)/k}$ 
in the right-hand side. A natural ansatz is 
\begin{align}
u_{\hat p}(z)= (z-b)^{k-l}(z-c)^l,\qquad 
v_{\hat q}(z)= (z-b)^{l}(z-c)^{k-l}, \label{CISYansatz}
\end{align}
which is characterized by the two indices $(k,l)$. Then the equation turns out to be 
\begin{align}
&(z-b)(z-c)\Bigl[\hat q\, \rho_{\hat p}'(z)\, \eta_{\hat q}(z)-\hat p\, \eta_{\hat q}'(z)\, \rho_{\hat p}(z) \Bigr]+ \nn\\
&+ \Bigl[(\hat q-\hat p) z + \hat p \gamma_{\hat p} - \hat q \gamma_{\hat q}\Bigr] \rho_{\hat p}(z)\, \eta_{\hat q}(z)=\hat p+\hat q-1, \label{CISYeq}
\end{align}
with
\begin{align}
\gamma_{\hat p}\equiv \frac{(k-l)b +lc}{k},\qquad  \gamma_{\hat q}\equiv  \frac{lb+(k-l)c}{k}. 
\end{align}
This solution can be realized in the matrix model as 
\begin{align}
\mathcal A(n,Z) =a^{\hat p/2} \lambda^{\hat p}\rho_{\hat p} (z) \Bigl[ \Gamma z - M\Bigr],\qquad 
\mathcal C^{\rm T}(n,Z) =-a^{\hat q/2} \lambda^{\hat q} \eta_{\hat q} (z) \Bigl[ \Gamma^{-1} z - \bar M\Bigr],
\end{align}
with
\begin{align}
M\equiv
\begin{pmatrix}
0 & c           &              &                 \cr
   &   0         & \ddots  &                  \cr 
   &              &  \ddots &               c      &                \cr 
  &              &             &                 0      &    b          &             &          \cr
   &              &             &                         &    \ddots & \ddots  &         \cr
   &              &             &                         &                &      0      &    b    \cr
b  &             &             &                          &                &             &     0  \cr 
\end{pmatrix},\ \, 
\bar M\equiv
\begin{pmatrix}
0 &             &             &            &             &            &     c \cr
b &   0         &            &            &            &                 \cr 
   &  \ddots &  \ddots &          &          &                &\cr 
   &              &     b      &       0      &          &                \cr
   &              &             &      c      &    \ddots        &             &             \cr
   &              &             &             & \ddots  &       0         &                \cr
  &             &             &            &              &      c          &       0        \cr 
\end{pmatrix}.
\end{align}
Note that there is a sequence of $l$ $c$'s in $M$ and a sequence of $l$ $b$'s in $\bar M$. 
These matrices satisfy $M^{k-1} = b^{k-l-1}c^{l-1} \bar M$ and $[\Gamma, M] = [\Gamma^{-1},\bar M]$
which result in (CCR0). 
Of course, there is some ambiguity of taking permutation among the positions of $b$ and $c$.%
\footnote{See Appendix \ref{DirectEvaluationThreeCutStringEq} where 
we explicitly study all possible solutions of the string equations 
in the three-cut cases.}
Also we have the constraint of eq.~\eq{HoneConst}, which means that 
\begin{align}
\rho_{\hat p}(z) = (-k)^{\hat p} \Bigl(z^{\hat p-1}+ \gamma_{\hat p} z^{\hat p-2}+\cdots\Bigr).
\end{align}
With the above setting, we can solve equation \eq{CISYeq}. 
From straightforward evaluation of eq.~\eq{CISYeq}, one observes that the parameters $b$ and $c$ satisfy 
\begin{align}
b+c=0,
\end{align}
except for $\hat p=1$ case; it can also be seen that $\gamma_{\hat q }$ is related to $\eta_{\hat q}(z)$ as 
\begin{align}
\eta_{\hat q}(z) = -k^{\hat q} \Bigl( z^{\hat q-1}+ \gamma_{\hat q}z^{\hat q-2}+\cdots \Bigr). 
\end{align}
For unitary series $\hat p=\hat q$, one can see that they satisfy 
\begin{align}
\eta_{\hat p}(z)= \rho_{\hat p}(-z),  \label{ExchangeABXP}
\end{align}
which leads to simple solutions of eq.~\eq{CISYeq}; for instance, we list first five solutions of $(k,l)=(3,2)$ in the following:%
\footnote{It should be stressed that 
the ansatz \eq{CISYansatz} can generate 
polynomials (i.e. solutions of eq.~\eq{CISYeq}) in all the $(\hat p,\hat q)$ critical points 
not only in the unitary cases $\hat p=\hat q$ listed below. }
\begin{align}
\rho_{2}(z)&=(-3)^2\Bigl(z+\frac{c}{3}\Bigr), \\
\rho_{3}(z)&=(-3)^3\Bigl(z^2+\frac{c}{3}z-\frac{8c^2}{27}\Bigr), \\
\rho_{4}(z)&=(-3)^4\Bigl(z^3+\frac{c}{3}z^2-\frac{5c^2}{9}z-\frac{7c^3}{81}\Bigr), \\
\rho_{5}(z)&=(-3)^5\Bigl(z^4+\frac{c}{3}z^3-\frac{17c^2}{21}z^2-\frac{97c^3}{567}z+\frac{128c^4}{1701}\Bigr),\\
&\  ... \nn
\end{align}
Our claim is that the solutions for the unitary cases $(\hat p=\hat q)$ 
can be expressed by the Jacobi polynomials, $P_n^{(\alpha,\beta)}(z)$.
In the diagonalized form \eq{DiagAC}, the solution is given as 
\begin{align}
\mathcal A(n,Z) 
&\simeq a^{\hat p/2}\, \frac{\Gamma(\hat p)^2}{2c \Gamma(2\hat p-1)} 
\bigl(-2c k\lambda\bigr)^{\hat p} 
P^{(\frac{2l-k}{k},-\frac{2l-k}{k})}_{\hat p-1}(z/c) \, 
\sqrt[k]{\bigl(z-c\bigr)^l\bigl(z+c\bigr)^{k-l}} \times \Omega \nn\\
\mathcal C^{\rm T}(n,Z) 
&\simeq  a^{\hat q/2}\, \frac{\Gamma(\hat p)^2}{2c \Gamma(2\hat p-1)} 
\bigl(2c k\lambda\bigr)^{\hat p} 
P^{(-\frac{2l-k}{k},\frac{2l-k}{k})}_{\hat p-1}(z/c) \, 
\sqrt[k]{\bigl(z-c\bigr)^{k-l}\bigl(z+c\bigr)^{l}}\times \Omega^{-1}, \label{CISYsol}
\end{align}
with
\begin{align}
(-1)^{\hat p} \bigl(2c k\bigr)^{\hat p+\hat q-1}
=-\frac{\pi}{k\hat p(2\hat p-1)\sin \bigl(\pi \frac{2l-k}{k}\bigr)B(\hat p+\frac{2l-k}{k},\hat p-\frac{2l-k}{k})B(\hat p,\hat p)}.
\end{align}
Here $B(x,y)=\Gamma(x)\Gamma(y)/\Gamma(x+y)$ is the beta function 
and also $\Omega$ is given in \eq{DefOmega}. 
Note that the relation \eq{ExchangeABXP} is essentially the identity \eq{ExchangeAB} of the Jacobi polynomials, $P^{(\alpha,\beta)}_n(z)=(-1)^n P^{(\beta,\alpha)}_n(-z)$. 
We show that the above expressions of $\mathcal A$ and $\mathcal C$ solve eq.~\eq{CISYeq} for general indices $(k,l)$ 
in Appendix \ref{JacobiPProof}. 

\subsection{Some other solutions for the $\hat p=1$ cases}

For the three-cut critical points of $(1,\hat q)$, 
some of the solutions can be written in terms of the Chebyshev polynomials.
Since this solution is very similar to the solutions of one-cut and two-cut cases, 
it provides simple examples to see what happens in multi-cut cases. 

In the case of $\hat p=1$, the constraints \eq{HoneConst} turns out to be 
\begin{align}
\gamma_{\hat p}= \frac{(k-l)b +lc}{k}=0.
\end{align}
When we have $(k,l)=(3,2)$, then $b+2c=0$.  A solution which satisfies this constraint is given by 
\begin{align}
\mathcal A(n,Z) 
&\simeq a^{1/2}\, (-3)\lambda \Bigl(2\cosh\bigl(\frac{3}{2}\tau\bigr)\Bigr)^{\frac{2}{3}}\times \Omega
=a^{1/2}\,(-3)\lambda \sqrt[3]{\bigl(z-c\bigr)^2\bigl(z+2c\bigr)}\times \Omega,\nn\\
\mathcal C^{\rm T}(n,Z) 
&\simeq a^{\hat q/2}\,c^{-1}\Bigl(\frac{3c \lambda}{2}\Bigr)^{\hat q} \Bigl(2\cosh\bigl(\frac{3}{2}\tau\bigr)\Bigr)^{\frac{1}{3}}\, 
\cosh \bigl(\frac{2\hat q-1}{2}\tau\bigr)\times \Omega^{-1}\nn\\
&= a^{\hat q/2}\,c^{-1}\Bigl(\frac{3c \lambda}{2}\Bigr)^{\hat q}\, W_{\hat q-1}\bigl(z/c\bigr)
\sqrt[3]{\bigl(z-c\bigr)\bigl(z+2c\bigr)^2}\times \Omega^{-1}, 
\label{Cheb3cutSol}
\end{align}
with $z=c \cosh \tau$. 
Here the polynomial $W_n(z)$ is the Chebyshev polynomial of the fourth kind:
\begin{align}
W_n\bigl(\cosh(\tau)\bigr)= \frac{\cosh(\frac{2n + 1}{2}\tau)}{\cosh(\frac{\tau}{2})},\qquad 
W_n(z) = T_n(z)-(z-1) U_{n-1}(z). 
\end{align}
This satisfies the deformed equation \eq{DeformedDKKeq} with the canonical pair $(w,\tilde \mu)$,%
\footnote{Note that there is no $(1,2)$ critical point in the $\mathbb Z_3$-symmetric 3-cut cases. }
\begin{align}
[w, \tilde \mu^n] = n \kappa \tilde\mu^n,\qquad 
w(z)\sim \cosh (\hat q-2)\tau \sim  T_{|\hat q-2|}(z/c),
\end{align}
with some proper normalization. 

\subsection{Geometry of the multi-cut solutions \label{secGeometryMultiCut}}

Here we consider geometry of the multi-cut solutions we have obtained. The purpose is two-folded; first, it serves as a check of our ansatz \eq{CISYansatz}; second, it provides some physical intuition to the multi-cut system. 
The algebraic equation of the $\hat p=1$ solution \eq{Cheb3cutSol} is easily written down as 
\begin{align}
F(\zeta,Q) 
&= 2 \mu^{3/4}\zeta^{3/2} \Bigl(T_3\bigl(Q/2\mu^{1/4}\zeta^{1/2}\bigr)-T_{2\hat q-1}\bigl(\zeta^{3/2}/2\sqrt{\mu}\bigr)\Bigr) 
= 4\mu^{3/2} \tilde x \Bigl(T_3(\tilde y) - T_{2\hat q-1}(\tilde x)\Bigr) \nn\\
&= Q^3 - 3\mu \zeta Q -2 \zeta^{3/2}\, T_{2\hat q-1}\bigl(\zeta^{3/2}/2\sqrt{\mu}\bigr)=0, 
\end{align}
where we define 
\begin{align}
\tilde x \equiv \frac{1}{2}\Bigl(\frac{\zeta}{\mu}\Bigr) ^{3/2}= \cosh\bigl(\frac{3}{2}\tau),\qquad 
\tilde y \equiv \frac{Q}{2( \mu\zeta)^{1/2}} = \cosh\bigl(\frac{2\hat q-1}{2}\tau\bigr),
\end{align}
and $(\zeta,Q)$ are the eigenvalues of 
the operators $\bP \sim \mathcal A$ and $\bQ \sim -\mathcal C^{\rm T}$
(see eq.~\eq{ToLax1}). 
One can easily see that this equation is an algebraic equation in $\zeta$ and $Q$. 
Note that there are three branches, $Q=Q_m(\zeta)$ $(m=0,1,2)$, of the algebraic equation, 
\begin{align}
F(\zeta,Q)=\bigl(Q-Q_0(\zeta)\bigr)\bigl(Q-Q_1(\zeta)\bigr)\bigl(Q-Q_2(\zeta)\bigr)=0,
\end{align}
and this defines the three sheets of the Riemann surface.  These branches are 
related by the following shift of the parameter $\tau$, 
\begin{align}
\frac{3}{2}\tau \to \frac{3}{2}\tau +2 \pi i m\qquad (m = 0,1,2), \label{ShiftBranches}
\end{align}
in the expression of eq.~\eq{Cheb3cutSol}, since this shift does not change the coordinate $\zeta$. 

This algebraic equation has singularities $(\zeta_*,Q_*)$, 
\begin{align}
0= F(\zeta_*,Q_*)=\frac{\del F(\zeta_*,Q_*)}{\del \zeta}= \frac{\del F(\zeta_*,Q_*)}{\del Q}, 
\end{align}
at the origin $(\zeta_*,Q_*)=(0,0)$ and at $(\zeta_*,Q_*)=(\zeta_{n,j},Q_{n,j})$ of
\begin{align}
\zeta_{n,j} 
= \omega^{j}\,\mu \Bigl(2 \cos \frac{n\pi}{2\hat q-1}\Bigr)^{2/3},\qquad
Q_{n,j}
= \omega^{-j}\,2\mu  \Bigl(2 \cos \frac{n \pi}{2\hat q-1}\Bigr)^{1/3} \cos\frac{n\pi}{3}, \label{SingularPointsPeq1}
\end{align}
with 
\begin{align}
3 \nmid n,\qquad 2\hat q-1 \nmid n,\qquad n \in \mathbb Z,
\end{align}
and $j=0,1,2$. The geometrical meaning of singular points is that 
different branches of Riemann surface intersect at the points. 
Since the singular point at the origin does not change 
under the shift of $\tau$ of eq.~\eq{ShiftBranches}, 
this singular point joins the three sheets at the same time. 
On the other hand, the singular points given in \eq{SingularPointsPeq1} only join two sheets at the same time. 
The reason is following: The invariance of the singular points \eq{SingularPointsPeq1} 
under the shift \eq{ShiftBranches} is essentially given by the invariance of the factor $\cos\frac{n\pi}{3}$ 
in $Q_{n,j}$, that is, 
\begin{align}
\cos\frac{n\pi}{3} = 
\cos\Bigl[\frac{n+(2\hat q-1)m}{3}\pi \Bigr] \qquad (m=0,1,2).
\end{align}
Since the $\mathbb Z_3$ symmetric $\hat p=1$ critical points 
satisfy $3 \nmid \hat q- 2 \, (\Rightarrow 3\nmid 2\hat q-1)$, the above condition 
is only satisfied by two different values of $m\,(=0,1,2)$. This means that the singular points only join 
two sheets at the same time. 

\begin{figure}[htbp]
 \begin{center}
  \includegraphics[scale=1]{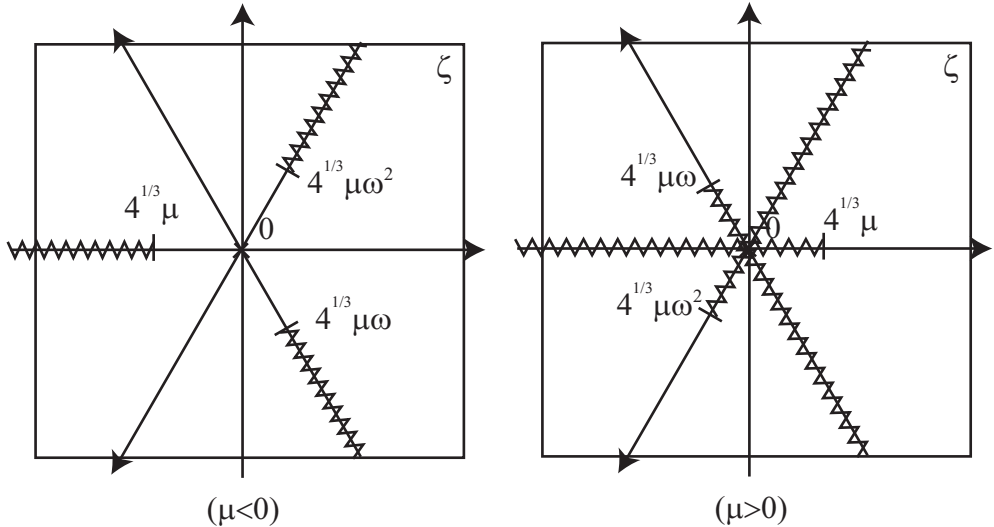}
 \end{center}
 \caption{\footnotesize 
The algebraic curve of the 3-cut $(1,1)$ case and $l=2$, which is given by 
$F(Q,\zeta)=Q^3-3\mu \zeta-\zeta^3=0$. 
The branch points are at $\zeta = 4^{1/3}\mu \omega^j$ $(j=0,1,2)$. 
The singularity is at the origin $\zeta=0$.}
 \label{3CutGeometry}
\end{figure}

\begin{figure}[htbp]
 \begin{center}
  \includegraphics[scale=1]{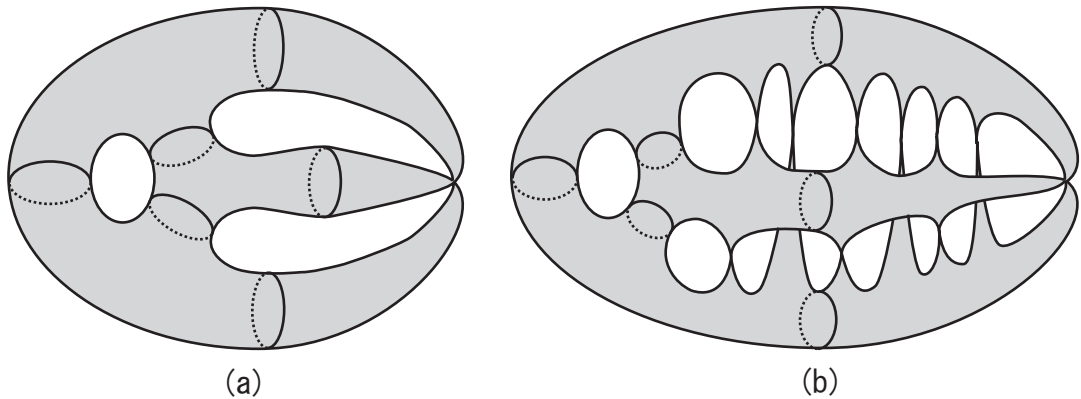}
 \end{center}
 \caption{\footnotesize 
The topology of the curves are shown in the case of (a) $(\hat p,\hat q)=(1,1)$ 
and (b) $(\hat p,\hat q)=(1,3)$. 
The junction of three sheets is the singular point at the origin. The torus comes from the cubic-root cut. 
Two-sheet junctions in (b) are the singular points of \eq{SingularPointsPeq1}, the number of which is nine. }
 \label{3CutTopology}
\end{figure}

The branch points $(\zeta_b,Q_b)$ of the function $Q(\zeta)$,
\begin{align}
0= F(\zeta_b,Q_b)=\frac{\del F(\zeta_b,Q_b)}{\del Q}, \qquad 0\neq \frac{\del F(\zeta_b,Q_b)}{\del \zeta}, 
\end{align}
appear when
\begin{align}
3 \nmid n,\qquad 2\hat q-1\mid n \qquad \Leftrightarrow \qquad 
\zeta =  2^{2/3}\mu \, \omega^{j} \qquad (j=0,1,2).
\end{align}
One can see that 
the singular points \eq{SingularPointsPeq1} appear between the origin and the branch points 
in the $\mathbb Z_3$ symmetric manner (i.e.~$\zeta \to \omega \zeta$). 
The simplest curve (the $(1,1)$ cases), 
\begin{align}
F(Q,\zeta)=Q^3-3\mu\zeta Q-\zeta^3=0, \label{AlgEqOneOne}
\end{align}
is drawn in Fig.~\ref{3CutGeometry}. The topology of the curves is also shown in Fig.~\ref{3CutTopology}. 
Note that there is a non-trivial handle in the algebraic curve. This comes from the cubic root in the curve. 
The appearance of non-trivial handle is explained in Appendix \ref{AppCubicRoots}, where 
some basic properties of cubic roots are summarized. 

We also write the algebraic curve of the $(2,2)$ cases as
\begin{align}
F(\zeta,Q)= (Q^3-\zeta^3)^2  +3f^8 \zeta Q -2 f^6 (Q^3+\zeta^3)=0,
\end{align}
with $f=3c/2$ and 
\begin{align}
\zeta =\mu^{2/3} \bigl(z+\frac{f}{2}\bigr)\,\sqrt[3]{\bigl(z-\frac{3f}{2}\bigr)^2\bigl(z+\frac{3f}{2}\bigr)}, \quad
Q=\mu^{2/3} \bigl(z-\frac{f}{2}\bigr)\,\sqrt[3]{\bigl(z-\frac{3f}{2}\bigr)\bigl(z+\frac{3f}{2}\bigr)^2}. 
\end{align}
One can see that this solution also has three branch points. 

\begin{figure}[htbp]
 \begin{center}
  \includegraphics[scale=1]{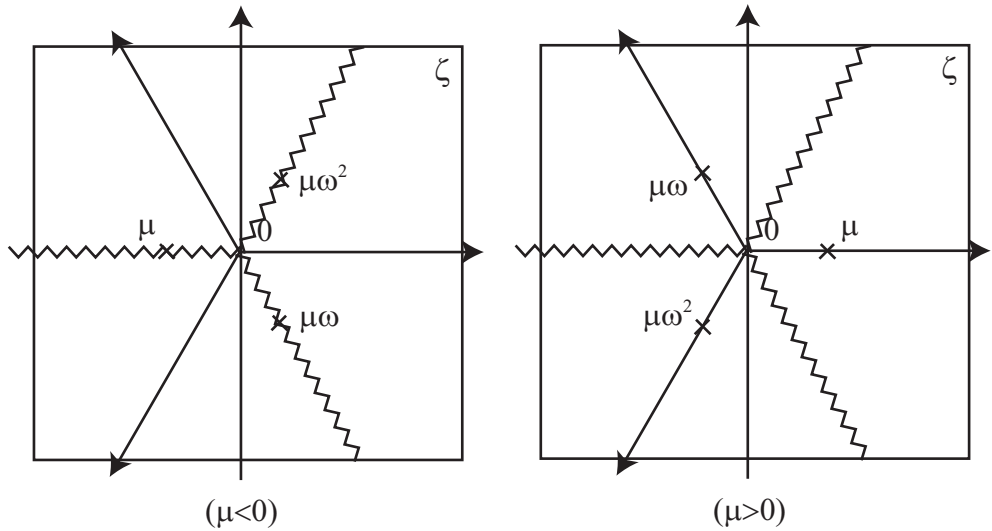}
 \end{center}
 \caption{\footnotesize 
The algebraic curve of the 3-cut $(1,1)$ case and $l=0$, which is given by 
$F(Q,\zeta)=Q^3-3\mu \zeta Q-\zeta^3-\mu^3=0$. 
The singularities are at $\zeta =\mu \omega^j$ $(j=0,1,2)$. 
There is no branch point, but we explicitly show the cut which indicates the eigenvalue condensation. 
This cut has a junction of three cuts, which is allowed in the cubic root. 
Since there is no branch point attached by this cut, 
one can also push this cut away to infinity. }
 \label{3CutonecutGeometry}
\end{figure}

\begin{figure}[htbp]
 \begin{center}
  \includegraphics[scale=1]{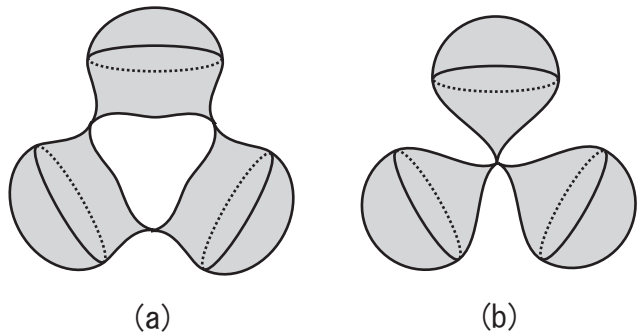}
 \end{center}
 \caption{\footnotesize 
The topology of the curve is shown in the case of $(\hat p,\hat q)=(1,1)$. 
(a) is the case of finite $\mu$ and (b) is the case of $\mu=0$. }
 \label{3CutonecutTopology}
\end{figure}

Next we consider the counterpart solution of ``one-cut phase'' 
in the two-cut matrix models, i.e. $l=0$. 
We here adopt the expression \eq{OneCutThreeCut} derived 
from direct analysis of the string equation, and leave the details in Appendix \ref{DirectEvaluationThreeCutStringEq}. 
The diagonalization of the matrix operators gives
\begin{align}
\mathcal A \sim \bP \simeq  \mu \,z\times \Omega,\qquad 
-\mathcal C^{\rm T}\sim \bQ \simeq \mu\,(z+1)\times \Omega^{-1}. 
\end{align}
Here we put $\mu = t$ and $z= \kappa \mu \del_\mu= \kappa t \del_t$ 
(see eq.~\eq{DefOfsmallz}). 
They give three algebraic equations:
\begin{align}
F^{(n)}(\zeta,Q)=\omega^{n}Q-\omega^{-n} \zeta- \mu=0 \qquad (n=0,1,2).  
\end{align}
Since the phase $\omega$ labels branches of the solution, 
the complete algebraic equation should be given by multiplication of these equations, 
\begin{align}
F(\zeta,Q)&\equiv F^{(0)}(\zeta,Q)\cdot F^{(1)}(\zeta,Q)\cdot F^{(2)}(\zeta,Q)
=Q^3-3\mu \zeta Q-\zeta^3 -\mu^3=0.  \label{ProdThreeAlgOneCut}
\end{align}
This curve has no branch point but has three singular points which are at 
$(\zeta_*,Q_*)=(\omega^j \mu,-\omega^{-j}\mu )$ $(j=0,1,2)$. 
One can also see that these singular points only join two sheets at the same time. 
The curve is drawn in Fig.~\ref{3CutonecutGeometry}. 
The topology of the curve is also shown in Fig.~\ref{3CutonecutTopology}. 

Here in Fig.~\ref{3CutonecutGeometry}, 
we draw a single connected cut which does not attach any branch points. 
This cut has a junction of three cuts. This kind of junction is always allowed in general $k$-th root 
(See Appendix \ref{AppCubicRoots}). Since there is no branch point here, 
one can also push this cut away to infinity. 

Although this cut can be pushed away, this has a physical meaning as the eigenvalue condensation. 
There is a similar phenomenon in the two-cut matrix models \cite{UniCom,SeSh}: 
In the one-cut phase of two-cut matrix models, the ``two cuts'' are completely connected with each other 
and form ``a single cut'', which run along the whole real axes. Since the branch points disappear 
in this procedure, one can also push this cut away to infinity. 
In our three-cut case, since the cubic-root system admits a $\mathbb Z_3$ symmetric cut which has 
a three-cut junction without any branch points, 
it is natural to interpret that this system still have this kind of cut which indicates the eigenvalue density. 

\section{Conclusion and discussion \label{secDiscussion}} 

In this paper, we have provided quantitative analysis of $(\hat p,\hat q)$ critical points 
in the multi-cut two-matrix models based on an extension of the prescription developed 
by Daul, Kazakov and Kostov in the one-cut two-matrix models \cite{DKK}. Right at the critical points, 
we identified the minimal construction of critical resolvents and potentials 
which can give $(\hat p,\hat q)$ critical points of the multi-cut two-matrix models. 
The hermiticity of the Lax operators realized in the multi-cut critical points is identified. 

We have also studied off-critical behavior of macroscopic loop amplitudes. 
In this study, we identified an ansatz for non-polynomial parts of amplitudes 
in the $\mathbb Z_k$ symmetric background. This ansatz enables us to 
generate several formulas for off-critical macroscopic loop amplitudes: 
The intriguing one is the formula written with the Jacobi polynomials 
in the unitary cases $\hat p=\hat q$. 


Several future issues about our results are in order:
\begin{itemize}
\item[1.] 
In identification of the critical points, we only consider the  system of equations \eq{primitiveCCR} 
and \eq{StEqMatrix}. In this investigation, there is no essential difference 
between the real potential models and the $\omega^{1/2}$-rotated models, 
in the sense of existence of $\mathbb Z_k$ symmetry breaking critical points. 
As an independent check, one can perform direct evaluations 
using the Monte Carlo approach \cite{monte}, 
since our real-potential solutions can avoid the difficulty of complex numbers in numerical analysis.
\item[2.]
Although we have found the formula for the unitary cases \eq{CISYsol}, 
the expression in terms of algebraic equation is missing. 
Since underlying structure is multi-component KP hierarchy \cite{fi1},
they should be expressed by the algebraic equation  of $(k\hat q,k\hat p)$ order, 
$F(x,Q)=0$ \cite{fim,fi1}.
\item[3.]
From our analysis, 
the system of Douglas equation in the $\mathbb Z_k$ symmetric background admits 
one discrete parameter $l$ in the expression \eq{CISYsol}. The off-critical system 
is, however, controlled only by a dimension-full perturbation $\mu$. 
This parameter $l$ seems mysterious because 
one may expect that there are at most two meaningful geometries 
which respect the sign of $\mu$.%
\footnote{For example, see the two-cut case \cite{UniCom}. }
The physical meaning of the parameter $l$, therefore, should be an interesting question to be investigated. 
\item[4.] 
Since our analysis only concentrates on the possible asymptotic (weak coupling) geometries 
of macroscopic loop amplitudes, the consideration about interpolation between different 
asymptotic geometries (which might be expressed by $l$) is necessary. This should be 
accomplished by non-perturbative analysis of the string equation 
as is in the two-cut one-matrix models \cite{UniCom,SeSh2}.  
\item[5.]
Since we have known about macroscopic loop amplitudes 
in the $\mathbb Z_k$ symmetric critical points, 
it should be interesting to identify the Liouville continuum formulation corresponding 
to the $\mathbb Z_k$ symmetric critical points. 
\item[6.]
To see the relation to minimal fractional superstring theory, we need to 
calculate the amplitudes in the $\mathbb Z_k$ symmetry breaking critical points. 
Since minimal fractional superstring theory is a natural generalization of 
bosonic and super string theories, 
the formula could be probably written by the Jacobi polynomials of the ultraspherical sequence, 
$P^{(\alpha,\alpha)}_n(z)$. 
\item[7.] 
One of the important features of the multi-cut matrix models is 
the $\mathbb Z_k$ charge of D-branes. In this sense, it should be important 
to see the annulus amplitude of macroscopic loop amplitudes \cite{Okuyama,Irie1}. 
To see the correlations among $\mathbb Z_k$ charges through annulus amplitudes of the matrix models 
is also an interesting problem to be studied. 
\end{itemize}

\vspace{1cm}
\noindent
{\bf \large Acknowledgment}  
\vspace{0.2cm}

\noindent
The authors would like to thank 
Shoichi Kawamoto for the useful discussion and comments on this work. 
We also would like to 
thank Chien-Ho Chen, Wei-Ming Chen, Kazuyuki Furuuchi, Pei-Ming Ho, Hiroshi Isono, 
Shoichi Kawamoto, Jen-Chi Lee, Feng-Li Lin, Tomohisa Takimi, Dan Tomino and 
Wen-Yu Wen for attending minimal string study meeting and sharing valuable discussions. 
Shih would like to show his gratitude to people in Taiwan string focus group 
where most of the work has been done. 
CT is supported by National Science Council of Taiwan under the contract 
No.~96-2112-M-021-002-MY3. The authors are also supported 
by National Center for Theoretical Science under NSC No.~98-2119-M-002-001.

\appendix

\section{The relation between the resolvents and orthonormal polynomials \label{RelationResolventOrthPoly}}

An intuitive understanding of 
the relation between the resolvents and orthonormal polynomials is following:
There is a useful expression for the general orthonormal polynomials \cite{GrossMigdal2}:
\begin{align}
\left\{
\begin{array}{l}
\alpha_n(x) = \dfrac{1}{\sqrt{h_n}}\Bigl<\det\bigl(x-X_{(n)}\bigr)\Bigr>_{(n)}
= \dfrac{1}{\sqrt{h_n}}\Bigl<e^{\tr \ln (x-X_{(n)})}\Bigr>_{(n)},
\cr
\beta_n(y) = \dfrac{1}{\sqrt{h_n}}\Bigl<\det\bigl(y-Y_{(n)}\bigr)\Bigr>_{(n)}
= \dfrac{1}{\sqrt{h_n}}\Bigl<e^{\tr \ln (y-Y_{(n)})}\Bigr>_{(n)}.  \label{OrthDet} 
\end{array}
\right.
\end{align}
Here the expectation value is defined by path integral over the $n\times n$ truncated matrices $X_{(n)}$ and $Y_{(n)}$ as 
\begin{align}
\Bigl<\det\bigl(x-X_{(n)}\bigr)\Bigr>_{(n)} \equiv 
\frac{1}{\mathcal Z_{(n)}} 
\int dX_{(n)} dY_{(n)} \, e^{-N\tr w(X_{(n)},Y_{(n)})} 
\Bigl( \det\bigl(x-X_{(n)}\bigr) \Bigr),
\end{align}
with 
$\mathcal Z_{(n)}=n! \prod_{l=0}^{n-1}h_l$. 
This implies the following powerful relation in the large $N\,(\sim n)$ limit, 
\begin{align}
A \sim x,\qquad B \sim \frac{1}{N} \vev{\tr \frac{1}{x-X}},\qquad 
C \sim y,\qquad D \sim \frac{1}{N} \vev{\tr \frac{1}{y-Y}}.
\end{align}
This relation has been argued and proved from several viewpoints 
\cite{DKK} \cite{Moore,MMSS} \cite{fim,fi1}. With this relation, solving 
the eigenvalue problems \eq{OrthRecAB} and \eq{OrthRecCD} in the large $N$ limit
is translated into obtaining the resolvents of the matrix model. 

\section{Examples of critical potentials \label{ListOfCriticalPotentials}}
\subsection{Two-cut one-matrix models ($\hat p=1$) \label{OneMatCritiPotentials}}

To compare our calculations with previous results in \cite{HMPN}, we 
make a list of the critical potentials (of real coefficients) $V_{\hat q}(x)$.%
\footnote{One-matrix models can have a natural embedding in the two-matrix-model setup 
by putting $V(y)=y^2/2$. The Gaussian integration over $y$ for the two-matrix model 
partition function gives back to the one-matrix models. 
However note that the potential 
$V_{\hat q}(x)$ here is defined as the usual one-matrix potential 
$Z= \int dX e^{-N\tr V(X)}$. } 
Note that our (real-coefficient) critical potentials are related to the (complex-coefficient) 
critical potentials in \cite{HMPN} 
by replacing $b \to i b$ and $g \to i g$.%
\footnote{Up to this replacement, we use the same notation as in \cite{HMPN}. 
The parameter $b$ has a different meaning in the main text. }
The parameter $b$ corresponds to the critical point of 
eigenvalue ($x_*=b$), which one can always put zero, $b=0$. The parameter $g$ is an irrelevant perturbation and 
essentially does not change the system but is necessary to obtain the critical points of odd $\hat q$ 
(one can always choose any finite non-zero number).
\begin{align}
V_1(x)&=
\left(-4 b+2 b^3-2 g+b^2 g\right) x
+\frac{1}{2}  \left(-2-3 b^2-2 b g\right) x^2
+\frac{g}{3}x^3 
+\frac{x^4}{4}\\
V_2(x)&=
\left(2 b-b^3\right) x+\frac{1}{2} \left(-2+3 b^2\right) x^2-b x^3+\frac{x^4}{4}, \\
V_3(x)&=
\left(-4 b-4 b^3+2 b^5-2 g-2 b^2 g+b^4 g\right)x +\nn\\
&\,\, +\frac{1}{4} \left(-2+14 b^2-15 b^4+8 b g-8 b^3 g\right) x^2 
-\frac{2}{3} \left(b-5 b^3+g-3 b^2 g\right) x^3 + \nn\\
&\,\, +\frac{1}{4} \left(-1-5 b^2-4 b g\right)x^4+\frac{ g}{5} x^5+
\frac{x^6}{12},\\
V_4(x)&=
\frac{1}{2} \left(2 b+2 b^3-b^5\right) x+\frac{1}{4} \left(-2-6 b^2+5 b^4\right) x^2
+\frac{1}{3} \left(3 b-5 b^3\right) x^3+\nn\\
&\,\,+\frac{1}{4} \left(-1+5 b^2\right) x^4-\frac{b x^5}{2}+\frac{x^6}{12}, \\
V_5(x)&=
\frac{1}{5} \left(-24 b-12 b^3-12 b^5+6 b^7-20 g-10 b^2 g-10 b^4 g+5 b^6 g\right) x+\nn\\
&\,\,+\frac{1}{10} \left(-4+22 b^2+46 b^4-35 b^6+20 b g+40 b^3 g-30 b^5 g\right) x^2+\nn\\
&\,\,+\frac{1}{15} \left(-8 b-64 b^3+84 b^5-10 g-60 b^2 g+75 b^4 g\right) x^3+\nn\\
&\,\,+\frac{1}{20} \left(-2+36 b^2-105 b^4+40 b g-100 b^3 g\right) x^4+\nn\\
&\,\,+\frac{1}{25} \left(-4 b+70 b^3-10 g+75 b^2 g\right) x^5+\nn\\
&\,\,+\left(-\frac{1}{15}-\frac{7 b^2}{10}-b g\right) x^6+\frac{g x^7}{7}+\frac{x^8}{40}.
\end{align}
The critical resolvents are $Q_{\hat q}$ defined by the operator 
$ \mathcal B^{(\rm real)}\, (\sim N^{-1} \del/\del x)$ as
\begin{align}
Q_{\hat q}(Z) \equiv \frac{1}{2}\Bigl(
\mathcal B^{(\rm real)}(*,Z)
+\mathcal B^{(\rm real)}{}^{\rm T}(*,Z)\Bigr),
\end{align}
and 
\begin{align}
Q_1(Z)&= \frac{1}{2}
\begin{pmatrix}
\dfrac{(1-Z^2)(1+Z^{2})}{Z^2} \cdot (3 b + g) &
\dfrac{(1 - Z^{2})^2  (1 + Z^{2}) }{Z^2} \cdot Z\cr
-\dfrac{(1 - Z^{2})^2  (1 + Z^{2}) }{Z^2} \cdot Z^{-1} &
\dfrac{(1-Z^2)(1+Z^{2})}{Z^2} \cdot (3 b + g)
\end{pmatrix} \\
Q_2(Z)&= \frac{1}{2}
\begin{pmatrix}
0 &
\dfrac{(1 - Z^2)^2 (1 + Z^2)}{Z^3} \cdot Z \cr
-\dfrac{(1 - Z^2)^2 (1 + Z^2)}{Z^3} \cdot Z^{-1}& 0
\end{pmatrix} \\
Q_3(Z)&= \frac{1}{2}
\begin{pmatrix}
-\dfrac{(1 -Z^2)^3 (1 + Z^2) }{2 Z^4}\cdot (5 b + 2 g) &
-\dfrac{(1 - Z^2)^4  (1 + Z^2)}{2 Z^5}\cdot Z \cr
\dfrac{(1 - Z^2)^4  (1 + Z^2)}{2 Z^5}\cdot Z^{-1} &
-\dfrac{(1 -Z^2)^3 (1 + Z^2) }{2 Z^4} \cdot (5 b + 2 g)
\end{pmatrix} \\
Q_4(Z)&= \frac{1}{2}
\begin{pmatrix}
0 &
-\dfrac{(1 - Z^2)^4  (1 + Z^2)}{2 Z^5}\cdot Z \cr
\dfrac{(1 - Z^2)^4  (1 + Z^2)}{2 Z^5}\cdot Z^{-1} &0
\end{pmatrix} \\
Q_5(Z)&= \frac{1}{2}
\begin{pmatrix}
\dfrac{(1 - Z^2)^5 (1 + Z^2)}{5Z^6} \cdot (7 b + 5 g) &
\dfrac{(1 - Z^2)^6  (1 + Z^2)}{5 Z^7}\cdot Z \cr
-\dfrac{(1 - Z^2)^6  (1 + Z^2)}{5 Z^7}\cdot Z^{-1} &
\dfrac{(1 - Z^2)^5 (1 + Z^2)}{5Z^6} \cdot (7 b + 5 g) 
\end{pmatrix}. 
\end{align}

\subsection{Two-cut two-matrix models ($\hat p\neq 1$)}
Here we turn on the coefficient $\theta $ in the potential 
\begin{align}
w(x,y)= V_1(x) +V_2(y) + \theta xy,
\end{align}
and put $R_*=1$. With suitable choices of the value of $\theta$ or $R_*$ 
(one of them is a free parameter), 
we can obtain critical potentials with smaller coefficients. 
For example, two-cut $(4,5)$ unitary critical potential with $\theta=1$ is given as 
\begin{align}
\tilde V^{(4)}(x)=-4 x^2+5670 x^4-176400 x^6+1157625 x^8 \qquad (R_* = 1/210),
\end{align}
and one can see that the coefficients are larger than those of the $R_*=1$ case \eq{twocut(4,5)critiWTheta}. 
Also the critical point in eigenvalue space $(x_*,y_*)$ are chosen to be the origin $(0,0)$. 
By making a shift of variable $x \to \alpha x + \beta$, 
we can eliminate the parameter $b$ like in Appendix \ref{OneMatCritiPotentials}. 

\paragraph{The unitary cases:} 
\ \vspace{0.3cm} \\  
The critical potential of the unitary cases 
$V_1(x)=V_2(x)=V^{(\hat p)}(x)$ (i.e. $(\hat p,\hat p+1)$ unitary minimal superstrings) are
\begin{align}
V^{(2)}(x)&=-2 x^2 + \frac{x^4}{4},\qquad (\theta=6), \\
V^{(3)}(x)&=3 x^2 - 3 x^4 + \frac{x^6}{6},\qquad (\theta=40), \\
V^{(4)}(x)&=-4 x^2 + 27 x^4 - 4 x^6 + \frac{x^8}{8},\qquad (\theta=210), \label{twocut(4,5)critiWTheta} \\
V^{(5)}(x)&=5 x^2 - 220 x^4 + 70 x^6 - 5 x^8 + \frac{x^{10}}{10},\qquad(\theta=1008), \\
V^{(6)}(x)&=-6 x^2+\frac{6825 x^4}{4}-\frac{3250 x^6}{3}+\frac{555 x^8}{4}-6 x^{10}
+\frac{x^{12}}{12},
\qquad (\theta=4620).
\end{align}
\paragraph{The $(\hat p,\hat q)$ cases:}
\ \vspace{0.3cm} \\  
The potentials $(V_1^{(\hat q)}(x),V_2^{(\hat p)}(y))$ of general $(\hat p, \hat q)$ cases are 
\begin{align}
&\underline{(\hat p,\hat q)=(2,3)\quad (\theta=14)}\quad    
V_1^{(3)}(x)=-\dfrac{7 x^4}{4}+\dfrac{x^6}{6},\qquad
V_2^{(2)}(y)=-\frac{7 y^2}{2}+\frac{y^4}{4},\qquad   \\
&\underline{(\hat p,\hat q)=(2,5)\quad (\theta=39)} \quad 
\left\{
\begin{array}{l}
V_1^{(5)}(x)=-\dfrac{13 x^4}{4}+\dfrac{13 x^6}{2}
                         -\dfrac{13 x^8}{8}+\dfrac{x^{10}}{10},\cr
V_2^{(2)}(y)=-\dfrac{13 y^2}{2}+\dfrac{y^4}{4}, 
\end{array}
\right. \\
&\underline{(\hat p,\hat q)=(2,7)\quad (\theta=76)} \quad 
\left\{
\begin{array}{l}
V_1^{(7)}(x)=\dfrac{95 x^6}{6}-\dfrac{57 x^8}{2}
+\dfrac{57 x^{10}}{5}-\dfrac{19 x^{12}}{12}+\dfrac{x^{14}}{14}, \cr
V_2^{(2)}(y)=-\dfrac{19 y^2}{2}+\dfrac{y^4}{4},
\end{array}
\right. \\
&\underline{(\hat p,\hat q)=(3,4)\quad (\theta=85)} \quad 
 \left\{
\begin{array}{l}
V_1^{(4)}(x)=\dfrac{51 x^4}{4}-\dfrac{17 x^6}{6}+\dfrac{x^8}{8}, \cr
V_2^{(3)}(y)=\dfrac{17 y^2}{2}-\dfrac{17 y^4}{4}+\dfrac{y^6}{6},
\end{array}
\right. \\
&\underline{(\hat p,\hat q)=(3,5)\quad (\theta=154)} \quad
 \left\{
\begin{array}{l}
V_1^{(5)}(x)=-\dfrac{55 x^4}{2}+\dfrac{121 x^6}{6}
                         -\dfrac{11 x^8}{4}+\dfrac{x^{10}}{10},\cr
V_2^{(3)}(y)=\dfrac{33 y^2}{2}-\dfrac{11 y^4}{2}+\dfrac{y^6}{6}.
\end{array}
\right. 
\end{align}

\paragraph{$\mathbb Z_2$ breaking cases:}
\ \vspace{0.3cm} \\  
The $\mathbb Z_2$ breaking critical potentials ($g$ is a breaking parameter) are 
\begin{align}
&\underline{(\hat p,\hat q)=(2,3)\quad (\theta=14)} \nn\\
&\qquad \left\{
\begin{array}{l}
V_1^{(3)}(x)=
-3 g x+3 g x^3+\dfrac{15 x^4}{4}-\dfrac{9 g x^5}{5}-\dfrac{5 x^6}{3}+\dfrac{g x^7}{7}+\dfrac{x^8}{8}, \cr
V_2^{(2)}(y)=
\left(g-g^3\right) y+\dfrac{1}{2} \left(-10+3 g^2\right) y^2-g y^3+\dfrac{y^4}{4}, 
\end{array}
\right. \\
&\underline{(\hat p,\hat q)=(2,4)\quad (\theta=25)} \nn\\
&\qquad \left\{
\begin{array}{l}
V_1^{(4)}(x)=
-3 g x-\dfrac{4 g x^3}{3}-\dfrac{13 x^4}{4}+6 g x^5+\dfrac{13 x^6}{2}-\dfrac{12 g x^7}{7}-\dfrac{13 x^8}{8}+\dfrac{g x^9}{9}+\dfrac{x^{10}}{10}, \cr
V_2^{(2)}(y)=
\left(g-g^3\right) y+\dfrac{1}{2} \left(-13+3 g^2\right) y^2-g y^3+\dfrac{y^4}{4}.
\end{array}
\right. 
\end{align}
Note that the critical potentials of $\omega^{1/2}(=i)$-rotated models 
are obtained by the analytic continuation, $g \to i g$.

\subsection{Multi-cut two-matrix models ($\mathbb Z_k$ symmetric)}

\subsubsection{Three-cut cases}

\paragraph{The unitary cases ($\hat p=\hat q$):}
\ \vspace{0.3cm} \\  
The potentials of ``unitary'' cases $V_1(x)=V_2(x)=V^{(\hat p)}(x)$ are
\begin{align}
V^{(1)}(x)&=\frac{1}{3}\left(1-4 \epsilon -5 \epsilon ^2\right) x^3+\frac{\epsilon x^6}{6},
\qquad (\theta=1+4 \epsilon +7 \epsilon ^2,\quad \epsilon \neq 1),\label{Peq1CriPotEx} \\
V^{(2)}(x)&=-\frac{8 x^3}{3}+\frac{x^6}{6},\qquad (\theta=12), \\
V^{(3)}(x)&=14 x^3-\frac{7 x^6}{2}+\frac{x^9}{9},\qquad (\theta=70), \\
V^{(4)}(x)&=-\frac{200 x^3}{3}+\frac{170 x^6}{3}-\frac{40 x^9}{9}+\frac{x^{12}}{12},\qquad (\theta=350), \\
V^{(5)}(x)&=\frac{910 x^3}{3}-\frac{5005 x^6}{6}+130 x^9-\frac{65 x^{12}}{12}+\frac{x^{15}}{15},\qquad(\theta=1638).
\end{align}

Note that, for the case of $\hat p=1$, 
we consider the potential of $m_2=\hat p+1=2$. Then the critical 
operator of $\tilde A$ is given as 
\begin{align}
\mathcal A(*,Z) = \sqrt{R_*}\frac{(1-Z^k)(1-\epsilon Z^k)}{Z}\, \Gamma(Z),\qquad \bigl(\epsilon \equiv A_{2k-1}(\infty)\bigr). \label{DefEpsilon}
\end{align}
From the construction, the vicinity of $\epsilon=1$ is the $\hat p=2$ critical point. 
In this class of solutions, we need to be careful about critical potentials with odd degrees. In general, 
negative roots for the critical potentials $V^{(2l+1)}(x)$ with $l\geq 1$, and the Fermi sea is filled 
from origin to some point around the first negative root. However for $\hat p=1$, the Fermi sea extends to 
negative infinity (See eq.~\eq{Peq1CriPotEx} with $\epsilon=0$). Hence we introduce a regularization parameter
$\epsilon$ by eq.~\eq{DefEpsilon}. This will make the matrix model well-defined and we will also adjust the 
normalization factor $\theta$ accordingly. 

\paragraph{The $(\hat p,\hat q)=(1,\hat q)$ cases:}
\ 
\vspace{0.3cm} 
\\  
The potentials $V_1(x)=V_1^{(\hat q)}(x)$ and $V_2(y)$ of the $(1,\hat q)$ case are
\begin{align}
&\left\{
\begin{array}{l} 
\ds
V_1^{(2)}(x)=\frac{1}{3} (-3-5 \epsilon ) x^3 +\frac{x^6}{6}, \cr 
\ds
\quad \! V_2(y)=\frac{1}{3} (1-9 \epsilon ) y^3 +\frac{\epsilon y^6}{6}, 
\qquad \bigl(\theta = 3 (1+3 \epsilon ),\quad \epsilon \neq 1\bigr),
\end{array}
\right. \\
&\left\{
\begin{array}{l} 
\ds
V_1^{(3)}(x)=   
\frac{1}{3} \epsilon  (1+12 \epsilon ) x^3 
+\frac{1}{6}(-5-8 \epsilon ) x^6 
+\frac{x^9}{9}, \cr 
\ds
\quad \! V_2(y)=   \frac{1}{3}(1-14 \epsilon ) y^3 +\frac{\epsilon y^6}{6}, 
\qquad (\theta = 5+21 \epsilon,
\quad \epsilon \neq 1),
\end{array}
\right. \\
&\left\{
\begin{array}{l} 
\ds
V_1^{(4)}(x)= \frac{1}{3} \left(\epsilon +15 \epsilon ^2-22 \epsilon ^3\right) x^3 
++\frac{1}{6}\left(7+23 \epsilon +33 \epsilon ^2\right) x^6 
+\frac{1}{9} (-7-11 \epsilon ) x^9 
+\frac{x^{12}}{12}, \cr 
\ds
\quad \! V_2(y)=   \frac{1}{3} (1-19 \epsilon ) y^3 +\frac{\epsilon y^6 }{6}, 
\qquad (   \theta=7+38 \epsilon 
,\quad \epsilon \neq 1),
\end{array}
\right. 
\end{align}
Since there is no $(1,2)$ critical point in three-cut cases, 
the above $(1,2)$ critical potential should result in the $(1,3)$ critical point. 

\paragraph{The $(\hat p,\hat q)$ cases:}
\ \vspace{0.3cm} \\  
The potentials $(V_1^{(\hat q)}(x),V_2^{(\hat p)}(y))$ of general $(\hat p, \hat q)$ cases are 
\begin{align}
&\underline{(\hat p,\hat q)=(2,3)\quad (\theta=26)}\quad   
\left\{
\begin{array}{l} \ds
V_1^{(3)}(x)=\dfrac{13 x^3}{3}-\dfrac{13 x^6}{6}+\dfrac{x^9}{9}, \cr
V_2^{(2)}(y)=-\dfrac{13 y^3}{3}+\dfrac{y^6}{6},
\end{array}
\right. \\
&\underline{(\hat p,\hat q)=(2,5)\quad (\theta=69)} \quad 
\left\{
\begin{array}{l} 
\ds
V_1^{(5)}(x)=-\frac{161 x^6}{6}+\frac{46 x^9}{3}
                         -\frac{23 x^{12}}{12}+\frac{x^{15}}{15}, \cr
\ds
V_2^{(2)}(x)=-\frac{23 y^3}{3}+\frac{y^6}{6},
\end{array}
\right. \\
&\underline{(\hat p,\hat q)=(3,4)\quad (\theta=145)} \quad 
 \left\{
\begin{array}{l}
\ds
V_1^{(4)}(x)=-\frac{58 x^3}{3}+29 x^6-\frac{29 x^9}{9}+\frac{x^{12}}{12},\cr
\ds
V_2^{(3)}(y)=29 y^3-\frac{29 y^6}{6}+\frac{y^9}{9},
\end{array}
\right. \\
&\underline{(\hat p,\hat q)=(3,5)\quad (\theta=259)} \quad
 \left\{
\begin{array}{l}
\ds
V_1^{(5)}(x)=\frac{37 x^3}{3}-\frac{407 x^6}{3}
                       +\frac{370 x^9}{9}-\frac{37 x^{12}}{12}+\frac{x^{15}}{15},\cr
\ds                       
V_2^{(3)}(y)=\frac{148 y^3}{3}-\frac{37 y^6}{6}+\frac{y^9}{9}.
\end{array}
\right. 
\end{align}

\subsubsection{Four-cut cases}

\paragraph{The unitary cases ($\hat p=\hat q$):}
\ 
\vspace{0.3cm} 
\\  
The potentials of ``unitary'' cases $V_1(x)=V_2(x)=V^{(\hat p)}(x)$ are
\begin{align}
V^{(1)}(x)&=\frac{1}{4} \left(1-6 \epsilon -7 \epsilon ^2\right)x^4+\frac{\epsilon x^8 }{8},
\qquad \bigl(\theta=2 \left(1+3 \epsilon +5 \epsilon ^2\right),\quad \epsilon\neq 1\bigr), \\
V^{(2)}(x)&=-3 x^4+\frac{x^8}{8},\qquad (\theta=18), \\
V^{(3)}(x)&=\frac{105 x^4}{4}-\frac{15 x^8}{4}+\frac{x^{12}}{12},\qquad (\theta=100), \\
V^{(4)}(x)&=-210 x^4+\frac{175 x^8}{2}-\frac{14 x^{12}}{3}+\frac{x^{16}}{16},\qquad (\theta=490), \\
V^{(5)}(x)&=\frac{6435 x^4}{4}-1860 x^8+\frac{765 x^{12}}{4}-\frac{45 x^{16}}{8}+\frac{x^{20}}{20},\qquad(\theta=2268).
\end{align}

\paragraph{The $(\hat p,\hat q)=(1,\hat q)$ cases:}
\ \vspace{0.3cm} \\  
The potentials $V_1(x)=V_1^{(\hat q)}(x)$ and $V_2(y)$ of the $(1,\hat q)$ case are
\begin{align}
&\left\{
\begin{array}{l} 
\ds
V_1^{(2)}(x)=\frac{1}{4} (-5-7 \epsilon ) x^4 +\frac{x^8}{8}, \cr 
\ds
\quad \! V_2(y)=  \frac{1}{4}(1-13 \epsilon ) y^4 +\frac{\epsilon y^8 }{8},  
\qquad (\theta =   5+13 \epsilon,
\quad \epsilon \neq 1),
\end{array}
\right. \\
&\left\{
\begin{array}{l} 
\ds
V_1^{(3)}(x)=  \frac{1}{2} \left(2+6 \epsilon +11 \epsilon ^2\right) x^4 
+\frac{1}{8}(-8-11 \epsilon ) x^8 
+\frac{x^{12}}{12}, \cr 
\ds
\quad \! V_2(y)=   \frac{1}{4} (1-20 \epsilon ) y^4 +\frac{\epsilon y^8}{8}, 
\qquad (\theta = 2 (4+15 \epsilon ),
\quad \epsilon \neq 1),
\end{array}
\right. \\
&\left\{
\begin{array}{l} 
\ds
V_1^{(4)}(x)=-\frac{1}{2} \left(\epsilon ^2+25 \epsilon ^3\right) x^4 
+\frac{1}{8} \left(22+61 \epsilon +60 \epsilon ^2\right) x^8 
 +\frac{1}{12}(-11-15 \epsilon ) x^{12} 
+\frac{x^{16}}{16}, \cr 
\ds
\quad \! V_2(y)=   \frac{1}{4} (1-27 \epsilon ) y^4 +\frac{\epsilon y^8 }{8}, 
\qquad (   \theta=      11+54 \epsilon,
\quad \epsilon \neq 1),
\end{array}
\right. 
\end{align}
Since there is no $(1,3)$ critical point in the four-cut cases, 
the above $(1,3)$ critical potential should result in the $(1,4)$ critical point. 

\paragraph{The $(\hat p,\hat q)$ cases:}
\ \vspace{0.3cm} \\  
The potentials $(V_1^{(\hat q)}(x),V_2^{(\hat p)}(y))$ of general $(\hat p, \hat q)$ cases are 
\begin{align}
&\underline{(\hat p,\hat q)=(2,3)\quad (\theta=38)}\quad   
\left\{
\begin{array}{l}
\ds
V_1^{(3)}(x)=\frac{19 x^4}{2}-\frac{19 x^8}{8}+\frac{x^{12}}{12}, \cr
\ds
V_2^{(2)}(y)=-\frac{19 y^4}{4}+\frac{y^8}{8},
\end{array}
\right. \\
&\underline{(\hat p,\hat q)=(2,4)\quad (\theta=65)} \quad 
\left\{
\begin{array}{l}
\ds
V_1^{(4)}(x)=-13 x^4+\frac{143 x^8}{8}-\frac{13 x^{12}}{6}+\frac{x^{16}}{16},\cr
\ds
V_2^{(2)}(y)=-\frac{13 y^4}{2}+\frac{y^8}{8},
\end{array}
\right. \\
&\underline{(\hat p,\hat q)=(2,5)\quad (\theta=99)} \quad 
\left\{
\begin{array}{l}
\ds
V_1^{(5)}(x)=\frac{33 x^4}{4}-\frac{605 x^8}{8}+\frac{99 x^{12}}{4}-\frac{33 x^{16}}{16}+\frac{x^{20}}{20},\cr
\ds
V_2^{(2)}(y)=-\frac{33 y^4}{4}+\frac{y^8}{8},
\end{array}
\right. \\
&\underline{(\hat p,\hat q)=(3,4)\quad (\theta=205)} \quad 
 \left\{
\begin{array}{l}
\ds
V_1^{(4)}(x)=-\frac{287 x^4}{4}+\frac{369 x^8}{8}
                         -\frac{41 x^{12}}{12}+\frac{x^{16}}{16},\cr
\ds
V_2^{(3)}(y)=\frac{205 y^4}{4}-\frac{41 y^8}{8}+\frac{y^{12}}{12},
\end{array}
\right. \\
&\underline{(\hat p,\hat q)=(3,5)\quad (\theta=364)} \quad
 \left\{
\begin{array}{l}
\ds
V_1^{(5)}(x)=\frac{455 x^4}{4}-\frac{663 x^8}{2}+\frac{377 x^{12}}{6}-\frac{13 x^{16}}{4}+\frac{x^{20}}{20},\cr
\ds
V_2^{(3)}(y)=\frac{169 y^4}{2}-\frac{13 y^8}{2}+\frac{y^{12}}{12}.
\end{array}
\right. 
\end{align}

\subsubsection{Six-cut cases}

\paragraph{The unitary cases ($\hat p=\hat q$):}
\ \vspace{0.3cm} \\  
The potentials of ``unitary'' cases $V_1(x)=V_2(x)=V^{(\hat p)}(x)$ are
\begin{align}
V^{(1)}(x)&=\frac{1}{6} \left(1-10 \epsilon -11 \epsilon ^2\right) x^6 
+\frac{\epsilon x^{12} }{12},\qquad \bigl(\theta=2 \left(2+5 \epsilon +8 \epsilon ^2\right)\bigr), \\
V^{(2)}(x)&=-\frac{10 x^6}{3}+\frac{x^{12}}{12},\qquad (\theta=30), \\
V^{(3)}(x)&=52 x^6-4 x^{12}+\frac{x^{18}}{18},\qquad (\theta=160), \\
V^{(4)}(x)&=-748 x^6+\frac{451 x^{12}}{3}-\frac{44 x^{18}}{9}+\frac{x^{24}}{24},\qquad (\theta=770).
\end{align}

\paragraph{The $(\hat p,\hat q)=(1,\hat q)$ cases:}
\ \vspace{0.3cm} \\  
The potentials $V_1(x)=V_1^{(\hat q)}(x)$ and $V_2(y)$ of the $(1,\hat q)$ case are
\begin{align}
&\left\{
\begin{array}{l} 
\ds
V_1^{(2)}(x)=\frac{1}{6}  (-9-11 \epsilon ) x^6+\frac{x^{12}}{12}, \cr 
\ds
\quad \! V_2(y)=  \frac{1}{6}(1-21 \epsilon ) y^6 +\frac{\epsilon  y^{12} }{12},  
\qquad \bigl(\theta =   3 (3+7 \epsilon ),
\quad \epsilon \neq 1\bigr),
\end{array}
\right. \\
&\left\{
\begin{array}{l} 
\ds
V_1^{(3)}(x)=  \frac{1}{6} \left(21+52 \epsilon +51 \epsilon ^2\right) x^6 
+\frac{1}{12} (-14-17 \epsilon ) x^{12} 
+\frac{x^{18}}{18}, \cr 
\ds
\quad \! V_2(y)=   \frac{1}{6} (1-32 \epsilon ) y^6 +\frac{\epsilon y^{12} }{12}, 
\qquad (\theta =2 (7+24 \epsilon ),
\quad \epsilon \neq 1),
\end{array}
\right. \\
&\left\{
\begin{array}{l} 
\ds
V_1^{(4)}(x)=\frac{1}{6} \left(-19-71 \epsilon -141 \epsilon ^2-161 \epsilon ^3\right) x^6 
+\frac{1}{12} \left(76+185 \epsilon +138 \epsilon ^2\right) x^{12}+ \cr
\ds 
\qquad\qquad \, +\frac{1}{18}  (-19-23 \epsilon ) x^{18}
+\frac{x^{24}}{24}, \cr 
\ds
\quad \! V_2(y)=  \frac{1}{6} (1-43 \epsilon ) y^6 +\frac{\epsilon y^{12} }{12}, 
\qquad (   \theta=    19+86 \epsilon,
\quad \epsilon \neq 1),
\end{array}
\right. 
\end{align}

\paragraph{The $(\hat p,\hat q)$ cases:}
\ \vspace{0.3cm} \\  
The potentials $(V_1^{(\hat q)}(x),V_2^{(\hat p)}(y))$ of general $(\hat p, \hat q)$ cases are 
\begin{align}
&\underline{(\hat p,\hat q)=(2,3)\quad (\theta=62)}\quad   
\left\{
\begin{array}{l}
\ds
V_1^{(3)}(x)=\frac{62 x^6}{3}-\frac{31 x^{12}}{12}+\frac{x^{18}}{18},\cr
\ds
V_2^{(2)}(y)=-\frac{31 y^6}{6}+\frac{y^{12}}{12},
\end{array}
\right. \\
&\underline{(\hat p,\hat q)=(2,4)\quad (\theta=105)} \quad 
\left\{
\begin{array}{l}
\ds
V_1^{(4)}(x)=-\frac{196 x^6}{3}+\frac{133 x^{12}}{4}
                         -\frac{7 x^{18}}{3}+\frac{x^{24}}{24},\cr
\ds
V_2^{(2)}(y)=-7 y^6+\frac{y^{12}}{12},
\end{array}
\right. \\
&\underline{(\hat p,\hat q)=(2,5)\quad (\theta=159)} \quad 
\left\{
\begin{array}{l}
\ds
V_1^{(5)}(x)=\frac{371 x^6}{3}-\frac{1007 x^{12}}{4}
                       +\frac{265 x^{18}}{6}-\frac{53 x^{24}}{24}+\frac{x^{30}}{30},\cr
\ds
V_2^{(2)}(y)=-\frac{53 y^6}{6}+\frac{y^{12}}{12},
\end{array}
\right. \\
&\underline{(\hat p,\hat q)=(3,4)\quad (\theta=325)} \quad 
 \left\{
\begin{array}{l}
\ds
V_1^{(4)}(x)=-\frac{845 x^6}{3}+\frac{325 x^{12}}{4}
                         -\frac{65 x^{18}}{18}+\frac{x^{24}}{24},\cr
\ds
V_2^{(3)}(y)=\frac{195 y^6}{2}-\frac{65 y^{12}}{12}+\frac{y^{18}}{18},
\end{array}
\right. \\
&\underline{(\hat p,\hat q)=(3,5)\quad (\theta=574)} \quad
 \left\{
\begin{array}{l}
\ds
V_1^{(5)}(x)=\frac{5863 x^6}{6}-\frac{5945 x^{12}}{6}
                       +\frac{1927 x^{18}}{18}-\frac{41 x^{24}}{12}+\frac{x^{30}}{30},\cr
\ds
V_2^{(3)}(y)=\frac{943 y^6}{6}-\frac{41 y^{12}}{6}+\frac{y^{18}}{18}.
\end{array}
\right. 
\end{align}

\subsection{Multi-cut two-matrix models ($\mathbb Z_k$ breaking)}
We concentrate on the $\mathbb Z_k$ breaking critical points of the multi-cut matrix models 
which is characterized by the operators $\mathcal A$ and $\mathcal C^{\rm T}$ of 
\begin{align}
\mathcal A(*,Z) &= \sqrt{R_*} \frac{(1-Z^k)^{\hat p}}{Z} \Gamma(Z), \\
\mathcal C^{\rm T}(*,Z) &=\sqrt{R_*} \frac{(1-Z^{-k})^{\hat q+1}}{Z^{-1}}\Gamma(Z)^{k-1}
+ \sqrt{R_*} g \frac{(1-Z^{-k})^{\hat q}}{Z} \Gamma(Z),
\end{align}
since they correspond to minimal fractional superstring theory \cite{irie2}. 
The parameter $g$ is the breaking parameter. The critical potentials of the $\omega^{1/2}$-rotated models 
are obtained by the analytic continuation, $g\, (=\bar C_{1}(*)) \to \omega^{-1}g$. We put $R_*=1$. 

\subsubsection{Three-cut cases}

\paragraph{The unitary models $(\hat q=\hat p+1)$:}
\ \vspace{0.3cm} \\  
The potentials $(V_1^{(\hat p+1)}(x),V_2^{(\hat p)}(y))$ 
of the unitary $(\hat p, \hat p+1)$ cases 
(unitary minimal fractional superstring theory) are 
\begin{align}
&\underline{(\hat p,\hat q)=(1,2)}\quad   
\left\{
\begin{array}{l}
\ds
V_1^{(2)}(x)=
\frac{1}{2} \epsilon  (-1+7 \epsilon ) g  x^2 
+\frac{1}{3}  \epsilon  (1+12 \epsilon ) x^3
+\frac{1}{5}  (-5  -7 \epsilon   ) g x^5 \cr
\ds \qquad\qquad 
+\frac{1}{6} (-5-8 \epsilon ) x^6 
+\frac{g x^8 }{8}
+\frac{x^9}{9}, \cr
\ds
V_2^{(1)}(y)=
 +(-2+3 \epsilon ) g y
+\frac{5}{2}\epsilon  g^2  y^2 
 +\frac{1}{3} (1-14 \epsilon ) y^3 
 -\frac{5}{4} \epsilon  g  y^4
+\frac{\epsilon y^6 }{6}, 
\end{array}
\right. \nn\\[2pt]
& \hspace{8cm}\bigl(\theta= 5+21 \epsilon,\quad \epsilon \neq 1\bigr), \\[2pt]
&\underline{(\hat p,\hat q)=(2,3)}\quad   
\left\{
\begin{array}{l}
\ds
V_1^{(3)}(x)=
3  g x^2 
+\frac{13 x^3}{3}
-\frac{12  g x^5 }{5}
-\frac{13 x^6}{6}
+\frac{ g x^8}{8}
+\frac{x^9}{9}, \cr
\ds
V_2^{(2)}(y)= 
g y
+\frac{5  g^2 y^2 }{2}
-\frac{13 y^3}{3}
-\frac{5 g y^4 }{4}
+\frac{y^6}{6},
\end{array}
\right. \nn\\
& \hspace{8cm}(\theta = 26  ). 
\end{align}

\paragraph{The $(\hat p,\hat q)=(1,\hat q)$ cases:}
\ \vspace{0.3cm} \\  
The critical potentials $V_1(x)=V^{(\hat q)}(x)$ and $V_2(y)$ are
\begin{align}
&\left\{
\begin{array}{l}
\ds
V_1^{(3)}(x)=
\frac{1}{2}  \left(-\epsilon  +11 \epsilon ^2  -10 \epsilon ^3  \right) g x^2
+\frac{1}{3} \left(\epsilon +15 \epsilon ^2-22 \epsilon ^3\right) x^3 + \cr
\ds\qquad \qquad 
+\frac{1}{5} \left(7+19 \epsilon +25 \epsilon ^2\right) g  x^5 
+\frac{1}{6}  \left(7+23 \epsilon +33 \epsilon ^2\right) x^6+ \cr
\ds \qquad \qquad 
+\frac{1}{8}  (-7  -10 \epsilon   )g x^8
+\frac{1}{9}  (-7-11 \epsilon ) x^9
+\frac{g x^{11} }{11}
+\frac{x^{12}}{12}, \cr
\ds
V_2(y)=  (-2+3 \epsilon ) g  y
+\frac{5}{2}  \epsilon  g ^2 y^2
+ \frac{1}{3} (1-19 \epsilon ) y^3 
-\frac{5}{4}  \epsilon  g y^4
+\frac{\epsilon y^6 }{6},
\end{array}
\right. \nn\\
& \hspace{8cm}(\theta = 7+38 \epsilon   ,\quad \epsilon \neq 1  ),  \\
&\left\{
\begin{array}{l}
\ds
V_1^{(5)}(x)=
-\frac{1}{2} \epsilon  \left(1-19 \epsilon +51 \epsilon ^2-49 \epsilon ^3+16 \epsilon ^4\right) g x^2 +\cr 
\ds \qquad\qquad
+\frac{1}{3} \left(\epsilon +21 \epsilon ^2-96 \epsilon ^3+125 \epsilon ^4-51 \epsilon ^5\right) x^3
+\epsilon ^3 (-1+28 \epsilon ) g x^5+ \cr
\ds \qquad\qquad
+\frac{7}{6} \left(\epsilon ^3+34 \epsilon ^4\right) x^6 
-\frac{1}{8}  \left(22+93 \epsilon +188 \epsilon ^2+192 \epsilon ^3\right) g x^8+\cr 
\ds \qquad\qquad
+\frac{1}{9} \left(-22-105 \epsilon -234 \epsilon ^2-255 \epsilon ^3\right) x^9 
+\frac{1}{11} \left(33+95 \epsilon +88 \epsilon ^2\right) g  x^{11} +\cr
\ds \qquad\qquad
+\frac{1}{12}  \left(33+103 \epsilon +102 \epsilon ^2\right) x^{12}
+\frac{1}{14} \left(-11 -16 \epsilon \right) g x^{14} +\cr
\ds \qquad\qquad
+\frac{1}{15}  (-11-17 \epsilon ) x^{15}
+\frac{g x^{17} }{17}
+\frac{x^{18}}{18}, \cr
\ds
V_2(y)= 
(-2+3 \epsilon ) g y
+\frac{5}{2}  \epsilon  g^2 y^2
+\frac{1}{3} (1-29 \epsilon ) y^3 
-\frac{5}{4} \epsilon  g y^4 
+\frac{\epsilon y^6 }{6},
\end{array}
\right. \nn\\
& \hspace{8cm}(\theta = 11+87 \epsilon,
\quad \epsilon \neq 1  ), 
\end{align}

\subsubsection{Four-cut cases}

\paragraph{The unitary models $(\hat q=\hat p+1)$:}
\ \vspace{0.3cm} \\  
The potentials $(V_1^{(\hat p+1)}(x),V_2^{(\hat p)}(y))$ 
of the unitary $(\hat p, \hat p+1)$ cases 
(unitary minimal fractional superstring theory) are 
\begin{align}
&\underline{(\hat p,\hat q)=(1,2)}\nn \\ 
&\left\{
\begin{array}{l}
\ds
V_1^{(2)}(x)=   
\frac{1}{3} \left(4+8 \epsilon +15 \epsilon ^2\right) g x^3 
+\frac{1}{2} \left(2+6 \epsilon +11 \epsilon ^2\right) x^4 
-\frac{2}{7}  (4 +5 \epsilon  ) g x^7 \cr
\ds \qquad\qquad 
+\frac{1}{8}  (-8-11 \epsilon ) x^8
+\frac{g x^{11} }{11}
+\frac{x^{12}}{12}, \cr
\ds
V_2^{(1)}(y)= 
(-3+4 \epsilon ) g y
+\frac{7}{2}  \epsilon  g^2 y^2
+\frac{1}{4}  (1-20 \epsilon ) y^4
-\frac{7}{5}  \epsilon  g  y^5
+\frac{ \epsilon }{8} y^8, 
\end{array}
\right. \nn\\[2pt]
& \hspace{8cm}\bigl(\theta=  2 (4+15 \epsilon ) ,\quad \epsilon \neq 1\bigr), \\[2pt]
&\underline{(\hat p,\hat q)=(2,3)}\nn \\
&\left\{
\begin{array}{l}
\ds
V_1^{(3)}(x)=
-\frac{25 g}{3} x^3 
-13 x^4
+\frac{125 g}{7} x^7 
+\frac{143 x^8}{8}
-\frac{25 g}{11} x^{11} 
-\frac{13 x^{12}}{6}
+\frac{g}{15} x^{15} 
+\frac{x^{16}}{16},\cr
\ds
V_2^{(2)}(y)=
g y 
+\frac{7 g^2}{2} y^2 
-\frac{13 y^4}{2}
-\frac{7 g}{5} y^5 
+\frac{y^8}{8}, 
\end{array}
\right.  \nn\\
& \hspace{8cm} (\theta=65).
\end{align}

\paragraph{The $(\hat p,\hat q)=(1,\hat q)$ cases:}
\ \vspace{0.3cm} \\  
The critical potentials $V_1(x)=V^{(\hat q)}(x)$ and $V_2(y)$ are
\begin{align}
&\left\{
\begin{array}{l}
\ds
V_1^{(3)}(x)=
\frac{1}{3} \left(3 \epsilon ^2  -28 \epsilon ^3  \right) g x^3 
-\frac{1}{2} \left(\epsilon ^2+25 \epsilon ^3\right) x^4 
+\frac{1}{7}\left(22+54 \epsilon +49 \epsilon ^2\right) g x^7 +\cr
\ds\qquad \qquad 
+\frac{1}{8}\left(22+61 \epsilon +60 \epsilon ^2\right) x^8 
+\frac{1}{11}(-11 -14 \epsilon  ) g x^{11} 
+\frac{1}{12} (-11-15 \epsilon ) x^{12} +\cr
\ds \qquad\qquad 
+\frac{g x^{15} }{15}
+\frac{x^{16}}{16}, \cr
\ds
V_2(y)= 
(-3+4 \epsilon ) g y
+\frac{7}{2}  \epsilon  g^2 y^2
+\frac{1}{4}  (1-27 \epsilon ) y^4
-\frac{7}{5}  \epsilon  g  y^5
+\frac{\epsilon y^8 }{8},
\end{array}
\right. \nn\\
& \hspace{8cm}(\theta = 11+54 \epsilon,\quad \epsilon \neq 1  ),  \\
&\left\{
\begin{array}{l}
\ds
V_1^{(4)}(x)= 
\frac{1}{3} \epsilon ^2 \left(3-40 \epsilon +45 \epsilon ^2\right) g x^3 
+\frac{1}{4} \epsilon ^2 \left(-2-60 \epsilon +95 \epsilon ^2\right) x^4 \cr
\ds \qquad\qquad 
-\frac{2}{7} \left(14  +51 \epsilon   +91 \epsilon ^2  +84 \epsilon ^3  \right) g x^7 
+\frac{1}{8} \left(-28-117 \epsilon -232 \epsilon ^2-228 \epsilon ^3\right) x^8 \cr
\ds \qquad\qquad 
+\frac{1}{11} \left(49+124 \epsilon +99 \epsilon ^2\right) g x^{11} 
+\frac{1}{12} \left(49+134 \epsilon +114 \epsilon ^2\right) x^{12} 
-\frac{2}{15} (7  +9 \epsilon   )g x^{15} \cr
\ds \qquad\qquad 
+\frac{1}{16} (-14-19 \epsilon ) x^{16} 
+\frac{g x^{19} }{19}
+\frac{x^{20}}{20}, \cr
\ds
V_2(y)=
(-3+4 \epsilon ) g y
+\frac{7}{2} \epsilon  g^2 y^2 
+\frac{1}{4} (1-34 \epsilon ) y^4 
-\frac{7}{5} \epsilon  g y^5 
+\frac{ \epsilon y^8}{8},
\end{array}
\right. \nn\\
& \hspace{8cm}(\theta = 14+85 \epsilon ,
\quad \epsilon \neq 1  ), 
\end{align}

\subsubsection{Six-cut cases}

\paragraph{The unitary models $(\hat q=\hat p+1)$:}
\ \vspace{0.3cm} \\  
The potentials $(V_1^{(\hat p+1)}(x),V_2^{(\hat p)}(y))$ of the unitary $(\hat p, \hat p+1)$ cases 
(unitary minimal fractional superstring theory) are 
\begin{align}
&\underline{(\hat p,\hat q)=(1,2)}\nn \\ 
&\left\{
\begin{array}{l}
\ds
V_1^{(2)}(x)=
\frac{1}{5}\left(21+44 \epsilon +40 \epsilon ^2\right) g x^5 
+\frac{1}{6}\left(21+52 \epsilon +51 \epsilon ^2\right) x^6 
-\frac{2}{11}(7 +8 \epsilon  ) g x^{11} +\cr
\ds \qquad\qquad 
+\frac{1}{12}(-14-17 \epsilon ) x^{12} 
+\frac{ g x^{17}}{17}
+\frac{x^{18}}{18},\cr
\ds
V_2^{(1)}(y)=
(-5+6 \epsilon ) g y
+\frac{11}{2} \epsilon  g^2 y^2 
+\frac{1}{6} (1-32 \epsilon ) y^6 
-\frac{11}{7} \epsilon g y^7 
+\frac{ \epsilon y^{12}}{12},
\end{array}
\right. \nn\\
& \hspace{8cm}(\theta = 2 (7+24 \epsilon ),
\quad \epsilon \neq 1  ),  \\
&\underline{(\hat p,\hat q)=(2,3)}\nn \\
&\left\{
\begin{array}{l}
\ds
V_1^{(3)}(x)=
-\frac{287 g}{5} x^5 
-\frac{196 x^6}{3}
+\frac{369 g}{11} x^{11} 
+\frac{133 x^{12}}{4}
-\frac{41 g}{17} x^{17} 
-\frac{7 x^{18}}{3}
+\frac{g}{23} x^{23} 
+\frac{x^{24}}{24}
,\cr
\ds
V_2^{(2)}(y)=
g y 
+\frac{11 g^2}{2} y^2 
-7 y^6
-\frac{11 g}{7} y^7 
+\frac{y^{12}}{12}.
\end{array}
\right. \nn\\
& \hspace{8cm}(\theta=105).
\end{align}

\paragraph{The $(\hat p,\hat q)=(1,\hat q)$ cases:}
\ \vspace{0.3cm} \\  
The critical potentials $V_1(x)=V^{(\hat q)}(x)$ and $V_2(y)$ are
\begin{align}
&\left\{
\begin{array}{l}
\ds
V_1^{(3)}(x)=
\frac{1}{5} \left(-19  -58 \epsilon  -100 \epsilon ^2 -110 \epsilon ^3  \right) g x^5 
+\frac{1}{6} \left(-19-71 \epsilon -141 \epsilon ^2-161 \epsilon ^3\right) x^6 +\cr
\ds \qquad\qquad 
+\frac{1}{11} \left(76+172 \epsilon +121 \epsilon ^2\right) g x^{11} 
+\frac{1}{12} \left(76+185 \epsilon +138 \epsilon ^2\right) x^{12} +\cr
\ds \qquad\qquad
+\frac{1}{17} (-19  -22 \epsilon  ) g x^{17} 
+\frac{1}{18} (-19-23 \epsilon ) x^{18} 
+\frac{ g x^{23}}{23}
+\frac{x^{24}}{24}, \cr
\ds
V_2(y)= 
(-5+6 \epsilon ) g y
+\frac{11}{2} \epsilon  g^2 y^2
+\frac{1}{6} (1-43 \epsilon ) y^6 
-\frac{11}{7} \epsilon  gy^7
+\frac{\epsilon y^{12} }{12},
\end{array}
\right. \nn\\
& \hspace{8cm}(\theta = 19+86 \epsilon,\quad \epsilon \neq 1  ),  \\
&\left\{
\begin{array}{l}
\ds
V_1^{(4)}(x)= 
\frac{1}{5} \left(6+24 \epsilon +60 \epsilon ^2+120 \epsilon ^3+245 \epsilon ^4\right) g x^5 +\cr
\ds \qquad\qquad 
+\frac{1}{3} \left(3+15 \epsilon +45 \epsilon ^2+105 \epsilon ^3+203 \epsilon ^4\right) x^6- \cr
\ds \qquad\qquad
-\frac{4}{11} \left(56  +189 \epsilon  +264 \epsilon ^2 +154 \epsilon ^3 \right) g x^{11}+ \cr 
\ds \qquad\qquad
+\frac{1}{12} \left(-224-819 \epsilon -1224 \epsilon ^2-754 \epsilon ^3\right) x^{12} +\cr
\ds \qquad\qquad
+\frac{2}{17} \left(78 +180 \epsilon  +119 \epsilon ^2 \right) g x^{17} 
+\frac{1}{6} \left(52+126 \epsilon +87 \epsilon ^2\right) x^{18} -\cr
\ds\qquad\qquad
-\frac{4}{23} (6 +7 \epsilon  )g  x^{23} 
+\frac{1}{24} (-24-29 \epsilon ) x^{24}
+\frac{ g x^{29}}{29}
+\frac{x^{30}}{30}, \cr
\ds
V_2(y)=
(-5+6 \epsilon ) g y
+\frac{11}{2} \epsilon  g^2 y^2 
+\frac{1}{6} (1-54 \epsilon ) y^6 
-\frac{11}{7} \epsilon  g  y^7 
+\frac{ \epsilon y^{12}}{12},
\end{array}
\right. \nn\\
& \hspace{8cm}(\theta =3 (8+45 \epsilon ),
\quad \epsilon \neq 1  ), 
\end{align}

\section{Direct evaluation of three-cut string equations \label{DirectEvaluationThreeCutStringEq}}

In this appendix, we make a detailed analysis of the string equation in the three-cut matrix models%
\footnote{The calculations in this appendix have been carried out 
mostly with the help of {\em Mathematica}${}^{\rm TM}$.}   
based on the formalism of $k$-component KP hierarchy \cite{kcKP}\cite{fi1}. The motivation behind 
this analysis is to check that our ansatz \eq{CISYansatz} corresponds to a special solution 
of Lax pairs. On the other hand, the complete analysis also provides useful hints for other 
solutions in the multi-cut matrix models. 

The realization of $\mathbb Z_k$ symmetric $(\hat p,\hat q)$ critical points have been shown 
in eq's \eq{ToLax1} and \eq{ToLax2}. Here we drop some irrelevant factors and the following simplified notation is used:
\begin{align}
\bP = \Gamma \del^{\hat p} + H_1\, \del^{\hat p-1}+\cdots + H_{\hat p}, \qquad
\bQ = \Gamma^{-1} \del^{\hat q} +\widetilde H_1\, \del^{\hat q-1}+\cdots +
\widetilde  H_{\hat q}. 
\end{align}
The coefficient matrices $H_i$ and $\widetilde H_i$ are all real functions. 
Then the Lax operators $\bGamma$ and $\bL$ 
are defined as $\bP= (\bGamma \bL^{\hat p})$ with $[\bP,\bGamma] = [\bP,\bL]=0$ 
and $\bGamma^k=I_k$, and 
\begin{align}
\bGamma = \Gamma + \sum_{n=1}^\infty S_n (t)\, \del^{-n},\qquad 
\bL = I_k\,\del + \sum_{n=1}^\infty X_n(t) \, \del^{-n}. 
\end{align}
In the following discussion, we put $k=3$ for simplicity. 
The relation $\bGamma^3=I_3$ can be easily solved 
if derivative terms are neglected:
\begin{align}
S_n^{(\rm diag)} = -\frac{\Gamma}{3} 
\sum_{\stackrel{n_1\leq n_2 \leq n_3\neq 0,}{n_1+n_2+n_3=n}}  
\bigl\{S_{n_1},S_{n_2},S_{n_3}\bigr\}+(\text{derivative terms}),
\end{align}
with $S_0=\Gamma$. The bracket $\{\,,\,,\}$ is a totally symmetric product which is defined by 
\begin{align}
\Bigl(\sum_{i=0}^\infty S_i\Bigr)^3 = \sum_{i_1\leq i_2\leq i_3} \bigl\{S_{i_1},S_{i_2},S_{i_3}\bigr\}
\end{align}
Here we define 
\begin{align}
F^{(\rm diag)} \equiv \frac{\Gamma}{3} \bigl\{\Gamma,\Gamma, F\bigr\},\qquad 
F^{(\rm off)} \equiv \frac{1}{3} \bigl[\Gamma^2,[\Gamma, F]\bigr],\qquad 
F=F^{(\rm diag)}+F^{(\rm off)}. 
\end{align}
In the KP hierarchy basis $(\Gamma \to \Omega)$, they are truly a diagonal part and an off-diagonal part 
of the matrix $F$. 

\subsection{The $\hat p=1$ cases}
Since we impose the $\mathbb Z_k$ symmetry here, the KP operator $\bP$ corresponding 
to the matrix operator $A$ 
is given in eq's \eq{ToLax1} and \eq{ToLax2} as
\begin{align}
\bP=\Gamma\,  \del + H 
\equiv (\bGamma \bL),
\end{align}
with 
\begin{align}
H\equiv 
\begin{pmatrix}
 & f_1 &  \cr
   &                &  f_2   \cr
f_3  &              &       \cr
\end{pmatrix},
\qquad 
f_1+f_2+f_3=0. \label{finHpeq1}
\end{align}
Then the equation $[\bP,\bGamma]=0$ gives 
\begin{align}
S_n^{(\rm off)} = -\frac{1}{3}\, \bigl[\Gamma^2,[H,S_{n-1}]\bigr] + (\text{derivative terms}). 
\end{align}
The operator $\bQ$ for the $(1,\hat q)$ case can be written \cite{SolDouglas,fkn} as
\begin{align}
\bQ= \sum_{n=1}^{\hat q+1} nt_n (\bGamma^{-1} \bL^{n-1})_+,
\end{align}
with $(\hat q+1)t_{\hat q+1} = 1$. 
Then the string equation can be easily derived%
\footnote{See Appendix D in \cite{fi1}, for example.} 
from
\begin{align}
0= \sum_{n=1}^{\hat q+1} nt_n \bigl[\bP,(\bGamma^{-1} \bL^{n-1})_+\bigr].
\end{align}
One can easily see that $\hat q \equiv 2 $ (mod. $3$) is not allowed as critical points. 
For example, in the first simplest case $(1,1)$, 
the string equation is written as 
\begin{align}
0=\Gamma\, \del H\, \Gamma + \bigl[H^2,\Gamma\bigr]+t \bigl[H,\Gamma^2\bigr]
\end{align}
with $t_1\equiv t$. 
Below, by imposing the information of the matrix model \eq{finHpeq1}, 
we solve this string equation at the first order of string coupling. 

\subsubsection{The case of $(1,1)$ \label{DirectAnalysis11Cases}}
There are four asymptotic solutions: 
The first one is generalization of ``one-cut solution'' in two-cut matrix models:
\begin{align}
\bP_0 = \Gamma \del,\qquad \bQ_0= \Gamma^{-1} (\del + t),\qquad (f_1=f_2=0). \label{OneCutThreeCut}
\end{align}
With a shift of $z$, this corresponds to our solution \eq{CISYsol} with the choice of $l=0$. 
The others $(i=1,2,3)$ are essentially expressed as 
\begin{align}
\bP_i = \Gamma \del - t M_i,\qquad \bQ_i= \Gamma^{-1} \del - t \bar M_i,
\end{align}
with 
\begin{align}
M_1 = 
\begin{pmatrix}
 & 1 & \cr
 &    & -2 \cr
 1 &  &  \cr
\end{pmatrix},
\qquad 
\bar M_1 = 
\begin{pmatrix}
 &  & -2\cr
-2 &    &  \cr
  & 1 &  \cr
\end{pmatrix}, \\
M_2 = 
\begin{pmatrix}
 & -2 & \cr
 &    & 1 \cr
 1 &  &  \cr
\end{pmatrix},
\qquad 
\bar M_2 = 
\begin{pmatrix}
 &  & -2\cr
1 &    &  \cr
  & -2 &  \cr
\end{pmatrix}, \\
M_3 = 
\begin{pmatrix}
 & 1 & \cr
 &    & 1 \cr
 -2 &  &  \cr
\end{pmatrix},
\qquad 
\bar M_3 = 
\begin{pmatrix}
 &  & 1\cr
-2 &    &  \cr
  & -2 &  \cr
\end{pmatrix}.
\end{align}
They all satisfy $M_i^2=\bar M_i$. 
Note that the forms of $\bP_i$ and $\bQ_i$ $(i=1,2,3)$ are different from each other but eigenvalues are the same. 
With a shift of $z$, this is equivalent to our ansatz \eq{CISYansatz} 
and the solution \eq{CISYsol} of $l=2$. 

We also note that the Douglas equation, $[\bP,\bQ]=g_{str}I_3$, requires 
the following first order correction of string coupling:
\begin{align}
\bP &= \Gamma \del -t M_i + g_{str}\, \frac{1}{t} J_i + O\bigl(g_{str}^2\bigr), 
\end{align}
with
\begin{align}
J_1= \frac{1}{3}
\begin{pmatrix}
 & -1 & \cr
 &      & 0\cr
1&     &
\end{pmatrix}, \qquad
J_2 = \frac{1}{3}
\begin{pmatrix}
 & 0 & \cr
 &      & 1\cr
-1&     &
\end{pmatrix},\qquad
J_3 = \frac{1}{3}
\begin{pmatrix}
 & 1 & \cr
 &      & -1\cr
0&     &
\end{pmatrix}.
\end{align}
This kind of first order correction cannot be realized by the ordering problem of $z$ and $t$ like 
in the two-cut cases \eq{2CutSolDKK}. 

\subsubsection{The case of $(1,3)$}
We consider the background $(b_1=t,b_2=t_2,b_3=0,b_4=1/4)$. 
There are five asymptotic solutions: 
The first one is a generalization of ``one-cut solution'' in two-cut matrix models:
\begin{align}
\bP_0 = \Gamma \del,\qquad \bQ_0= \Gamma^{-1} (\del ^3+t_2 \del + t),\qquad (f_1=f_2=0). 
\end{align}
Other three solutions $(i=1,2,3)$ are essentially expressed as 
\begin{align}
\bP_i = \Gamma \del +f M_i,\qquad 
\bQ_i=(\del^2+ f\del -f^2+t_2) \bigl(\Gamma^{-1} \del +f \bar M_i \bigr),
\end{align}
with 
\begin{align}
3 t= (t_2-\frac{5}{3}f^2) f. 
\end{align}
These solutions precisely correspond to the ansatz \eq{CISYansatz}. 

The last solution is different from the ansatz \eq{CISYansatz}, and can be expressed as 
\begin{align}
\bP_4 = \Gamma \del + H
,\qquad 
\bQ_4= \bigl(\Gamma \del + H\bigr)^2 \,\del = \bP_4^2\, \del .
\end{align}
The functions $f_1$ and $f_2$ in $H$ (see \eq{finHpeq1}) are related as 
\begin{align}
t= \frac{1}{3}f_1f_2(f_1+f_2),\qquad t_2 = \frac{1}{3}(f_1^2+f_1f_2+f_2^2). 
\end{align}
It would be of interest to make further study of these solutions. 

\subsection{The case of $(2,2)$: $\hat p=2$}

The KP operator $\bP$ which corresponds to the matrix operator $\mathcal A$ 
is given in eq's \eq{ToLax1} and \eq{ToLax2} as
\begin{align}
\bP=\Gamma\,  \del^2 + H_1(t) \del +H_2(t)
\equiv (\bGamma \bL^2),
\end{align}
with 
\begin{align}
H_1\equiv 
\begin{pmatrix}
 & f_1 &  \cr
   &                &  f_2   \cr
f_3  &              &       \cr
\end{pmatrix},
\qquad 
H_2\equiv 
\begin{pmatrix}
 & g_1 &  \cr
   &                &  g_2   \cr
g_3  &              &       \cr
\end{pmatrix},
\qquad 
f_1+f_2+f_3=0. \label{finHpeq2}
\end{align}
Then the equation $[\bP,\bGamma]=0$ gives 
\begin{align}
S_n^{(\rm off)} = -\frac{1}{3}\, 
\bigl[\Gamma^2,
\Bigl(
[H_1,S_{n-1}]+[H_2,S_{n-2}]
\Bigr)
\bigr] + (\text{derivative terms}). 
\end{align}
The operator $\bQ$ and the string equation for the $(2,2)$ case can be written as
\begin{align}
\bQ= (\bGamma^{-1} \bL^{2})_+,\qquad [\Gamma \bL^2,\bGamma^{-1} \bL^2]
=g_{str} I_3.
\end{align}
There are also several equivalent solutions like the 2nd, 3rd and 4th solutions of the 
$(1,1)$ and $(1,3)$ cases. Here we only list representative solutions. 

The first solution is the counterpart of the one-cut solution: There are three equivalent expressions. 
One of them is given as
\begin{align}
\bP = 
\begin{pmatrix}
 & \Bigl(\del + \dfrac{f}{2}\Bigr)^2 & \cr
 &                                      & \Bigl(\del-\dfrac{f}{2}\Bigr)\Bigl(\del+\dfrac{f}{2}\Bigr) \cr
\Bigl(\del - \dfrac{f}{2}\Bigr)^2 & & 
\end{pmatrix}, \nn\\
\bQ = 
\begin{pmatrix}
 & &\Bigl(\del + \dfrac{f}{2}\Bigr)^2 \cr
\Bigl(\del - \dfrac{f}{2}\Bigr)^2 &   &  \cr
 & \Bigl(\del-\dfrac{f}{2}\Bigr)\Bigl(\del+\dfrac{f}{2}\Bigr) & 
\end{pmatrix}.
\end{align}
The parameter $f$ is fixed by the Douglas equation $[\bP,\bQ]=g_{str}I_3$. 
This case is also related to our solution \eq{CISYsol} of $l=0$. 

The second solution corresponds to our ansatz \eq{CISYansatz}:
There are also three expressions and one of them is given as
\begin{align}
\bP &=  \Bigl(\del- \frac{f}{2}\Bigr)
\begin{pmatrix}
 & \del + \dfrac{3}{2} f  & \cr
 &                                      & \del + \dfrac{3}{2} f  \cr
\del - \dfrac{3}{2} f  & & 
\end{pmatrix}, \nn\\
\bQ &= \Bigl(\del+ \frac{f}{2}\Bigr)
\begin{pmatrix}
 &  & \del + \dfrac{3}{2} f  \cr
\del - \dfrac{3}{2} f &     &   \cr
 &  \del - \dfrac{3}{2} f &
\end{pmatrix}.
\end{align}
The parameter $f$ is fixed by the Douglas equation $[\bP,\bQ]=g_{str}I_3$. 
This case is also related to our solution \eq{CISYsol} of $l=2$.

\section{The Jacobi polynomials \label{JacobiP}}
Here we write some basic facts about the Jacobi polynomials, which we have used in this paper. 
Some standard reference is \cite{Jacobi}, for example. 

\subsection{A basic definition and the differential equation}
The Jacobi polynomial $P_n^{(\alpha,\beta)}(z)$ is a solution of the hypergeometric equation, 
\begin{align}
\Bigl[(1-z^2) \frac{d^2}{dz^2}
+ \Bigl(\beta-\alpha-(\alpha+\beta+2)\,z\Bigr) \frac{d}{dz}
+ n(n+\alpha+\beta+1)\Bigr]P_n^{(\alpha,\beta)}(z)=0,\label{JacobiEq}
\end{align}
and is defined by 
\begin{align}
P_n^{(\alpha,\beta)}(z)= \frac{(-1)^n}{2^n n!} 
(1-z)^{-\alpha}(1+z)^{-\beta} \frac{d^n}{dz^n} \Bigl[(1-z)^{\alpha+n}(1+z)^{\beta+n}\Bigr].
\end{align}
They are orthogonal polynomials with respect to the following inner product:
\begin{align}
\int_{-1}^1 dz P_n^{(\alpha,\beta)}(z)  P_m^{(\alpha,\beta)}(z) (1-z)^{\alpha}(1+z)^{\beta} 
= \frac{2^{\alpha+\beta+1}}{2n+\alpha+\beta+1}
\frac{\Gamma(n+\alpha+1)\Gamma(n+\beta+1)}{n! \Gamma(n+\alpha+\beta+1)} \delta_{n,m}.
\end{align}

\subsection{Connection with other polynomials \label{secJacobiPConnectionWithOtherPol}}

The Jacobi polynomials are related to other famous polynomials. 
The Gegenbauer polynomials, $C_n^{(\lambda)}(z)$, and 
the Legendre polynomials, $P_n(z)$, are 
\begin{align}
P_n^{(\lambda -1/2,\lambda-1/2)}(z) = \frac{(\lambda+1/2)_n}{(2\lambda)_n}\, C_n^{(\lambda)} (z), 
\qquad P_n^{(0,0)}(z)=P_n(z). 
\end{align}
The Chebyshev polynomials of the first kind, $T_n(z)$, and the second kind, $U_n(z)$, are 
\begin{align}
P_n^{(-1/2,-1/2)}(z) = \frac{(2n)!}{2^{2n} (n!)^2}\, T_n(z), \qquad
P_n^{(1/2,1/2)}(z) = \frac{(2n+1)!}{2^{2n+1} [(n+1)!]^2}\, U_n(z).
\end{align}
They are also called the ultraspherical polynomials $P^{(\alpha,\alpha)}(z)$. 
The twisted type of ultraspherical polynomials, $P^{(\alpha,-\alpha)}(z)$ \cite{Cheb34}, can 
include the Chebyshev polynomials of the third kind, $V_n(z)$, and the forth kind, $W_n(z)$:
\begin{align}
P_n^{(-1/2,1/2)}(z) = \frac{(2n)!}{2^{2n} (n!)^2}\, V_n(z),\qquad
P_n^{(1/2,-1/2)}(z) = \frac{(2n)!}{2^{2n} (n!)^2}\, W_n(z). 
\end{align}
So we can see that $W_n(z)=(-1)^n V_n(-z)$. 

\subsection{Useful relations}
Some useful relations are in order: 

\  \\
\noindent 
\underline{Reflection relation:}
\begin{align}
P^{(\alpha,\beta)}_n(z) = (-1)^n P^{(\beta,\alpha)}_n(-z). \label{ExchangeAB}
\end{align}
\underline{The normalization:}
\begin{align}
P_n^{(\alpha,\beta)}(1)= \binom{n+\alpha}{n},\qquad 
P_n^{(\alpha,\beta)}(-1)= (-1)^{n}\binom{n+\beta}{n}. \label{JacobiNormalization1}
\end{align}
\underline{The leading coefficient:}
\begin{align}
P_n^{(\alpha,\beta)}(z) = \frac{\Gamma(2n+\alpha+\beta+1)}{2^n n! \Gamma(n+\alpha+\beta+1)} z^n+\cdots. \label{JacobiNormalization2}
\end{align}
\underline{Derivative formula:}
\begin{align}
\frac{d}{dz} P^{(\alpha,\beta)}_n(z) = \frac{n+\alpha+\beta+1}{2} P_{n-1}^{(\alpha+1,\beta+1)}(z). 
\end{align}

\subsection{Proof of the solution \eq{CISYsol} \label{JacobiPProof}}

Here we check that our formula \eq{CISYsol}
\begin{align}
\pi_{\hat p}(z) 
&= P^{(\frac{2l-k}{k},-\frac{2l-k}{k})}_{\hat p-1}(z) \, 
\sqrt[k]{\bigl(z-1\bigr)^l\bigl(z+1\bigr)^{k-l}},  \nn\\
\xi_{\hat q}(z)
&= P^{(-\frac{2l-k}{k},\frac{2l-k}{k})}_{\hat q-1}(z) \, 
\sqrt[k]{\bigl(z-1\bigr)^{k-l}\bigl(z+1\bigr)^{l}}, 
\end{align}
generally satisfies eq.~\eq{DKKeq}
\begin{align}
\hat q\,\pi'_{\hat p}(z)\, \xi_{\hat q}(z)-\hat p\,\xi'_{\hat q}(z)\,\pi_{\hat p}(z)
=\text{const.}, \label{DKKeq2}
\end{align}
in the case of $\hat p=\hat q$. 
For sake of simplicity, we neglect normalization factors. 
What we need to show is that the derivative of eq.~\eq{DKKeq2} vanishes, 
\begin{align}
\pi''_{\hat p}(z)\, \xi_{\hat p}(z)-\xi''_{\hat p}(z)\,\pi_{\hat p}(z)=0. \label{DerDKKeq}
\end{align}
We first calculate $\pi''_{\hat p}(z)\, \xi_{\hat p}(z)$ as follows:
\begin{align}
\pi''_{\hat p}(z)\, \xi_{\hat p}(z)
&= \xi_{\hat p}(z)
\biggl[
\del_z^2 P^{(\frac{2l-k}{k},-\frac{2l-k}{k})}_{\hat p-1}(z) \times
\sqrt[k]{\bigl(z-1\bigr)^l\bigl(z+1\bigr)^{k-l}}\,+\nn\\
&\qquad\quad\quad  + 
2\, \del_z P^{(\frac{2l-k}{k},-\frac{2l-k}{k})}_{\hat p-1}(z) \times
\del_z
\Bigl(\sqrt[k]{\bigl(z-1\bigr)^l\bigl(z+1\bigr)^{k-l}}\Bigr)+\nn\\[3pt]
&\qquad\quad\quad +
P^{(\frac{2l-k}{k},-\frac{2l-k}{k})}_{\hat p-1}(z) \times
\del_z^2\Bigl(\sqrt[k]{\bigl(z-1\bigr)^l\bigl(z+1\bigr)^{k-l}}\Bigr)
\biggr] \nn\\
&= P^{(-\frac{2l-k}{k},\frac{2l-k}{k})}_{\hat p-1}(z)
\biggl[
(z^2-1)\, \del_z^2 P^{(\frac{2l-k}{k},-\frac{2l-k}{k})}_{\hat p-1}(z) + \nn\\
&\qquad+
2\Bigl(z+\dfrac{l}{k}-\dfrac{k-l}{k}\Bigr)\,
\del_z P^{(\frac{2l-k}{k},-\frac{2l-k}{k})}_{\hat p-1}(z)
-
\dfrac{4\dfrac{l}{k}\Bigl(\dfrac{k-l}{k}\Bigr)}{z^2-1} 
P^{(\frac{2l-k}{k},-\frac{2l-k}{k})}_{\hat p-1}(z)
\biggr] \nn\\[3pt]
&=
\biggl[
-\dfrac{4\dfrac{l}{k}\Bigl(\dfrac{k-l}{k}\Bigr)}{z^2-1} +\hat p(\hat p-1)
\biggr]
P^{(-\frac{2l-k}{k},\frac{2l-k}{k})}_{\hat p-1}(z)\times P^{(\frac{2l-k}{k},-\frac{2l-k}{k})}_{\hat p-1}(z)
\end{align}
Here we use the following relations,
\begin{align}
\sqrt[k]{\bigl(z-1\bigr)^{k-l}\bigl(z+1\bigr)^{l}}\times
\sqrt[k]{\bigl(z-1\bigr)^l\bigl(z+1\bigr)^{k-l}}
&=z^2-1,  \\[5pt]
\sqrt[k]{\bigl(z-1\bigr)^{k-l}\bigl(z+1\bigr)^{l}}\,
\frac{\del}{\del z} 
\Bigl(\sqrt[k]{\bigl(z-1\bigr)^l\bigl(z+1\bigr)^{k-l}}\Bigr)
&=z+\dfrac{l}{k}-\dfrac{k-l}{k}, \\
\sqrt[k]{\bigl(z-1\bigr)^{k-l}\bigl(z+1\bigr)^{l}}\,
\frac{\del^2}{\del z^2} 
\Bigl(\sqrt[k]{\bigl(z-1\bigr)^l\bigl(z+1\bigr)^{k-l}}\Bigr)
&=-\dfrac{4\dfrac{l}{k}\Bigl(\dfrac{k-l}{k}\Bigr)}{z^2-1},
\end{align}
and the differential equation \eq{JacobiEq},
\begin{align}
\Bigl[(z^2-1) \frac{d^2}{dz^2}
+ 2\Bigl(z +\frac{l}{k}-\frac{k-l}{k}\Bigr) \frac{d}{dz}\Bigr]P^{(\frac{2l-k}{k},-\frac{2l-k}{k})}_{\hat p-1}(z) 
=\hat p(\hat p-1)\,P^{(\frac{2l-k}{k},-\frac{2l-k}{k})}_{\hat p-1}(z). 
\end{align}
Since the calculation of $\xi''_{\hat p}(z)\pi_{\hat p}(z)$ is similar to that of 
$\pi''_{\hat p}(z)\, \xi_{\hat p}(z)$ by exchanging 
the two indices:
\begin{align}
\frac{l}{k} \quad \leftrightarrow \quad \frac{k-l}{k},
\end{align}
and one concludes
\begin{align}
\xi''_{\hat p}(z)\,\pi_{\hat p}(z) = 
\pi''_{\hat p}(z)\, \xi_{\hat p}(z).
\end{align}
This is the equation \eq{DerDKKeq}. 
One can calculate the constant in eq.~\eq{DKKeq2} and fix
the normalization of eq.~\eq{CISYsol} by using 
eq's \eq{JacobiNormalization1} and \eq{JacobiNormalization2}.

\section{Topology of cubic roots \label{AppCubicRoots}}
\begin{figure}[htbp]
 \begin{center}
  \includegraphics[scale=1]{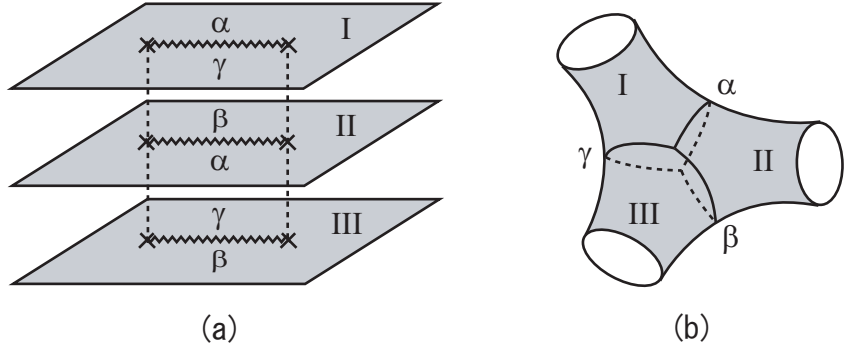}
 \end{center}
 \caption{\footnotesize 
The first type cubic cut of finite length. 
a) How to connect the three sheets, I, II and III. 
The symbols $\alpha,\beta,\gamma$ indicate the way of connection. 
b) After cutting the cuts out and gluing them, 
one can observe that the topological role of the first type cut, which is a junction of the three sheets. 
}
 \label{cubicroot1}
\end{figure}

Here we summarize basic properties of cubic roots of analytic functions and the topology of 
associated Riemann surface, 
which serves as the simplest prototype of general $k$-th root cases. 
In the case of cubic root of a single complex variable, 
\begin{align}
\sqrt[3]{z^l}\, \to\, \sqrt[3]{(e^{2\pi i}z)^l} = \omega^{l} \sqrt[3]{z^l} \qquad (l=0,1,2,\cdots),
\end{align}
we can define the charge $\rmq$ of the branch point $z=0$, according to the power of the fundamental 
phase $l$, as $\rmq\equiv  l \mod 3$. 
One can have branch cuts in the complex plane which connect branch points with vanishing 
total charges, $\rmq_1+\rmq_2+\cdots\equiv 0\mod 3$. For cubic roots of an analytic function, there are two types of branch cuts: 

(i) Type I branch cut connects two branch points of opposite charges. 
One concrete example is given by 
\begin{align}
f_{1}(z) = \sqrt[3]{(z-a)(z-b)^2}.
\end{align}
Here the branch point $z=a$ carries charge $\rmq=1$, and the branch point $z=b$ carries charge $\rmq=2$. 
The geometry of the associated Riemann surface is given in Fig.~\ref{cubicroot1}(a). 
By cutting the three complex planes along the branch cuts and gluing the six edges according to 
the identification rule, we can see clearly that the topology of the type I branch cut is given by a simple 
junction of the three sheets, as shown in Fig.~\ref{cubicroot1}(b). 

\begin{figure}[htbp]
 \begin{center}
  \includegraphics[scale=1]{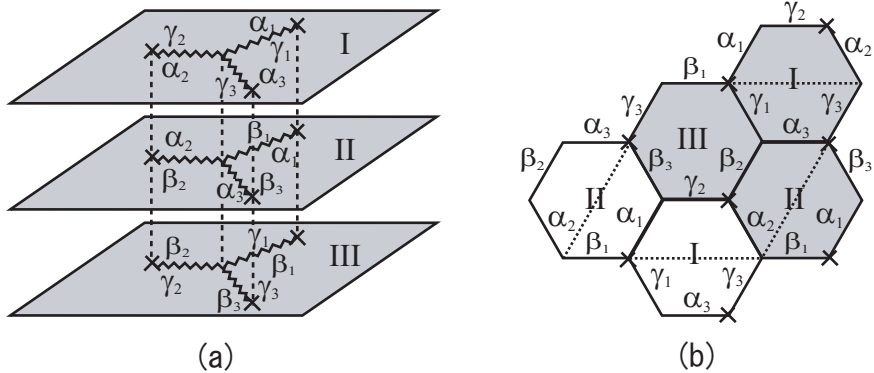}
 \end{center}
 \caption{\footnotesize 
The second type cubic cut of finite length. 
a) How to connect the three sheets, I, II and III. 
The symbols $\alpha_i,\beta_i,\gamma_i$ indicate the way of connection around the branch point $i\,(=1,2,3)$. 
b) Cut the cuts out and open them up. After gluing the sheets, one can observe that the topology is torus. That is, 
the topological role of the second type cut is a junction of the three sheets with a torus. 
}
 \label{cubicroot2}
\end{figure}

(ii) Type II branch cut connects three branch points of identical charges. 
One concrete example is given by
\begin{align}
f_{2}(z)= \sqrt[3]{(z-a)(z-a\omega)(z-a\omega^2)} = \sqrt[3]{z^3-a^3}
\end{align}
Here all three branch points $\zeta=a,a\omega,a\omega^2$ carry charge $\rmq=1$. 
The geometry of the associated Riemann surface is given in Fig.~\ref{cubicroot2}(a). 
By cutting the three complex planes along the branch cuts and flipping the resulting 
complex plane into a hexagon (including the point at infinity), 
one can transform the identification rule in Fig.~\ref{cubicroot2}(a) for three complex planes into
a gluing rule for three hexagons. Hence, the topology of the type II branch cut is the same as that of 
a torus (Fig.~\ref{cubicroot2}(b)). 

\begin{thebibliography}{99}

\bibitem{Polyakov}
  A.~M.~Polyakov,
   ``Quantum geometry of bosonic strings,''
  Phys.\ Lett.\ B {\bf 103} (1981) 207; 
   ``Quantum geometry of fermionic strings,''
  Phys.\ Lett.\ B {\bf 103} (1981) 211.

\bibitem{KPZ}
  V.~G.~Knizhnik, A.~M.~Polyakov and A.~B.~Zamolodchikov,
   ``Fractal structure of 2d-quantum gravity,''
  Mod.\ Phys.\ Lett.\ A {\bf 3} (1988) 819.

\bibitem{DDK}
  F.~David,
   ``Conformal field theories coupled to 2-D gravity in the conformal gauge,''
  Mod.\ Phys.\ Lett.\ A {\bf 3} (1988) 1651;\\
  J.~Distler and H.~Kawai,
   ``Conformal field theory and 2-D quantum gravity, 
    or who's afraid of Joseph Liouville?,''
  Nucl.\ Phys.\ B {\bf 321} (1989) 509.

\bibitem{DSL}
  E.~Brezin and V.~A.~Kazakov,
   ``Exactly solvable field theories of closed strings,''
  Phys.\ Lett.\  B {\bf 236} (1990) 144;\\
  M.~R.~Douglas and S.~H.~Shenker,
  Nucl.\ Phys.\  B {\bf 335} (1990) 635;\\
  D.~J.~Gross and A.~A.~Migdal,
   ``Nonperturbative Two-Dimensional Quantum Gravity,''
  Phys.\ Rev.\ Lett.\  {\bf 64} (1990) 127.

\bibitem{TwoMatString}
  E.~Brezin, M.~R.~Douglas, V.~Kazakov and S.~H.~Shenker,
   ``The Ising model coupled to 2-d Gravity: A nonperturbative analysis,''
  Phys.\ Lett.\  B {\bf 237} (1990) 43;\\
  D.~J.~Gross and A.~A.~Migdal,
   ``Nonperturbative Solution of the Ising Model on a Random Surface,''
  Phys.\ Rev.\ Lett.\  {\bf 64} (1990) 717.

\bibitem{GrossMigdal2}
  D.~J.~Gross and A.~A.~Migdal,
   ``A nonperturbative treatment of two-dimensional quantum gravity,''
  Nucl.\ Phys.\  B {\bf 340} (1990) 333.

\bibitem{MatrixReloaded}
  J.~McGreevy and H.~L.~Verlinde,
  ``Strings from tachyons: The $c = 1$ matrix reloaded,''
  JHEP {\bf 0312} (2003) 054
  [arXiv:hep-th/0304224].

\bibitem{GaiottoRastelli}
  D.~Gaiotto and L.~Rastelli,
  ``A paradigm of open/closed duality: Liouville D-branes and the  Kontsevich
  model,''
  JHEP {\bf 0507} (2005) 053
  [arXiv:hep-th/0312196].


 \bibitem{GrossWitten}
   D.~J.~Gross and E.~Witten,
    ``Possible Third Order Phase Transition In The Large N Lattice Gauge
   Theory,''
   Phys.\ Rev.\  D {\bf 21} (1980) 446.

 \bibitem{PeShe}
   V.~Periwal and D.~Shevitz,
    ``Unitary matrix models as exactly solvable string theories,''
   Phys.\ Rev.\ Lett.\  {\bf 64} (1990) 1326;
    ``Exactly solvable unitary matrix models: multicritical potentials 
       and correlations,''
   Nucl.\ Phys.\ B {\bf 344} (1990) 731.

 \bibitem{DSS}
   M.~R.~Douglas, N.~Seiberg and S.~H.~Shenker,
    ``Flow and instability in quantum gravity,''
   Phys.\ Lett.\  B {\bf 244} (1990) 381.

 \bibitem{Nappi}
   C.~R.~Nappi,
    ``Painleve-II And Odd Polynomials,''
   Mod.\ Phys.\ Lett.\  A {\bf 5} (1990) 2773.

 \bibitem{CDM}
   C.~Crnkovic, M.~R.~Douglas and G.~W.~Moore,
    ``Loop equations and the topological phase of multi-cut matrix models,''
   Int.\ J.\ Mod.\ Phys.\ A {\bf 7} (1992) 7693
   [arXiv:hep-th/9108014].

\bibitem{HMPN}
  T.~J.~Hollowood, L.~Miramontes, A.~Pasquinucci and C.~Nappi,
   ``Hermitian Versus Anti-Hermitian One Matrix Models And Their Hierarchies,''
  Nucl.\ Phys.\  B {\bf 373} (1992) 247
  [arXiv:hep-th/9109046].

\bibitem{TT}
  T.~Takayanagi and N.~Toumbas,
  ``A matrix model dual of type 0B string theory in two dimensions,''
  JHEP {\bf 0307} (2003) 064
  [arXiv:hep-th/0307083].

\bibitem{NewHat}
  M.~R.~Douglas, I.~R.~Klebanov, D.~Kutasov, J.~M.~Maldacena, E.~J.~Martinec and N.~Seiberg,
  ``A new hat for the $c = 1$ matrix model,''
  arXiv:hep-th/0307195.
  
\bibitem{UniCom}
  I.~R.~Klebanov, J.~M.~Maldacena and N.~Seiberg,
   ``Unitary and complex matrix models as 1-d type 0 strings,''
  Commun.\ Math.\ Phys.\  {\bf 252} (2004) 275
  [arXiv:hep-th/0309168].
  
\bibitem{MultiCut}
  C.~Crnkovic and G.~W.~Moore,
   ``Multicritical multicut matrix models,''
  Phys.\ Lett.\  B {\bf 257} (1991) 322.
  
\bibitem{irie2}
  H.~Irie,
   ``Fractional supersymmetric Liouville theory and the multi-cut matrix models,''
  Nucl.\ Phys.\ B {\bf 819} (2009) 351 
  [arXiv:0902.1676 [hep-th]]. 

\bibitem{BIPZ}
  E.~Brezin, C.~Itzykson, G.~Parisi and J.~B.~Zuber,
  ``Planar Diagrams,''
  Commun.\ Math.\ Phys.\  {\bf 59} (1978) 35.

\bibitem{KazakovSeries}
  V.~A.~Kazakov,
  ``The Appearance of Matter Fields from Quantum Fluctuations of 2D Gravity,''
  Mod.\ Phys.\ Lett.\  A {\bf 4} (1989) 2125.

\bibitem{Kostov1}
  I.~K.~Kostov,
  ``Strings embedded in Dynkin diagrams,''
  Cargese 1990, Proceedings, Random surfaces and quantum gravity, pp.135-149. 

\bibitem{Kostov2}
  I.~K.~Kostov,
  ``Loop amplitudes for nonrational string theories,''
  Phys.\ Lett.\  B {\bf 266} (1991) 317.

\bibitem{Kostov3}
  I.~K.~Kostov,
  ``Strings with discrete target space,''
  Nucl.\ Phys.\  B {\bf 376} (1992) 539
  [arXiv:hep-th/9112059].

\bibitem{BDSS}
  T.~Banks, M.~R.~Douglas, N.~Seiberg and S.~H.~Shenker,
  ``Microscopic and macroscopic loops in nonperturbative two-dimensional gravity,''
  Phys.\ Lett.\  B {\bf 238} (1990) 279.

\bibitem{MSS}
  G.~W.~Moore, N.~Seiberg and M.~Staudacher,
  ``From loops to states in 2-D quantum gravity,''
  Nucl.\ Phys.\  B {\bf 362} (1991) 665.

\bibitem{DKK}
  J.~M.~Daul, V.~A.~Kazakov and I.~K.~Kostov,
   ``Rational theories of 2-D gravity from the two matrix model,''
  Nucl.\ Phys.\  B {\bf 409} (1993) 311
  [arXiv:hep-th/9303093].

\bibitem{fy3}
  M.~Fukuma and S.~Yahikozawa,
  ``Comments on D-instantons in c $<$ 1 strings,''
  Phys.\ Lett.\  B {\bf 460} (1999) 71
  [arXiv:hep-th/9902169].

\bibitem{MMSS}
  J.~M.~Maldacena, G.~W.~Moore, N.~Seiberg and D.~Shih,
   ``Exact vs. semiclassical target space of the minimal string,''
  JHEP {\bf 0410} (2004) 020
  [arXiv:hep-th/0408039].

\bibitem{DouglasGeneralizedKdV}
  M.~R.~Douglas,
   ``Strings in less than one-dimension and the generalized KdV hierarchies,''
  Phys.\ Lett.\  B {\bf 238} (1990) 176.

\bibitem{TadaYamaguchiDouglas}
  T.~Tada and M.~Yamaguchi,
   ``$P$ and $Q$ operator analysis for two matrix model,''
  Phys.\ Lett.\  B {\bf 250} (1990) 38; \\
  M.~R.~Douglas,
{\it  In *Cargese 1990, Proceedings, Random surfaces and quantum gravity* 77-83. (see HIGH ENERGY PHYSICS INDEX 30 (1992) No. 17911)}; \\
  T.~Tada,
   ``$(Q,P)$ Critical Point From Two Matrix Models,''
  Phys.\ Lett.\  B {\bf 259} (1991) 442.

\bibitem{fim}
  M.~Fukuma, H.~Irie and Y.~Matsuo,
   ``Notes on the algebraic curves in $(p,q)$ minimal string theory,''
  JHEP {\bf 0609} (2006) 075
  [arXiv:hep-th/0602274].

\bibitem{fi1}
  M.~Fukuma and H.~Irie,
   ``A string field theoretical description of $(p,q)$ minimal superstrings,''
  JHEP {\bf 0701} (2007) 037
  [arXiv:hep-th/0611045].

\bibitem{SeSh}
  N.~Seiberg and D.~Shih,
   ``Branes, rings and matrix models in minimal (super)string theory,''
  JHEP {\bf 0402} (2004) 021
  [arXiv:hep-th/0312170].


\bibitem{KazakovKostov}
  V.~A.~Kazakov and I.~K.~Kostov,
  ``Instantons in non-critical strings from the two-matrix model,''
  arXiv:hep-th/0403152.

\bibitem{SatoTsuchiya}
  A.~Sato and A.~Tsuchiya,
  ``ZZ brane amplitudes from matrix models,''
  JHEP {\bf 0502} (2005) 032
  [arXiv:hep-th/0412201].

\bibitem{SeSh2}
  N.~Seiberg and D.~Shih,
   ``Flux vacua and branes of the minimal superstring,''
  JHEP {\bf 0501} (2005) 055
  [arXiv:hep-th/0412315].

\bibitem{fis}
  M.~Fukuma, H.~Irie and S.~Seki,
  ``Comments on the D-instanton calculus in $(p,p+1)$ minimal string theory,''
  Nucl.\ Phys.\  B {\bf 728} (2005) 67
  [arXiv:hep-th/0505253].
  
\bibitem{iky}
  N.~Ishibashi, T.~Kuroki and A.~Yamaguchi,
  ``Universality of nonperturbative effects in $c < 1$ noncritical string theory,''
  JHEP {\bf 0509} (2005) 043
  [arXiv:hep-th/0507263].

\bibitem{David}
  F.~David,
  ``Phases of the large N matrix model and nonperturbative effects in 2-d gravity,''
  Nucl.\ Phys.\  B {\bf 348} (1991) 507;
  ``Nonperturbative effects in matrix models and vacua of two-dimensional
  Phys.\ Lett.\  B {\bf 302} (1993) 403
  [arXiv:hep-th/9212106].

\bibitem{GinspargZinnJustin}
  P.~H.~Ginsparg and J.~Zinn-Justin,
  ``Action principle and large order behavior of nonperturbative gravity,''

\bibitem{EynardZinnJustin}
  B.~Eynard and J.~Zinn-Justin,
  ``Large order behavior of 2-D gravity coupled to $d < 1$ matter,''
  Phys.\ Lett.\  B {\bf 302} (1993) 396
  [arXiv:hep-th/9301004].

\bibitem{fy12}
  M.~Fukuma and S.~Yahikozawa,
  ``Nonperturbative effects in noncritical strings with soliton 
  backgrounds,''
  Phys.\ Lett.\ B {\bf 396} (1997) 97
  [arXiv:hep-th/9609210];
  ``Combinatorics of solitons in noncritical string theory,''
  Phys.\ Lett.\ B {\bf 393} (1997) 316
  [arXiv:hep-th/9610199];
  %


\bibitem{FZZT}
V.~Fateev, A.~B.~Zamolodchikov and Al.~B.~Zamolodchikov,
``Boundary Liouville field theory. I: 
Boundary state and boundary two-point function,''
arXiv:hep-th/0001012;\\
%
J.~Teschner,
``Remarks on Liouville theory with boundary,''
arXiv:hep-th/0009138.

\bibitem{ZZ}
  A.~B.~Zamolodchikov and Al.~B.~Zamolodchikov,
  ``Liouville field theory on a pseudosphere,''
  arXiv:hep-th/0101152.

\bibitem{Mehta}
  M.~L.~Mehta,
  ``A Method Of Integration Over Matrix Variables,''
  Commun.\ Math.\ Phys.\  {\bf 79}, 327 (1981).



\bibitem{Martinec}
  E.~J.~Martinec,
   ``On the origin of integrability in matrix models,''
  Commun.\ Math.\ Phys.\  {\bf 138} (1991) 437.

\bibitem{fkn}
  M.~Fukuma, H.~Kawai and R.~Nakayama,
  ``Continuum Schwinger-Dyson equations and universal structures 
   in two-dimensional quantum gravity,''
  Int.\ J.\ Mod.\ Phys.\ A {\bf 6} (1991) 1385;
  ``Infinite dimensional Grassmannian structure 
   of two-dimensional quantum gravity,''
  Commun.\ Math.\ Phys.\  {\bf 143} (1992) 371;
  ``Explicit solution for $p$--$q$ duality
   in two-dimensional quantum gravity,''
  Commun.\ Math.\ Phys.\  {\bf 148} (1992) 101.

\bibitem{monte}
  N.~Kawahara, J.~Nishimura and A.~Yamaguchi,
  ``Monte Carlo approach to nonperturbative strings -- demonstration in
  noncritical string theory,''
  JHEP {\bf 0706} (2007) 076
  [arXiv:hep-th/0703209].

\bibitem{Okuyama}
  K.~Okuyama,
  ``Annulus amplitudes in the minimal superstring,''
  JHEP {\bf 0504} (2005) 002
  [arXiv:hep-th/0503082]. 

\bibitem{Irie1}
  H.~Irie,
  ``Notes on D-branes and dualities in $(p,q)$ minimal superstring theory,''
Nucl.\ Phys.\  B {\bf 794} [PM] (2008) 402
  [arXiv:0706.4471 hep-th].

\bibitem{Moore}
  G.~W.~Moore,
  ``Geometry Of The String Equations,''
  Commun.\ Math.\ Phys.\  {\bf 133} (1990) 261;
  ``Matrix Models Of 2-D Gravity And Isomonodromic Deformation,''
  Prog.\ Theor.\ Phys.\ Suppl.\  {\bf 102} (1990) 255.

\bibitem{kcKP}
M.~Sato, 
RIMS Kokyuroku {\bf 439} (1981) 30; \\
%
  E.~Date, M.~Jimbo, M.~Kashiwara and T.~Miwa,
   ``Transformation groups for soliton equations. 3. 
     Operator approach to the Kadomtsev-Petviashvili equation,''
RIMS-358; \\
  M.~Jimbo and T.~Miwa,
  ``Solitons and infinite dimensional Lie algebras,''
  Publ.\ Res.\ Inst.\ Math.\ Sci.\ Kyoto {\bf 19} (1983) 943; \\
  V.~G.~Kac and J.~W.~van de Leur,
  ``The $n$-component KP hierarchy and representation theory,''
  J.\ Math.\ Phys.\  {\bf 44} (2003) 3245
  [arXiv:hep-th/9308137].

\bibitem{SolDouglas}
  I.~Krichever,
  ``The dispersionless Lax equations and topological minimal models,''
  Commun.\ Math.\ Phys.\  {\bf 143} (1992) 415.

\bibitem{Jacobi}
  G.~E.~Andrews, R.~Askey and R.~Roy,
  ``Special Functions,
  Encyclopedia of Mathematics and its Applications 71,''
  Cambridge University Press (1999) 684 p. 

\bibitem{Cheb34}
 K.~Aghigh, M.~Masjed-Jamei and M.~Dehghan,
  ``A survey on third and fourth kind of Chebyshev polynomials and their applications,''
  Applied Mathematics and Computation Vol.199 (2008) 2. 

  

  






  
  






\end{thebibliography}
\end{document}